\documentclass[oldversion]{aa}
\usepackage{graphicx}
\usepackage{txfonts}
\usepackage{natbib}
\usepackage{amssymb}
\usepackage{url}
\newfont{\gwpfont}{cmssq8 scaled 1000}
\newcommand{\rexcess}{{\gwpfont REXCESS}}
\begin{document}
%
\def\etal{et al.}

\def\Mgv{M_{\rm g,500}}
\def\Mg{M_{\rm g}}
\def\YX {Y_{\rm X}}
\def\TX {T_{\rm X}}
\def\fgv {f_{\rm g,500}}
\def\fg  {f_{\rm g}}
\def\kT {{\rm k}T}
\def\Mv {M_{\rm 500}}
\def \Rv {R_{500}}
\def\keV {\rm keV}

\def\MT {$M_{500}$--$T_{\rm X}$}
\def\MY {$M_{500}$--$Y_{\rm X}$}
\def\MMg {$M_{500}$--$M_{\rm g,500}$}
\def\MgT {$M_{\rm g,500}$--$T_{\rm X}$}
\def\MgY {$M_{\rm g,500}$--$Y_{\rm X}$}
\def\LxT {$L$--$T$}
\def\LxYx {$L-Y_X$}
\def\YxM {$Y_X-\Mv$}
\def\LxM {$L-\Mv$}
\def\fgas {$f_{\rm gas}$}

\def\msol {{\rm M_{\odot}}}

\def\lesssim{\mathrel{\hbox{\rlap{\hbox{\lower4pt\hbox{$\sim$}}}\hbox{$<$}}}}
\def\gtrsim{\mathrel{\hbox{\rlap{\hbox{\lower4pt\hbox{$\sim$}}}\hbox{$>$}}}}

\newcommand{\propsim}{\lower 3pt \hbox{$\, \buildrel {\textstyle
       \propto}\over {\textstyle \sim}\,$}}

\def \xmm {\hbox{\it XMM-Newton}}
\def \chandra {\hbox{\it Chandra }}

\title{Galaxy cluster X-ray luminosity scaling relations from a representative local sample (REXCESS)}   
\author{G.W. Pratt \inst{1}, J.H. Croston \inst{2}, M. Arnaud
  \inst{3} and H. B\"ohringer \inst{1} }
\offprints{G.W. Pratt, \email{gwp@mpe.mpg.de}}
\authorrunning{G.W. Pratt et al.}
\titlerunning{X-ray luminosity scaling relations of the \rexcess}
 \institute{
 $^1$ Max-Planck-Institut f\"ur extraterrestriche Physik, Giessenbachstra{\ss}e, 85748 Garching, Germany \\
 $^2$ Centre for Astrophysics Research, Science and Technology Research Institute, University of
 Hertfordshire, College Lane, Hatfield AL10 9AB, UK \\
 $^3$ Laboratoire AIM, DAPNIA/Service d'Astrophysique - CEA/DSM - CNRS
 - Universit\'{e} Paris Diderot, B\^{a}t. 709, CEA-Saclay, F-91191
 Gif-sur- Yvette Cedex, France \\ 
 }

  \date{Received 22 Sept 2008; accepted 20 Feb 2009}
  \abstract 
   {We examine the X-ray luminosity scaling relations of 31 nearby galaxy clusters from the Representative {\it
       XMM-Newton\/} 
     Cluster Structure Survey (\rexcess). The objects are selected only in X-ray
     luminosity, optimally sampling the cluster luminosity
     function. Temperatures range from 2 to 9 keV, and there is no bias toward any particular morphological type. To reduce measurement
     scatter we extract pertinent values in
     an aperture corresponding to $\Rv$, estimated using the tight correlation between $Y_X$ (the product of gas mass and temperature) and total mass. The data exhibit power law
     relations between bolometric X-ray luminosity and temperature, $Y_X$ and total mass, all with slopes that are significantly steeper than self-similar
     expectations. We examine the possible causes for the steepening, finding that structural variations have little effect and that the primary driver appears to be a systematic variation of the gas content with mass.
     Scatter about the relations is dominated in all cases by
     the presence of cool cores. The natural logarithmic scatter about the raw X-ray luminosity-temperature relation is about 70 per cent, and about the X-ray luminosity-$Y_X$ relation it is 40 per cent. Systems with more morphological substructure show similar scatter about scaling relations than clusters with less substructure, due to the preponderance of cool core systems in the regular cluster subsample. Cool core and morphologically disturbed systems occupy distinct regions in the residual space with respect to the best fitting mean relation, the former lying systematically at the high luminosity side, the latter lying systematically at the low luminosity side. Simple exclusion of the central regions serves to reduce the scatter about the scaling relations by more than a factor of two. The scatter reduces by a similar amount
with the use of the central gas density as a third parameter. Using $Y_X$ as a total mass proxy, we derive a Malmquist bias-corrected local luminosity-mass relation and compare with other recent determinations. Our results indicate that luminosity can be a reliable mass proxy with controllable scatter, which has important implications for upcoming all-sky cluster surveys, such as those to be undertaken with {\it Planck} and {\it eROSITA}, and ultimately for the use of the cluster population for cosmological purposes. }    
 \keywords{Cosmology: observations,  Cosmology: dark
      matter, Galaxies: cluster: general, (Galaxies) Intergalactic  
medium, X-rays: galaxies: clusters}

   \maketitle
%

\section{Introduction}

The X-ray luminosity is an observationally attractive quantity because
of the relative ease with which it can be measured, and thus it is a key parameter for cosmological applications of the galaxy cluster population. 
For a fully virialised cluster formed through pure
gravitational collapse, the X-ray luminosity $L$ is determined
solely by the mass and distribution of gas in the intracluster medium (ICM), and the
X-ray temperature $T$ is determined by the depth of the potential well
in which the ICM rests. Correlations between these two basic
quantities were found in the very early days of X-ray observations of
clusters, even while the thermal nature of the emission was still under
debate \citep{mitchell77,mushotzky78,ht79}. Initial results from these works suggested that the slope of the luminosity temperature relation was steeper than expected from gravitational collapse alone. The launch of the {\it
  EXOSAT} and {\it Einstein} observatories enabled the first
systematic studies of large samples of clusters
\citep{es91,david93}, and the subsequent launch of {\it ROSAT, ASCA} and
{\it Ginga} allowed further investigation with increasingly better
quality data. 

The density squared ($n_e^2$) dependence of the X-ray emission means that
luminosity measurements are very sensitive to the exact physics of the
gas near the cluster core. Mechanisms such as rapid radiative
cooling or merging can change the thermodynamic state of this core
gas, introducing scatter into the various luminosity scaling
relations. Since our knowledge of the absolute extent of the
scatter limits the constraints that can be put on cosmological models
with the cluster population, the magnitude of the scatter, its source(s),
and how to correct for it constitute some of the most important
open issues in the study of clusters \citep[see e.g.][]{lh05}. {\citet{fabian94} were the first to note that the offset of a cluster from the mean relation was connected to the presence of a cool
core, motivating examination of methods to correct for this
effect. \citet{mark98} derived quantities corrected for the presence
of cooling cores by exclusion of the emission from the central region
and the introduction of a second spectral component in the temperature estimation; \citet{ae99} determined the luminosity temperature relation
using clusters specifically chosen to have weak or non-existent cool
cores. More recent efforts have aimed at reducing scatter by using the peak surface brightness as a third parameter \citep{ohara06}. At the same time it had long been suspected that merging events also contributed to scatter about the mean relation. This effect has been  investigated with increasingly sophisticated numerical simulations, which have shown that while major mergers can indeed boost both luminosity and temperature, the boosting appears to be short lived and the net movement in the luminosity
temperature plane is approximately parallel to the mean relation
\citep{rs01,rt02, hart07}. 

In the
present paper we re-investigate the luminosity scaling relations with
\rexcess\ \citep{boehringer07}, a sample of 33 local ($z < 0.2$)
clusters drawn from the REFLEX catalogue \citep{reflex}, all of which
have been observed with a single satellite, \xmm, with the aim of minimising effects due to instrumental cross calibration uncertainties. The unique sample
selection strategy, in which clusters have been
selected by luminosity only, in such a 
way as to have close to homogeneous coverage of luminosity space,
delivers an optimal sampling of the luminosity function of the
cluster population with no bias towards any particular morphological type. Moreover, distances were optimised so that
the angular scale of the objects is such that $\sim R_{500}$ falls
well within the \xmm\ field of view, allowing detailed local background
modelling to be undertaken, increasing the precision of measurements
at large radii as compared to more nearby clusters which often fill
the field of view. Since the basic selection criterion is X-ray
luminosity, \rexcess\ should be representative of any local, unbiased
high-quality X-ray survey, of the type applicable to testing of
cosmological models. 

In the following, we first use the representative nature of the \rexcess\ sample to investigate the raw luminosity scaling relations, finding that the slopes are steeper than expected if the gas is heated purely by gravitational processes and that cooling cores are the dominant contributor to scatter about them. Dividing the data into subsamples, we investigate the effect of cool cores and morphological disturbance on a cluster's position with respect to the mean relation, finding that the former lie systematically at the high luminosity side, and the latter lie systematically at the low luminosity side. We then investigate two different methods to minimise scatter: simple exclusion of the central region, and use of the central gas density as a third parameter in scaling law fitting, finding that both methods result in a significant reduction in the dispersion about the best fitting relations. Lastly, we examine the physical causes of the steep slope of the luminosity scaling relations, concluding that variations of gas content with total mass are most likely the dominant reason why these are steeper than expected. Our Appendix details the \rexcess\ survey volume calculations and a first attempt at correcting for Malmquist bias in the luminosity-mass relation. 

\begin{table*}
\begin{center}
\caption{{\footnotesize Cluster properties. 
The luminosity is the bolometric
    [0.01-100] keV luminosity.  All quantities are
    calculated assuming $\Omega_M=0.3$, $\Omega_{\Lambda}=0.7$, and $h_0 =
    0.7$.}}\label  
{tab:tab1}
\centering
\begin{tabular}{l l l r l r l r r r c c}
\hline
\hline

\multicolumn{1}{c}{Cluster} & \multicolumn{1}{c}{$z$} &
\multicolumn{1}{c}{$T_1$} & \multicolumn{1}{c}{$L_1$} &
\multicolumn{1}{c}{$T_2$} & \multicolumn{1}{c}{$L_2$} &
\multicolumn{1}{c}{$T_3$} &
\multicolumn{1}{c}{$Y_X$} & \multicolumn{1}{c}{$\Rv$} &
\multicolumn{1}{c}{$R_{\rm det}$} &
\multicolumn{1}{c}{CC} &
\multicolumn{1}{c}{Disturbed} \\

\multicolumn{1}{c }{(1)} & \multicolumn{1}{c}{(2)} & 
\multicolumn{1}{c}{(3)} & \multicolumn{1}{c}{(4)} & 
\multicolumn{1}{c }{(5)} & \multicolumn{1}{c}{ (6)} & 
\multicolumn{1}{c }{(7)} &
\multicolumn{1}{c}{(8)} & 
\multicolumn{1}{c}{(9)} & \multicolumn{1}{c}{(10) } &
\multicolumn{1}{c}{(11) } & \multicolumn{1}{c}{(12) } 
\\

\hline
\\

RXC\,J0003.8+0203 & 0.0924 & $3.85_{-0.09}^{+0.09}$ & $ 1.88_{-0.01}^{+0.01}$ & $3.64_{-0.09}^{+0.09}$ & $ 1.16_{-0.01}^{+0.01}$ & $3.87_{-0.10}^{+0.10}$ & $ 7.69_{-0.26}^{+0.26}$ &  876.7 & 0.84 & \ldots & \ldots \\
RXC\,J0006.0-3443 & 0.1147 & $5.03_{-0.19}^{+0.19}$ & $ 4.13_{-0.05}^{+0.05}$ & $4.60_{-0.16}^{+0.21}$ & $ 3.18_{-0.05}^{+0.05}$ & $5.18_{-0.20}^{+0.20}$ & $22.74_{-1.21}^{+1.22}$ & 1059.3 & 0.93 & \ldots & \checkmark \\
RXC\,J0020.7-2542 & 0.1410 & $5.69_{-0.11}^{+0.11}$ & $ 6.52_{-0.04}^{+0.04}$ & $5.24_{-0.15}^{+0.15}$ & $ 4.07_{-0.04}^{+0.04}$ & $5.55_{-0.13}^{+0.13}$ & $22.41_{-0.63}^{+0.65}$ & 1045.3 & 1.07 & \ldots & \ldots\\
RXC\,J0049.4-2931 & 0.1084 & $3.09_{-0.10}^{+0.10}$ & $ 1.78_{-0.02}^{+0.02}$ & $2.79_{-0.11}^{+0.11}$ & $ 1.00_{-0.02}^{+0.02}$ & $3.03_{-0.12}^{+0.12}$ & $ 5.09_{-0.24}^{+0.25}$ &  807.8 & 0.93 & \ldots & \ldots \\
RXC\,J0145.0-5300 & 0.1168 & $5.53_{-0.13}^{+0.13}$ & $ 5.00_{-0.03}^{+0.03}$ & $5.51_{-0.16}^{+0.16}$ & $ 3.88_{-0.03}^{+0.03}$ & $5.63_{-0.14}^{+0.14}$ & $26.61_{-0.87}^{+0.89}$ & 1089.3 & 1.23 & \ldots & \checkmark \\
RXC\,J0211.4-4017 & 0.1008 & $2.07_{-0.00}^{+0.07}$ & $ 0.81_{-0.01}^{+0.01}$ & $2.02_{-0.06}^{+0.06}$ & $ 0.48_{-0.01}^{+0.01}$ & $2.07_{-0.05}^{+0.05}$ & $ 2.03_{-0.06}^{+0.06}$ &  685.0 & 1.33 & \ldots & \ldots \\
RXC\,J0225.1-2928 & 0.0604 & $2.47_{-0.06}^{+0.15}$ & $ 0.51_{-0.01}^{+0.01}$ & $2.61_{-0.16}^{+0.16}$ & $ 0.31_{-0.01}^{+0.01}$ & $2.67_{-0.13}^{+0.13}$ & $ 2.00_{-0.12}^{+0.12}$ &  693.9 & 0.91 & \ldots & \checkmark \\
RXC\,J0345.7-4112 & 0.0603 & $2.19_{-0.04}^{+0.04}$ & $ 0.77_{-0.01}^{+0.01}$ & $2.15_{-0.08}^{+0.08}$ & $ 0.37_{-0.01}^{+0.01}$ & $2.30_{-0.06}^{+0.09}$ & $ 1.91_{-0.06}^{+0.09}$ &  688.4 & 0.89 & \checkmark & \ldots \\
RXC\,J0547.6-3152 & 0.1483 & $6.02_{-0.11}^{+0.11}$ & $ 8.97_{-0.04}^{+0.04}$ & $5.68_{-0.11}^{+0.11}$ & $ 5.76_{-0.04}^{+0.04}$ & $6.06_{-0.14}^{+0.14}$ & $35.54_{-0.99}^{+1.02}$ & 1133.7 & 1.32 & \ldots & \ldots \\
RXC\,J0605.8-3518 & 0.1392 & $4.56_{-0.05}^{+0.05}$ & $ 9.54_{-0.04}^{+0.04}$ & $4.81_{-0.12}^{+0.12}$ & $ 4.26_{-0.04}^{+0.04}$ & $4.91_{-0.11}^{+0.11}$ & $22.39_{-0.63}^{+0.66}$ & 1045.9 & 1.17 & \checkmark & \ldots \\
RXC\,J0616.8-4748 & 0.1164 & $4.22_{-0.10}^{+0.10}$ & $ 2.38_{-0.02}^{+0.02}$ & $4.16_{-0.12}^{+0.12}$ & $ 1.88_{-0.02}^{+0.02}$ & $4.17_{-0.11}^{+0.11}$ & $11.81_{-0.41}^{+0.39}$ &  939.2 & 1.12 & \ldots & \checkmark \\
RXC\,J0645.4-5413 & 0.1644 & $6.95_{-0.13}^{+0.13}$ & $18.88_{-0.10}^{+0.10}$ & $6.97_{-0.19}^{+0.19}$ & $11.39_{-0.09}^{+0.09}$ & $7.27_{-0.18}^{+0.18}$ & $71.61_{-2.33}^{+2.35}$ & 1280.0 & 1.28 & \ldots & \ldots \\
RXC\,J0821.8+0112 & 0.0822 & $2.68_{-0.09}^{+0.09}$ & $ 0.77_{-0.01}^{+0.01}$ & $2.44_{-0.12}^{+0.12}$ & $ 0.54_{-0.01}^{+0.01}$ & $2.84_{-0.10}^{+0.10}$ & $ 3.34_{-0.15}^{+0.15}$ &  755.9 & 0.93 & \ldots & \ldots \\
RXC\,J0958.3-1103 & 0.1669 & $5.34_{-0.21}^{+0.21}$ & $11.56_{-0.15}^{+0.15}$ & $5.85_{-0.40}^{+0.45}$ & $ 5.25_{-0.16}^{+0.16}$ & $6.30_{-0.44}^{+0.50}$ & $28.04_{-2.30}^{+2.67}$ & 1077.4 & 0.78 & \checkmark & \ldots \\
RXC\,J1044.5-0704 & 0.1342 & $3.41_{-0.03}^{+0.03}$ & $ 7.42_{-0.02}^{+0.02}$ & $3.52_{-0.05}^{+0.05}$ & $ 3.00_{-0.02}^{+0.02}$ & $3.57_{-0.05}^{+0.05}$ & $11.77_{-0.19}^{+0.19}$ &  931.9 & 1.09 & \checkmark & \ldots \\
RXC\,J1141.4-1216 & 0.1195 & $3.31_{-0.03}^{+0.03}$ & $ 3.75_{-0.01}^{+0.01}$ & $3.40_{-0.06}^{+0.06}$ & $ 1.70_{-0.01}^{+0.01}$ & $3.54_{-0.05}^{+0.05}$ & $ 8.60_{-0.15}^{+0.16}$ &  885.2 & 1.25 & \checkmark & \ldots \\
RXC\,J1236.7-3354 & 0.0796 & $2.70_{-0.05}^{+0.05}$ & $ 1.03_{-0.01}^{+0.01}$ & $2.57_{-0.03}^{+0.11}$ & $ 0.61_{-0.01}^{+0.01}$ & $2.73_{-0.01}^{+0.09}$ & $ 3.27_{-0.02}^{+0.15}$ &  753.5 & 0.99 & \ldots & \ldots \\
RXC\,J1302.8-0230 & 0.0847 & $2.97_{-0.07}^{+0.06}$ & $ 1.38_{-0.01}^{+0.01}$ & $2.92_{-0.07}^{+0.09}$ & $ 0.83_{-0.01}^{+0.01}$ & $3.44_{-0.07}^{+0.07}$ & $ 6.07_{-0.18}^{+0.19}$ &  842.1 & 1.22 & \checkmark & \checkmark \\
RXC\,J1311.4-0120 & 0.1832 & $8.91_{-0.08}^{+0.08}$ & $36.06_{-0.08}^{+0.08}$ & $8.24_{-0.13}^{+0.13}$ & $15.13_{-0.07}^{+0.07}$ & $8.44_{-0.12}^{+0.12}$ & $88.18_{-1.50}^{+1.51}$ & 1319.2 & 1.31 & \checkmark & \ldots \\
RXC\,J1516.3+0005 & 0.1181 & $4.51_{-0.06}^{+0.06}$ & $ 4.12_{-0.02}^{+0.02}$ & $4.18_{-0.08}^{+0.08}$ & $ 2.77_{-0.02}^{+0.02}$ & $4.48_{-0.07}^{+0.07}$ & $15.81_{-0.31}^{+0.30}$ &  989.9 & 1.29 & \ldots & \ldots \\
RXC\,J1516.5-0056 & 0.1198 & $3.55_{-0.07}^{+0.07}$ & $ 2.31_{-0.02}^{+0.02}$ & $3.40_{-0.08}^{+0.08}$ & $ 1.77_{-0.02}^{+0.02}$ & $3.74_{-0.09}^{+0.10}$ & $11.08_{-0.36}^{+0.41}$ &  927.0 & 1.37 & \ldots & \checkmark \\
RXC\,J2014.8-2430 & 0.1538 & $4.78_{-0.05}^{+0.05}$ & $21.06_{-0.07}^{+0.07}$ & $5.63_{-0.11}^{+0.11}$ & $ 7.52_{-0.07}^{+0.07}$ & $5.73_{-0.10}^{+0.10}$ & $39.89_{-0.82}^{+0.78}$ & 1155.3 & 1.09 & \checkmark & \ldots \\
RXC\,J2023.0-2056 & 0.0564 & $2.71_{-0.09}^{+0.09}$ & $ 0.61_{-0.01}^{+0.01}$ & $2.46_{-0.12}^{+0.12}$ & $ 0.40_{-0.01}^{+0.01}$ & $2.72_{-0.09}^{+0.09}$ & $ 2.81_{-0.12}^{+0.13}$ &  739.5 & 0.86 & \ldots & \checkmark \\
RXC\,J2048.1-1750 & 0.1475 & $4.65_{-0.07}^{+0.13}$ & $ 5.13_{-0.03}^{+0.03}$ & $4.59_{-0.08}^{+0.08}$ & $ 4.40_{-0.03}^{+0.03}$ & $5.01_{-0.11}^{+0.11}$ & $26.91_{-0.80}^{+0.81}$ & 1078.0 & 1.48 & \ldots & \checkmark \\
RXC\,J2129.8-5048 & 0.0796 & $3.81_{-0.15}^{+0.15}$ & $ 1.46_{-0.02}^{+0.02}$ & $3.64_{-0.12}^{+0.16}$ & $ 1.19_{-0.02}^{+0.02}$ & $3.88_{-0.14}^{+0.14}$ & $ 8.67_{-0.41}^{+0.40}$ &  900.6 & 0.93 & \ldots & \checkmark \\
RXC\,J2149.1-3041 & 0.1184 & $3.26_{-0.04}^{+0.04}$ & $ 3.56_{-0.02}^{+0.02}$ & $3.40_{-0.08}^{+0.08}$ & $ 1.58_{-0.02}^{+0.02}$ & $3.50_{-0.07}^{+0.07}$ & $ 8.65_{-0.32}^{+0.32}$ &  886.6 & 1.26 & \checkmark & \ldots\\
RXC\,J2157.4-0747 & 0.0579 & $2.46_{-0.08}^{+0.08}$ & $ 0.45_{-0.01}^{+0.01}$ & $2.30_{-0.06}^{+0.10}$ & $ 0.37_{-0.01}^{+0.01}$ & $2.76_{-0.07}^{+0.07}$ & $ 3.07_{-0.11}^{+0.11}$ &  751.5 & 0.97 & \ldots & \checkmark \\
RXC\,J2217.7-3543 & 0.1486 & $4.86_{-0.09}^{+0.09}$ & $ 6.12_{-0.03}^{+0.03}$ & $4.45_{-0.09}^{+0.09}$ & $ 3.70_{-0.03}^{+0.03}$ & $4.65_{-0.08}^{+0.10}$ & $20.32_{-0.47}^{+0.54}$ & 1022.6 & 1.33 & \ldots & \ldots \\
RXC\,J2218.6-3853 & 0.1411 & $5.84_{-0.11}^{+0.11}$ & $ 9.43_{-0.06}^{+0.06}$ & $5.88_{-0.15}^{+0.20}$ & $ 5.60_{-0.06}^{+0.06}$ & $6.16_{-0.19}^{+0.19}$ & $34.36_{-1.33}^{+1.30}$ & 1130.1 & 1.04 & \ldots & \checkmark \\
RXC\,J2234.5-3744 & 0.1510 & $7.78_{-0.15}^{+0.15}$ & $19.15_{-0.11}^{+0.11}$ & $6.95_{-0.14}^{+0.14}$ & $12.36_{-0.10}^{+0.10}$ & $7.30_{-0.12}^{+0.12}$ & $70.43_{-1.54}^{+1.51}$ & 1283.2 & 1.15 & \ldots & \ldots \\
RXC\,J2319.6-7313 & 0.0984 & $2.22_{-0.03}^{+0.03}$ & $ 2.00_{-0.02}^{+0.02}$ & $2.48_{-0.08}^{+0.08}$ & $ 0.97_{-0.01}^{+0.01}$ & $2.52_{-0.07}^{+0.07}$ & $ 4.37_{-0.16}^{+0.16}$ &  788.7 & 1.11 &\checkmark & \checkmark \\

\\
\hline
\end{tabular}
\end{center}

Columns: (1) Cluster name; (2) $z$: cluster redshift; (3) $T_1$: spectroscopic  
temperature of the $R < \Rv$ region in keV; (4) $L_1$: luminosity in the $R < \Rv$ region in units of $10^{44}$ erg s$^{-1}$; (5) $T_2$: spectroscopic temperature in the $[0.15-1]\,\Rv$ region in keV; (6) $L_2$: luminosity in the $[0.15-1]\, \Rv$ region in units of $10^{44}$ erg s$^{-1}$; (7) $T_3$: spectroscopic temperature in the $[0.15-0.75]\,\Rv$ region in keV; (8) $Y_X$ in units of $10^{13}\, M_\odot\ {\rm keV}$; (9) $\Rv$ in kpc; (10) ratio of the
    detection radius of the surface brightness profile at $3\sigma$
    significance, $R_{\rm det}$, to $\Rv$; (11) systems classified as cool cores on the basis of central density vs. cooling time (see Sect.~\ref{sec:subsamples}); (12) systems classified as disturbed on the basis of the centre shift parameter $\langle w \rangle$ (see Sect.~\ref{sec:subsamples}).
    \end{table*}

We adopt a $\Lambda$CDM cosmology with $H_0=70$
km s$^{-1}$ Mpc$^{-1}$, $\Omega_M=0.3$ and $\Omega_\Lambda=0.7$, and all uncertainties are quoted at the 68 per cent confidence level. All logarithmic quantities are given to base $e$, and the quantity $L$ refers to the bolometric [0.01-100 keV] X-ray luminosity.


\section{Data analysis}


Full details of the sample,
including \xmm\ observation details, can be found in
\citet{boehringer07}. 
Two of the \rexcess\ clusters, RXC\,J0956.4-1004 (the Abell 901/902
supercluster) and J2157.4-0747 (a bimodal cluster),
display complex morphology and are excluded from the present
analysis. The basic characteristics of the clusters discussed in this
paper are given in Table~\ref{tab:tab1}.


\subsection{Scaling}

In order to estimate cluster quantities consistently, we define them
in terms of $R_{500}$, the radius at which the mean mass density is
500 times the critical density at the cluster
redshift\footnote{$M_{500} = 500 \rho_c (z) (4\pi/3) R_{500}^3$, where
$\rho_{\rm c}(z)= h^{2}(z) 3 H_0^2 / 8 \pi G$ and $h^{2}(z) = \Omega_{\rm
M}(1 + z)^3 + \Omega_\Lambda$.}. While $\Rv$ can be estimated from the total mass profile
derived under the assumption of hydrostatic equilibrium (HE), the
present sample contains clusters in a wide variety of dynamical states and consequently the HE assumption may not be valid in all cases \citep[see the discussion in][]{pratt07}. Instead we estimate $\Rv$
using $Y_X$ as a mass proxy. This quantity, defined as the product of
$M_{\rm g,500}$, the gas mass within $\Rv$, and the
spectroscopic temperature in the $[0.15-1]\,\Rv$ region, is the X-ray
analogue of the integrated SZ signal $Y_{SZ}$, and has been shown to
be a low scatter mass proxy in the numerical simulations of
\citet{kvn06} even in the presence of significant dynamical activity. Recent observational investigations using a variety of
cluster samples have demonstrated that $Y_X$ is indeed a low-scatter mass proxy
\citep{maughan07,app07}, and the theoretical results
have been verified in independent numerical simulations 
\citep{poole07,yang08}. We estimate $\Rv$ iteratively from the $\Mv$--$Y_X$
relation derived from \xmm\ observations of a sample of 10 nearby
morphologically relaxed local clusters by 
\citet{app07}, viz.,  

{\small
\begin{equation}
h(z)^{2/5}\Mv = 10^{14.556 \pm 0.015} \left[\frac{\YX}{2\times10^{14}\,{\msol}\,\keV}\right]^{0.548 \pm 0.027}\,{\rm h_{70}^{-1}\,\msol},
\end{equation}
}

\noindent which was derived using substantially similar methods to those described in this paper. The \rexcess\ gas density profiles from which $M_{\rm gas}$ is derived are discussed in \citet{croston08}. 

Note that there is an $\sim 8$ per cent normalisation offset of the observed relation when compared to the relation derived by \citet{nkv07} from numerical simulations. However, an iterative measurement of $\Rv$ from the simulated $\Mv$--$Y_X$ relation changes the values of the temperature and luminosity by less than 1.5 per cent on average, due to the steep drop of emission with radius. Simulations also suggest a $\pm 8$ per cent scatter about the $\Mv$--$Y_X$. Using randomisation assuming a 1.5 per cent relative change in the measured quantities due to this scatter, we have verified that the slopes and normalisations of the scaling laws do not change, and that the maximum change in the scatter about the relations is only 7 per cent.


\begin{table*}[]
\begin{center}
\caption{{\footnotesize Observed bolometric X-ray luminosity scaling relations. For each set of observables ($L,A$), we fitted a power law relation of the form 
$h(z)^n L = C\, (A/A_0)^{\alpha}$, with $A_0=5$ keV, $2\times10^{14}\,
M_{\odot}$ keV and $2\times10^{14}\,
M_{\odot}$, and $n= -1$, $-9/5$ and $-7/3$ for $T$, $Y_X$ and $M$, respectively. Results are given for the BCES (Y$|$X) and BCES orthogonal fitting methods (see Section~\ref{sec:fitting}). The intrinsic natural logarithmic scatter about
the best fitting relation in the ln-ln plane is also given for each fit. $^a$ Since $M$ is derived from $Y_X$, the values of the scatter in the $L-M$ relation are identical to those for the $L-Y_X$ relation. $^b$ Corrected for Malmquist bias (see Appendix~\ref{sec:surveylx}).
}}\label{tab:lxrel}
\begin{tabular}{l r l l | r l l }
\hline
\hline

\multicolumn{1}{l}{Relation } &
\multicolumn{6}{c}{Fitting Method} \\

\multicolumn{1}{c}{ } &
\multicolumn{3}{c|}{BCES (Y$|$X)} & \multicolumn{3}{c}{BCES Orthogonal} \\

\cline{2-7}

\multicolumn{1}{c }{ } & \multicolumn{1}{c}{$C\,(10^{44}$ erg s$^{-1}$)} & 
\multicolumn{1}{c}{$\alpha$} & \multicolumn{1}{c|}{$\sigma_{\rm ln\,L, intrinsic}$ } & \multicolumn{1}{c}{$C\,(10^{44}$ erg s$^{-1}$)} & 
\multicolumn{1}{c}{$\alpha$} & \multicolumn{1}{c}{$\sigma_{\rm ln\,L, intrinsic}$ } \\

\hline

\\
\multicolumn{1}{c}{} & \multicolumn{6}{c}{$R < \Rv$} \\

All \\
\cline{1-1}
$L_1$--$T_1$ & $6.07\pm0.58$ & $2.70\pm0.24$ & $0.663\pm0.116$ & 
				$7.13\pm1.03$ & $3.35\pm0.32$ & $0.733\pm0.135$ \\ 

$L_1$--$T_3$ & $5.62\pm0.46$ & $2.88\pm0.23$ & $0.525\pm0.097$ & 
				$6.27\pm0.67$ & $3.42\pm0.27$ & $0.560\pm0.115$ \\ 

$L_1$--$Y_X$ & $5.20\pm0.36$ & $0.99\pm0.05$ & $0.384\pm0.060$ & 
				$5.35\pm0.38$ & $1.04\pm0.06$ & $0.383\pm0.061$ \\  

$L_1$--$M_Y$ & $1.81\pm0.13$ & $1.81\pm0.10$ & $^a$\ldots &
				$1.74\pm0.13$ & $1.96\pm 0.11$ & $^a$\ldots \\

$L_1$--$M_Y$ MB$^b$ & $1.45\pm0.12$ & $1.90\pm0.11$ & $^a$\ldots &
				$1.38\pm0.12$ & $2.08\pm 0.13$ & $^a$\ldots \\

\\
Cool core \\
\cline{1-1}
$L_1$--$T_1$ & $11.15\pm2.42$ & $2.71\pm0.48$ & $0.432\pm0.108$ & 
				$12.79\pm3.80$ & $3.15\pm0.63$ & $0.479\pm0.135$ \\

$L_1$--$Y_X$ &  $7.71\pm0.58$ & $1.04\pm0.07$ & $0.234\pm0.103$ & 
				$7.84\pm0.65$ & $1.06\pm0.09$ & $0.236\pm0.107$ \\

\\
Non-cool core \\
\cline{1-1} 
$L_1$--$T_1$ &  $4.78\pm0.29$ & $2.89\pm0.21$ & $0.267\pm0.058$ & 
				$4.97\pm0.29$ & $3.06\pm0.19$ & $0.285\pm0.068$ \\

$L_1$--$Y_X$ &  $4.27\pm0.20$ & $0.96\pm0.05$ & $0.214\pm0.035$ & 
				$4.32\pm0.20$ & $0.98\pm0.05$ & $0.214\pm0.036$ \\

\\

Disturbed \\
\cline{1-1} 
$L_1$--$T_1$ &  $4.18\pm0.59$ & $2.49\pm0.56$ & $0.497\pm0.215$ & 
				$5.43\pm2.74$ & $3.19\pm0.78$ & $0.646\pm0.346$ \\

$L_1$--$Y_X$ &  $3.72\pm0.27$ & $0.92\pm0.09$ & $0.245\pm0.120$ & 
				$3.85\pm0.32$ & $0.96\pm0.08$ & $0.249\pm0.123$ \\

\\
Regular \\
\cline{1-1}
$L_1$--$T_1$ &  $7.26\pm0.86$ & $2.62\pm0.21$ & $0.578\pm0.118$ & 
				$7.97\pm1.28$ & $3.13\pm0.33$ & $0.634\pm0.142$ \\

$L_1$--$Y_X$ &  $6.15\pm0.42$ & $0.97\pm0.05$ & $0.302\pm0.058$ & 
				$6.21\pm0.44$ & $1.00\pm0.05$ & $0.303\pm0.059$ \\

\\
\hline
\\
\multicolumn{1}{c}{} & \multicolumn{6}{c} {$0.15 < R < \Rv$} \\

All \\
\cline{1-1}
$L_2$--$T_2$ & $3.89\pm0.18$ & $2.78\pm0.13$ & $0.269\pm0.055$ & 
				$4.06\pm0.22$ & $2.94\pm0.15$ & $0.279\pm0.059$  \\ 

$L_2$--$T_3$ & $3.31\pm0.16$ & $2.84\pm0.17$ & $0.331\pm0.068$ & 
				$3.48\pm0.21$ & $3.07\pm0.18$ & $0.346\pm0.075$ \\

$L_2$--$Y_X$ & $3.05\pm0.07$ & $0.97\pm0.03$ & $0.156\pm0.038$ & 
				$3.06\pm0.07$ & $0.98\pm0.03$ & $0.156\pm0.038$ \\ 

$L_2$--$M_Y$ & $1.09\pm 0.05$ & $1.77\pm0.05$ & $^a$\ldots & 
				$1.08\pm0.04$ & $1.80\pm 0.05$ & $^a$\ldots \\

\\
Cool core  \\
\cline{1-1}
$L_2$--$T_2$ & $4.31\pm0.42$ & $2.58\pm0.23$ & $0.242\pm0.110$ & 
				$4.46\pm0.56$ & $2.70\pm0.26$ & $0.247\pm0.113$ \\

$L_2$--$Y_X$ & $3.36\pm0.16$ & $0.96\pm0.04$ & $0.144\pm0.098$ & 
				$3.38\pm0.17$ & $0.97\pm0.05$ & $0.145\pm0.098$\\
\\

Non-cool core \\
\cline{1-1}
$L_2$--$T_2$ & $3.74\pm0.21$ & $2.89\pm0.18$ & $0.231\pm0.035$ & 
				$3.88\pm0.22$ & $3.02\pm0.19$ & $0.237\pm0.039$ \\

$L_2$--$Y_X$ & $2.91\pm0.06$ & $0.97\pm0.03$ & $0.114\pm0.027$ & 
				$2.92\pm0.06$ & $0.98\pm0.03$ & $0.114\pm0.027$  \\
\\

Disturbed  \\
\cline{1-1}
$L_2$--$T_2$ & $3.58\pm0.41$ & $2.88\pm0.37$ & $0.295\pm0.080$ & 
				$4.00\pm0.73$ & $3.18\pm0.38$ & $0.312\pm0.090$ \\

$L_2$--$Y_X$ & $2.77\pm0.07$ & $0.99\pm0.04$ & $0.111\pm0.096$ & 
				$2.79\pm0.08$ & $0.99\pm0.04$ & $0.111\pm0.096$ \\
\\

Regular  \\
\cline{1-1}
$L_2$--$T_2$ & $4.13\pm0.21$ & $2.68\pm0.11$ & $0.225\pm0.070$ & 
				$4.20\pm0.23$ & $2.76\pm0.11$ & $0.231\pm0.075$ \\

$L_2$--$Y_X$ & $3.24\pm0.08$ & $0.94\pm0.02$ & $0.115\pm0.045$ & 
				$3.24\pm0.08$ & $0.94\pm0.02$ & $0.115\pm0.045$  \\
\\
\hline
\end{tabular}
\end{center}
$L_1/T_1$: luminosity/temperature interior to $\Rv$; 
$L_2/T_2$: luminosity/temperature in the $[0.15-1]\,\Rv$ aperture; 
$T_3$: temperature in the $[0.15-0.75]\,\Rv$ aperture; 
$M_Y$: mass measured from the $M_{500}$--$Y_X$ relation of \citet{app07}. 

\end{table*}

\begin{figure*}[]
\begin{centering}
\includegraphics[scale=1.,angle=0,keepaspectratio,width=0.32\textwidth]{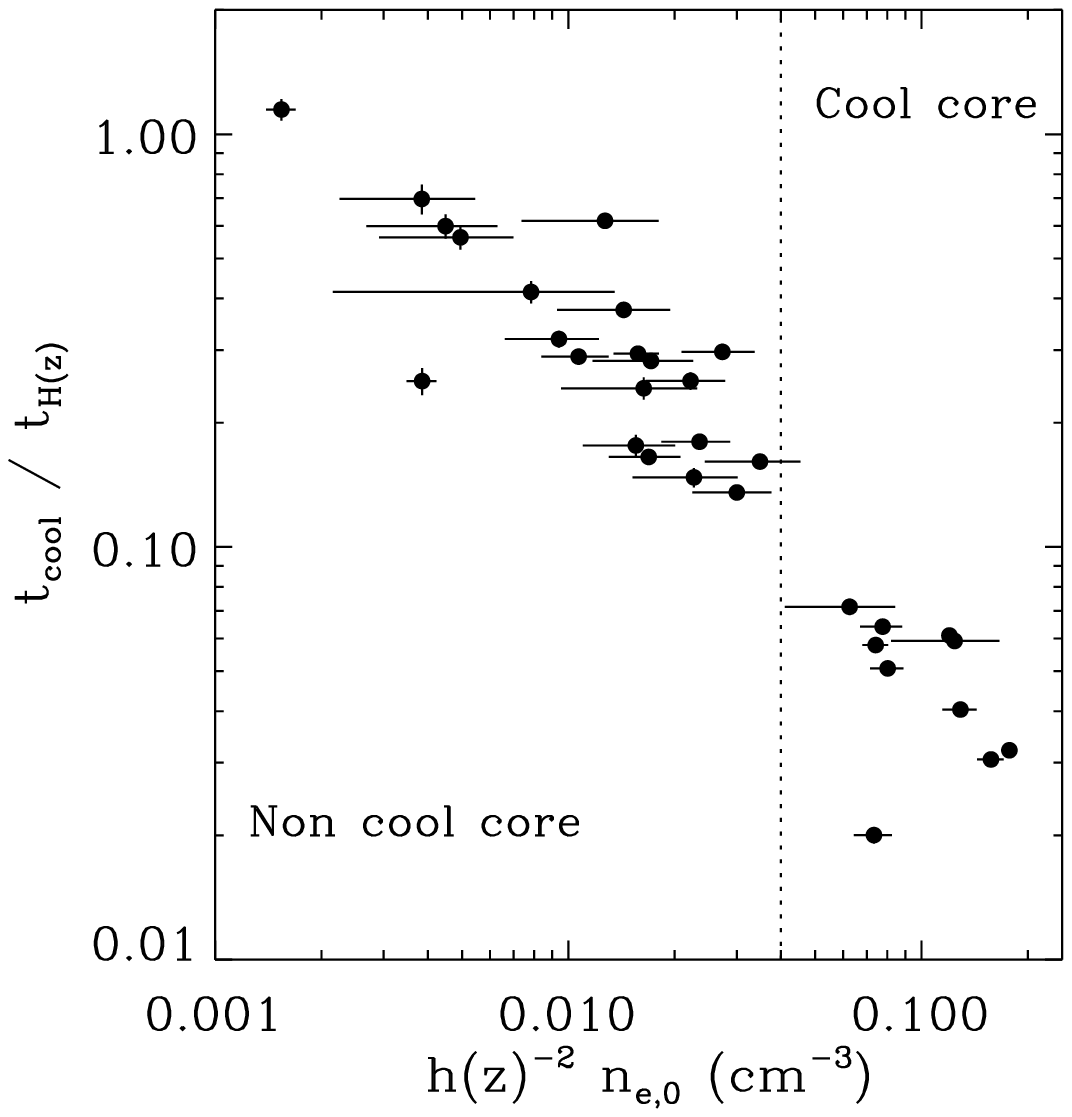}
\hfill
\includegraphics[scale=1.,angle=0,keepaspectratio,width=0.315\textwidth]{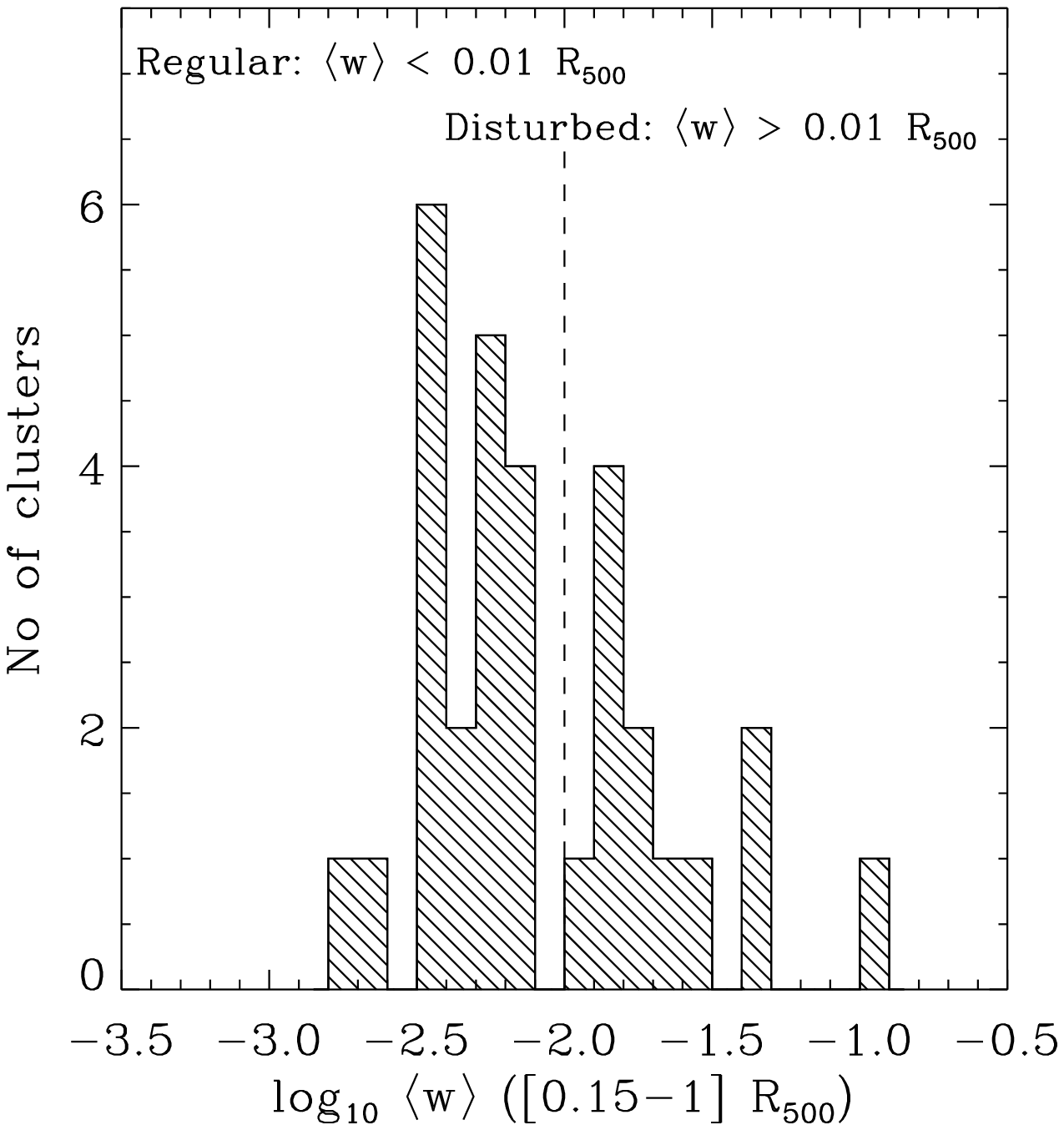}
\hfill
\includegraphics[scale=1.,angle=0,keepaspectratio,width=0.335\textwidth]{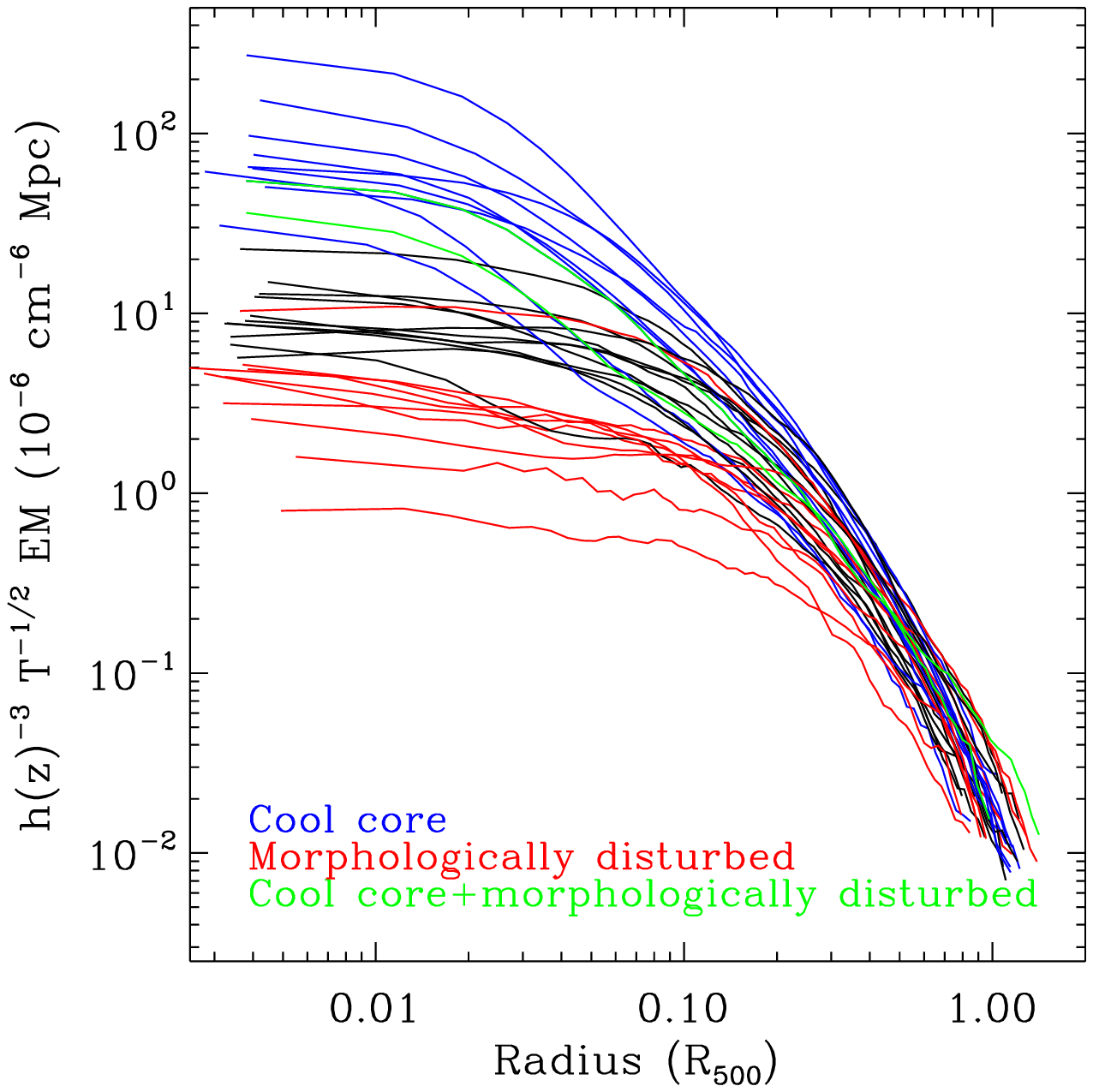}
\caption{{\footnotesize Definition of cluster subsamples. {\it Left:} Central cooling time vs. central gas density, $n_{e,0}$. The dotted line
    delineates the threshold we use to define cool core systems: $h(z)^{-2} n_{e,0} > 4 \times 10^{-2}$ cm$^{-3}$, $t_{\rm cool} < 10^9$ years. {\it Centre:} Histogram of centre shift parameter $\langle w \rangle$, evaluated in the [0.15-1]\,$\Rv$ aperture. Clusters with $\langle w \rangle > 0.01\,\Rv$ are classified as morphologically disturbed. {\it Right:} Emission measure profiles of the
  \rexcess\ sample, scaled according to the standard dependence on temperature and expected evolution with redshift. Systems classified as cool core and as morphologically disturbed are indicated (see Sect.~\ref{sec:subsamples}).}}\label{fig:tci0}   
\end{centering}
\end{figure*}


\subsection{Luminosities and temperatures}
\label{sec:lumin}

Bolometric X-ray luminosities, referred to as $L$ throughout the remainder of this paper, were derived for two apertures: (i) the entire
cluster emission interior to $\Rv$ (hereafter $L_1$) and (ii) in the [0.15-1]\,$\Rv$
aperture (hereafter $L_2$). We estimated the count rates from surface brightness
profiles in the [0.3-2] keV band, and used the best fitting spectral
model estimated in the same aperture to convert the count rate to
bolometric ([0.01-100] keV) luminosity. At $3\sigma$ significance, the
surface brightness profiles are detected out to at least $0.8\Rv$ for
all clusters. For 11 clusters, we extrapolated the surface brightness
profile using a power law with a slope measured from that of the data
at large radius. As can be seen in Table~\ref{tab:tab1}, in most cases
the need for extrapolation is minimal. Measured luminosities for both
apertures are given in Table~\ref{tab:tab1}. Errors take into account
statistical factors, uncertainties in $\Rv$ and extrapolation uncertainties. These were estimated from Monte Carlo realisations in which the above procedure, including extrapolation,
was repeated for 100 surface brightness profiles, the profiles and
$\Rv$ values each being randomised according to the observed
uncertainties. A PSF correction was implemented by using the gas density profile derived from regularised deprojection of the surface brightness as described in 
\cite{croston06}; the correction was obtained from the ratio of the observed to PSF-corrected count rates in each aperture. The correction is negligible in the full aperture but can be up to 13 per cent for strong cooling core systems in the core-excluded aperture. 

We have also calculated luminosities in the $[0.1-2.4]$ keV and $[0.5-2]$ keV bands for ease of comparison with previous soft X-ray survey results from {\it ROSAT}. Luminosities and their associated scaling relations are given in Appendix~\ref{sec:surveylx}.

Spectroscopic temperatures were measured in the $[0.15-0.75]\,\Rv$ aperture\footnote{Note that this aperture is more appropriate for comparison with distant clusters, which often have poor signal to noise in the outer regions.} (referred to hereafter as $T_3$) from iteration about the $M_{500}-Y_X$ relation of \citet[][]{app07}. Temperatures in the full aperture and in the  $[0.15-1]\,\Rv$ aperture (hereafter $T_1$ and $T_2$, respectively) were then calculated by re-extraction of spectra in the relevant regions. In all cases the spectra were fitted with a {\sc MeKaL} model with an absorption fixed at the HI value (excepting RXC\,J2014.8-2430, which was found to have a significantly higher absorption than that indicated from the HI value). The three EPIC cameras were fitted simultaneously in the [0.3-10] keV band, with the regions around the instrumental lines (1.4-1.6 keV for all cameras and 7.45-9 keV for the pn camera) excluded from the fit. Temperatures for all three apertures are listed in Table~\ref{tab:tab1}.


\subsection{Subsamples}
\label{sec:subsamples}

We further subdivide the sample to elucidate the effects of cool cores
and merger-related phenomena on the scaling relations.

\subsubsection{Cooling time classification}

In an approach similar to that used by \citet{ohara06}, we use gas density and cooling time profiles to classify cooling core systems. \citet{croston08} describe the gas density and cooling time profiles of the present sample, which are fully deprojected and PSF-corrected using the non-parametric method described in \citet{croston06}. We estimate the central gas density $n_{e,0}$ from a $\beta$ model fit to the deconvolved, deprojected gas density profiles interior to $0.03\,\Rv$. Figure~\ref{fig:tci0} shows the central cooling time versus $n_{e,0}$, which exhibits a strong correlation, as expected since the cooling time is derived from the gas density. We classify clusters according to their central gas density $n_{e,0}$, such that those with
$h(z)^{-2}\ n_{e,0} > 4 \times 10^{-2}$ cm$^{-3}$ (equivalent to those with a central $EM$ value $EM_{\rm cen} \gtrsim 20 \times 10^{-6}$ cm$^{-6}$ Mpc in the right hand panel of Fig.~\ref{fig:tci0}) are defined 
as cool core systems; 10/31 clusters are classified as such. Figure~\ref{fig:tci0} shows that these systems
have central cooling times $t_{\rm cool,0} < 10^{9}$ years. 

\subsubsection{Morphological classification}

The sample also contains clusters in a wide variety of dynamical
states \citep{boehringer07,pratt07}. To investigate the effect of
dynamical state on the relations, we have calculated values of centroid shift
$\langle w \rangle$, defined as the standard deviation of the
projected separations between the X-ray peak and the centroid at each
radius in the $[0.1 -1]\, \Rv$ region:

\begin{equation}
\langle w \rangle = \left[ \frac{1}{N-1} \sum \left(\Delta_i - \langle \Delta \rangle \right)^2 \right]^{1/2} \times \frac{1}{R_{500}}, 
\end{equation}

\noindent where $\Delta_i$ is the projected distance between the X-ray peak and centroid in the $i$\,th aperture.

Introduced by \citet{mfg93}, this quantity was found to be the most sensitive indicator of dynamical activity in the numerical simulations of \citet{poole07}. We calculate centroids in circular apertures of radii $n \times 0.1 \times \Rv$ with $n=2, 3\ldots10$, excluding the central regions to avoid biases associated with enhanced emission from cool cores (although exclusion of the central region does not have a significant effect on the results). The centroid shift $\langle w \rangle$ is then defined as the standard deviation of the projected separations between the X-ray surface brightness peak and the centroid in units of $\Rv$. A forthcoming paper will discuss these results in more detail. For the current analysis, the distribution of $\langle w \rangle$ for the present sample is shown in the central panel of Figure~\ref{fig:tci0}. We classify clusters with $\langle w \rangle > 0.01\,\Rv$ as morphologically disturbed, and
clusters with $\langle w \rangle < 0.01\,\Rv$ as morphologically
regular. In total, 12/31 clusters are defined as morphologically disturbed. 

The different subsample classifications are indicated in Table~\ref{tab:tab1} and illustrated in Figure~\ref{fig:tci0}. In general, the presence of a cool core is anti-correlated with indications for morphological disturbance. However, two clusters possess both a cool core and display evidence for morphological disturbance (RXC\,J1302.8 -0230 and RXC\,J2319.6 -7313). A gallery of the cool core and non-cool core systems, sorted by $\langle w \rangle$,  can be found in Figures~\ref{fig:gallery1} and ~\ref{fig:gallery2}, respectively, in Appendix~\ref{app:app1}.


\subsection{Fitting procedure}
\label{sec:fitting}

For each set of observables $(B,A)$, we fitted a power law relation of
the form $h(z)^n B = C (A/A_0)^\alpha$, where $h(z)$ is the Hubble
constant normalised to its present day value and $n$ was fixed to the
expected scaling with $z$. The fit was undertaken using linear
regression in the log-log plane, taking the uncertainties in both
variables into account. Assuming a linear relation of the form $Y= aX + b$, and a sample of $N$ data points ($Y_i,X_i$) with errors $\sigma_{Y_i}$ and $\sigma_{X_i}$, the raw scatter was estimated using the
error weighted orthogonal distances to the regression line:

\begin{equation}
\sigma^2_{\rm raw} = \frac{1}{N-2} \sum_{i=1}^{N} w_i\, (Y_i -aX_i -b)^2
\end{equation}

\noindent where

\begin{equation}
w_i = \frac{1/\sigma_i^2}{(1/N) \sum_{i=1}^N 1/\sigma_i^2}\ \ \ \ \ \ \ {\rm and}\ \ \ \ \ \ \sigma_i^2 = \sigma^2_{Y_i} + a^2 \sigma^2_{X_i}.
\end{equation}

\noindent  The intrinsic scatter was computed from the quadratic difference between the raw scatter and that expected from the statistical uncertainties. 

As Figures~\ref{fig:LxTraw}-~\ref{fig:YxLxcorr} show, the uncertainties in the present data set are entirely negligible compared to the intrinsic scatter, so that error weighting of individual data points will have no effect on the resulting fits. In the following we use the BCES regression method \citep{bces}, which takes into account measurement errors in both coordinates and intrinsic scatter in the data and is  widely used in astronomical regression, giving results that may easily be compared with other data sets fitted using the same method.

It is well-known that different regression methods give different slopes even at the population level \citep[e.g.,][]{isobe,bces}.  It is therefore of paramount importance to choose the regression method best suited to the data in hand. With the present data set, there is no easy answer to the question of which quantity to treat as the dependent variable and which to treat as the independent variable. In cosmological and theoretical applications, the mass of a cluster is its most fundamental property. Given the tight mass-temperature relation \citep[e.g.,][]{app05}, it is reasonable to assume that $T$ is closely coupled to the mass. However, as will be seen below, there is a large intrinsic scatter in $L$, presumably due to baryon physics. One possible minimisation method would thus treat $L$ as the dependent variable. A second possible minimisation method would be to assume that {\it both} variables are quasi-independent, and to treat them symmetrically. 

In the following, we thus give the results from the BCES (Y$|$X) fitting method, which minimises the residuals in $L$, and from the BCES orthogonal fitting method, which minimises the squared orthogonal distances. In the case of maximum scatter (the raw, uncorrected $L_1$--$T_1$ relation), the  BCES (Y$|$X) method typically gives slightly shallower slopes than the orthogonal BCES method\footnote{The BCES (Y$|$X) method gives precisely the same results as the modified weighted least squares (WLSS) method described in \citet{pap06}.}. As the scatter decreases, the various regression methods give results which agree very well within their $1\sigma$ uncertainties (Table~\ref{tab:lxrel}). Uncertainties on all fit parameters and associated scatter are determined from 10\,000 bootstrap resamples of the data. Since measurement errors are at the $1-3$ per cent level, we give only estimates of the intrinsic dispersion about the best fitting relations.


\section{Results}

\begin{figure*}[]
\centering
\includegraphics[width=0.67\textwidth]{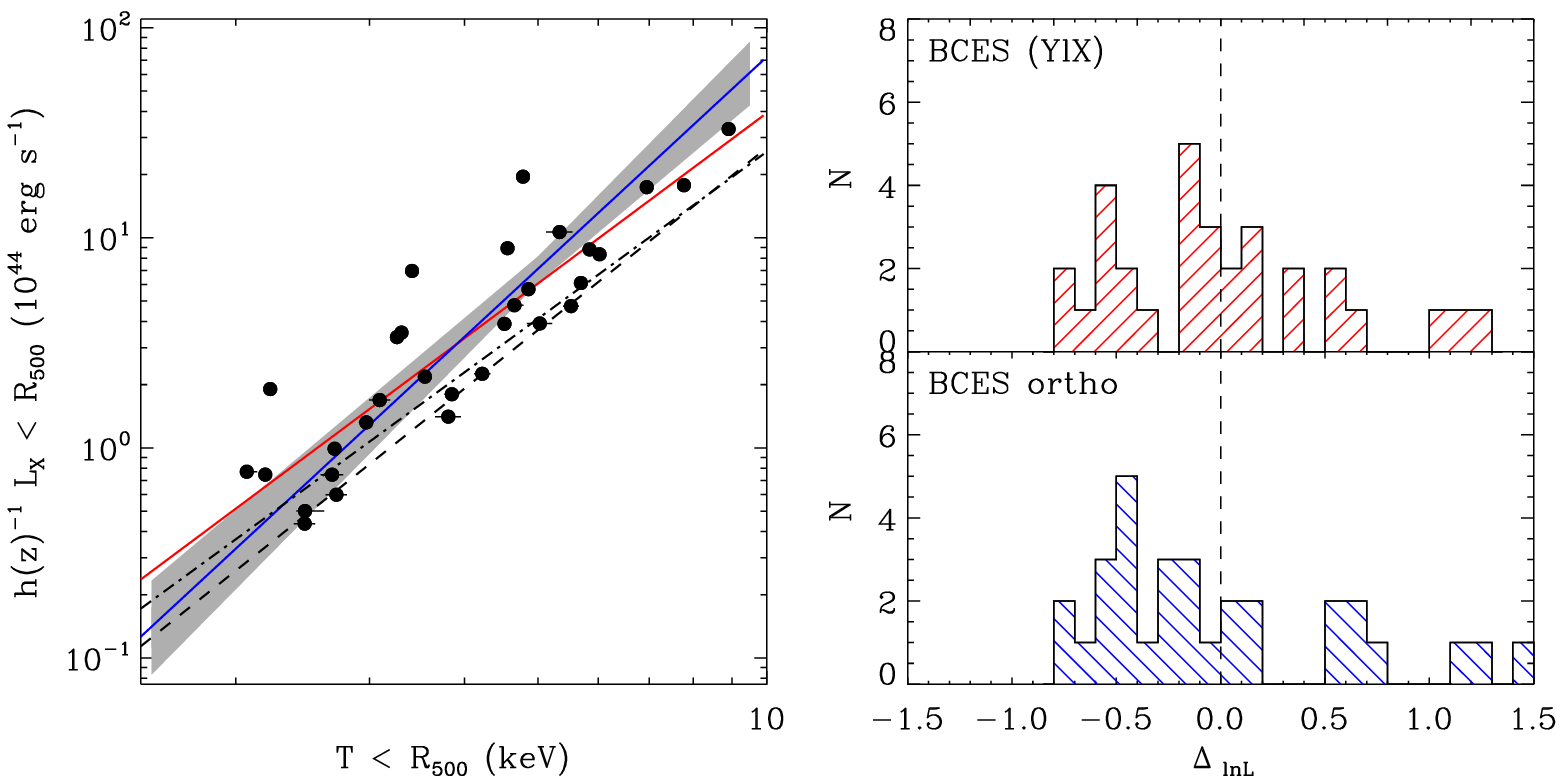}
\includegraphics[width=0.32\textwidth]{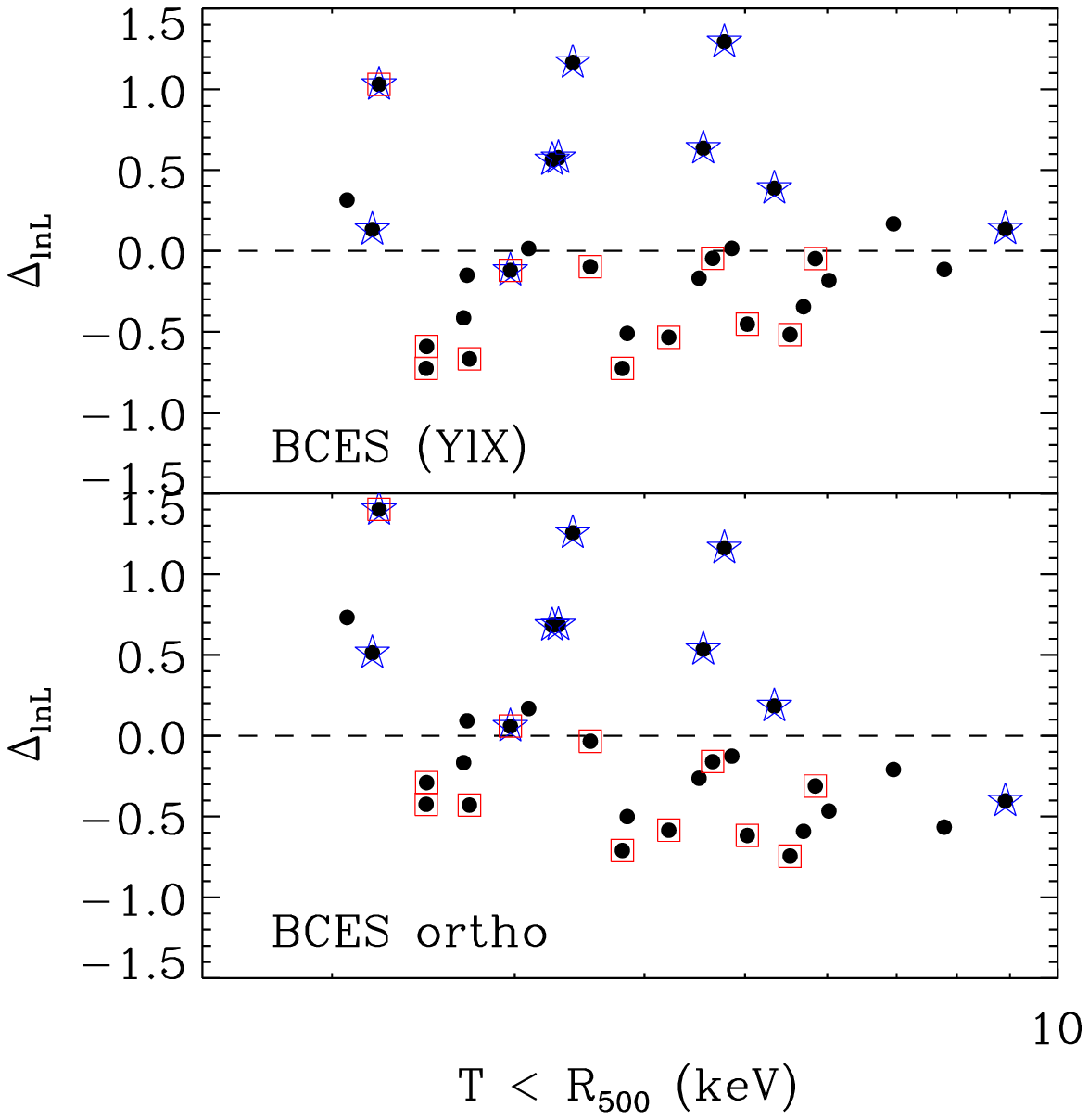}
\caption{{\footnotesize {\it Left}: $L_1$--$T_1$ relation for the
    \rexcess\ sample (quantities derived from all
    emission interior to $\Rv$). The error bars are smaller than the
    points in many cases. The best fitting power law relation
    derived from the BCES (Y$|$X) (red line) and BCES orthogonal (blue line) are overplotted; the shaded region corresponds to the $1\sigma$ uncertainty on the latter. The dashed and dot-dashed lines are the relations of \citet{ae99} and \citet{mark98}, respectively. {\em Centre:} Histogram of the log
    space residuals from the best 
    fitting $L$-$T$ relation, derived from each fitting method as indicated. {\em Right:} Log space residuals for both fitting methods as indicated. Cooling core clusters
    (blue stars) and morphologically disturbed clusters (red squares)
    occupy two distinct regions in the plot in both cases.}}\label{fig:LxTraw} 
   \end{figure*}
\begin{figure*}[]
\centering
\includegraphics[width=0.67\textwidth]{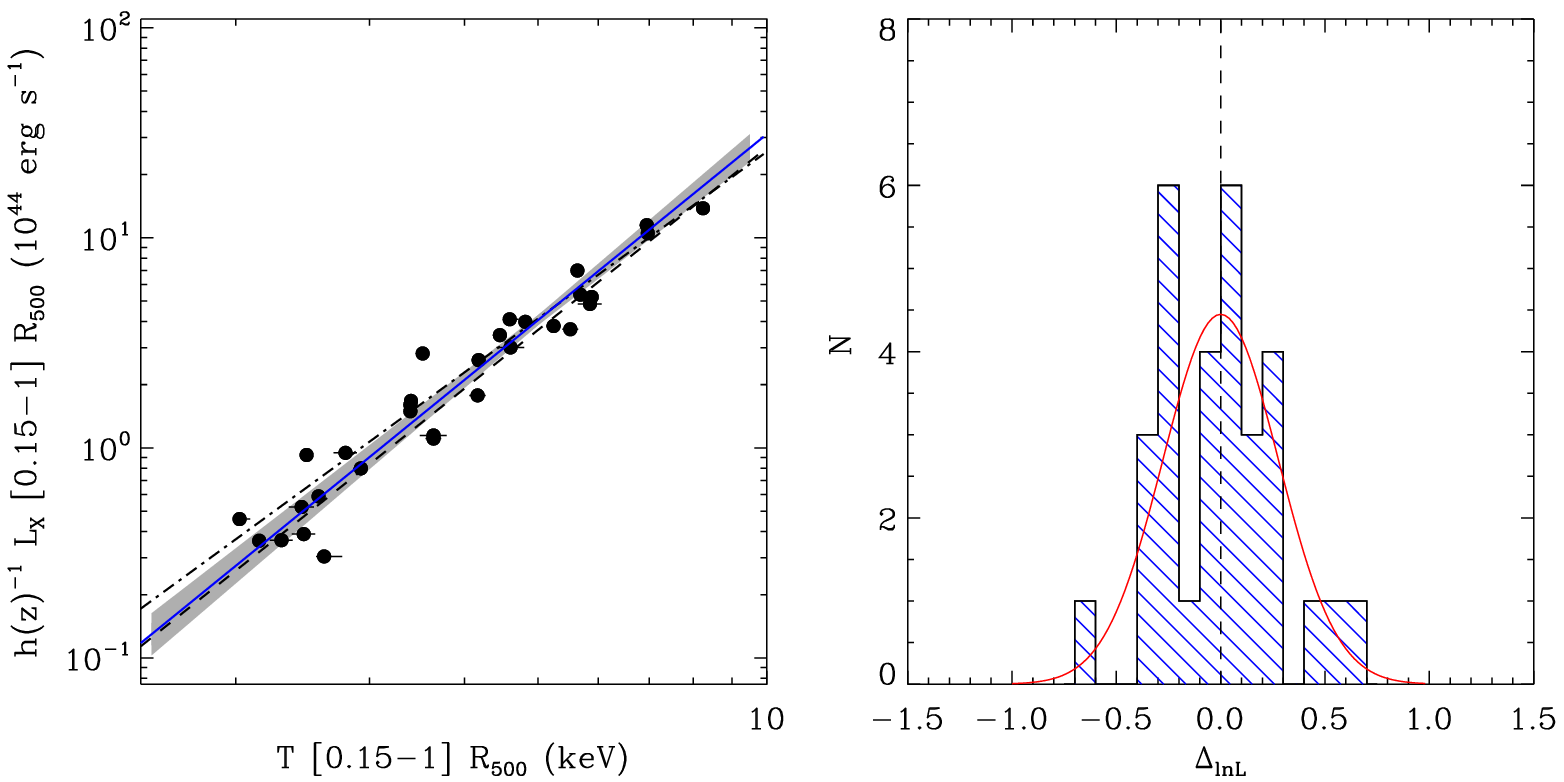}
\includegraphics[width=0.32\textwidth]{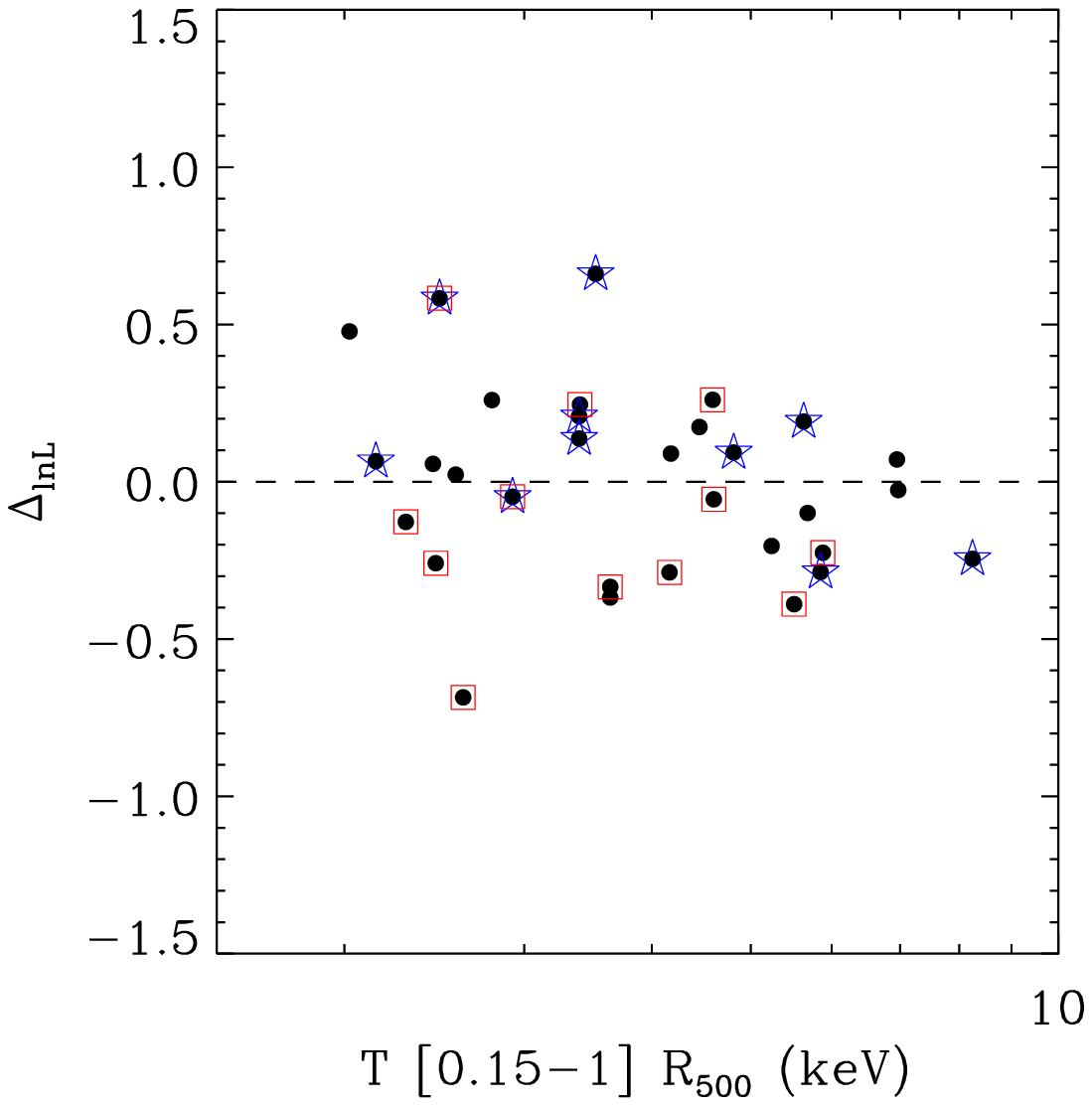}
\caption{{\footnotesize {\it Left}: $L_2$--$T_2$ relation for the
    \rexcess\ sample (quantities derived from emission in the 
    $0.15\,\Rv < R < \Rv$ aperture). The best fitting power law relation
    derived from the BCES orthogonal fitting method is 
    overplotted as a solid line (the BCES (Y$|$X) results are very similar); the shaded region corresponds to the $1\sigma$ uncertainty on the fit. The dashed and dot-dashed lines are the relations of \citet{ae99} and \citet{mark98}, respectively. {\em Centre:} Histogram of the log
    space residuals from the best 
    fitting $L$-$T$ relation, derived from the BCES orthogonal fit
    method. The solid curve is a Gaussian with $\sigma_{\ln{L}} =
    0.28$, corresponding to the scatter about the best fitting
    relation. {\em Right:} Log space residuals. Cooling core clusters
    (blue stars) and morphologically disturbed clusters (red squares)
    are less obviously segregated once the central region is
    excised. }}\label{fig:LxTcorr}   
   \end{figure*}


\subsection{Scaled emission measure profiles}

The emission measure ($EM$) was calculated from the surface
brightness profiles extracted in the [0.3-2] keV band via:  

\begin{equation}
EM(r) = \frac{4 \pi\, (1 + z)^4\,  S(\theta(x))}{\epsilon(T,z)} \\;\ \ \ \ \ \ \ r =  d_A(z) \theta,
\end{equation}

\noindent where $S(\theta)$ is the surface brightness, $d_A (z)$ is the
angular distance at redshift $z$, and $\epsilon(T,z)$ is the emissivity,
which has been calculated taking into account absorption and the
instrument response \citep[e.g.][]{na99}. We then scaled the $EM$ profiles according to their expected evolution
with redshift and dependence on temperature, $EM \propto
h(z)^{-3} T^{-1/2}$, shown in 
the right hand panel of Figure~\ref{fig:tci0}. The behaviour of the scaled
profiles is very similar
to that seen in the gas density profiles discussed in
\citet{croston08}: outside the central regions, the dispersion rapidly
decreases and the profiles begin to show indications of
similarity. The relative dispersion in scaled profiles 
shows a broad minimum of $\sigma/\langle EM(r) \rangle \sim 0.35$ from
$0.2 - 0.9\,\Rv$, with a maximum of 1.56 in the central regions
and a minimum of 0.32 at $0.5\,\Rv$. The latter is somewhat smaller
than that found by \citet{na99}, who used a relation taken from numerical simulations to calculate $\Rv$. 

The two subsamples form distinct classes in the plot. The cool core systems, unsurprisingly, show very strong central emission and also appear to scale somewhat more tightly in the outer regions. In contrast, all of the clusters with the lowest central emission measure are classed as disturbed.


\subsection{The $L-T$ relation}


In the left-hand panel of Figure~\ref{fig:LxTraw} we show the $L_1$--$T_1$
relation obtained with $T_1$ and bolometric $L_1$ derived from all emission
interior to $\Rv$ (i.e., equivalent to a raw, uncorrected
relation). In many cases the errors
are smaller than the points, a testament to the exceptional quality of
the data. The best fitting power law relations derived from the BCES fits are overplotted; fits are listed in Table~\ref{tab:lxrel}. The BCES (Y$|$X) slope, $2.70\pm 0.23$, is consistent with previous
determinations such as those of \citet[][$2.64\pm0.16$]{mark98},
\citet[][$2.88\pm0.15$]{ae99},
\citet[][$2.9\pm0.3$]{allenfab98} and \citet[][$2.82\pm0.32$]{nsh02}. The slope derived from the BCES orthogonal fit, $3.35\pm0.32$ is somewhat steeper, although only at slightly more than $2\sigma$, a consequence of the very large scatter in the data. The relations of \citet{ae99} and~\cite{mark98} are also plotted in the Figure: their normalisations are notably lower that that found in the present work, due to their being a non-cool core cluster sample and cool core-corrected sample respectively.

The central panel of Fig.~\ref{fig:LxTraw} shows the histogram of
the log space residuals from the best fitting relations. The logarithmic 
scatter about the $L_1$--$T_1$ relation is $\sigma_{\ln{L}} \sim 0.7\pm0.1$ in both cases, and is dominated by the intrinsic component, the statistical scatter
being negligible. Although the residuals present a clear skew towards higher luminosity systems, the KS probability that the residuals are drawn from a Gaussian distribution are 0.15 and 0.10
for the BCES (Y$|$X) and orthogonal fits, respectively. This result does not strongly exclude the Gaussian hypothesis, underlining the need for a larger sample to better understand the scatter (although note that \citealt{nsh02} find that the log space residuals of a larger sample are consistent with a Gaussian distribution with a similar $\sigma$ to that found for the present sample).

\begin{figure*}[]
\centering
\includegraphics[width=0.67\textwidth]{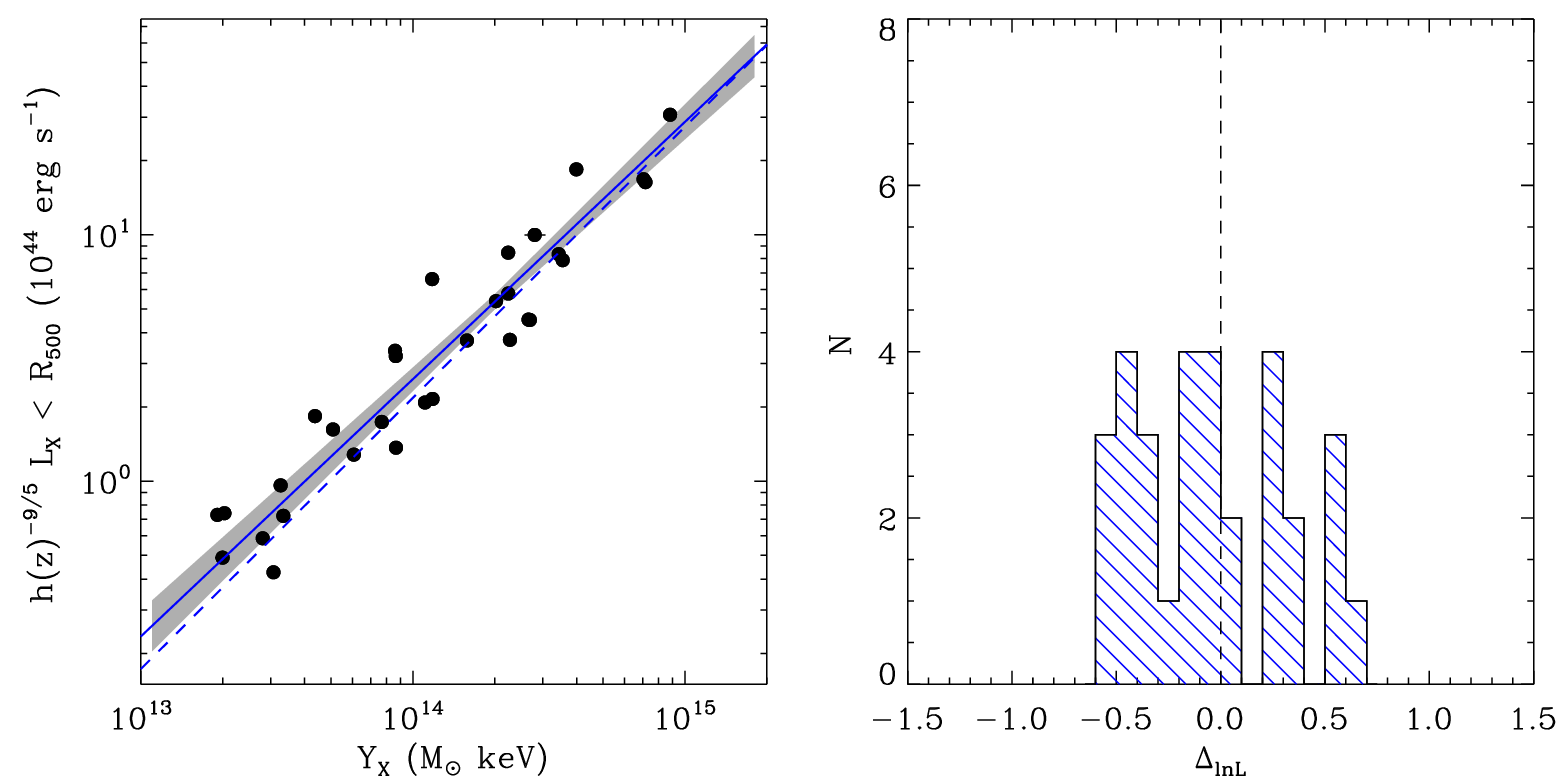}
\includegraphics[width=0.32\textwidth]{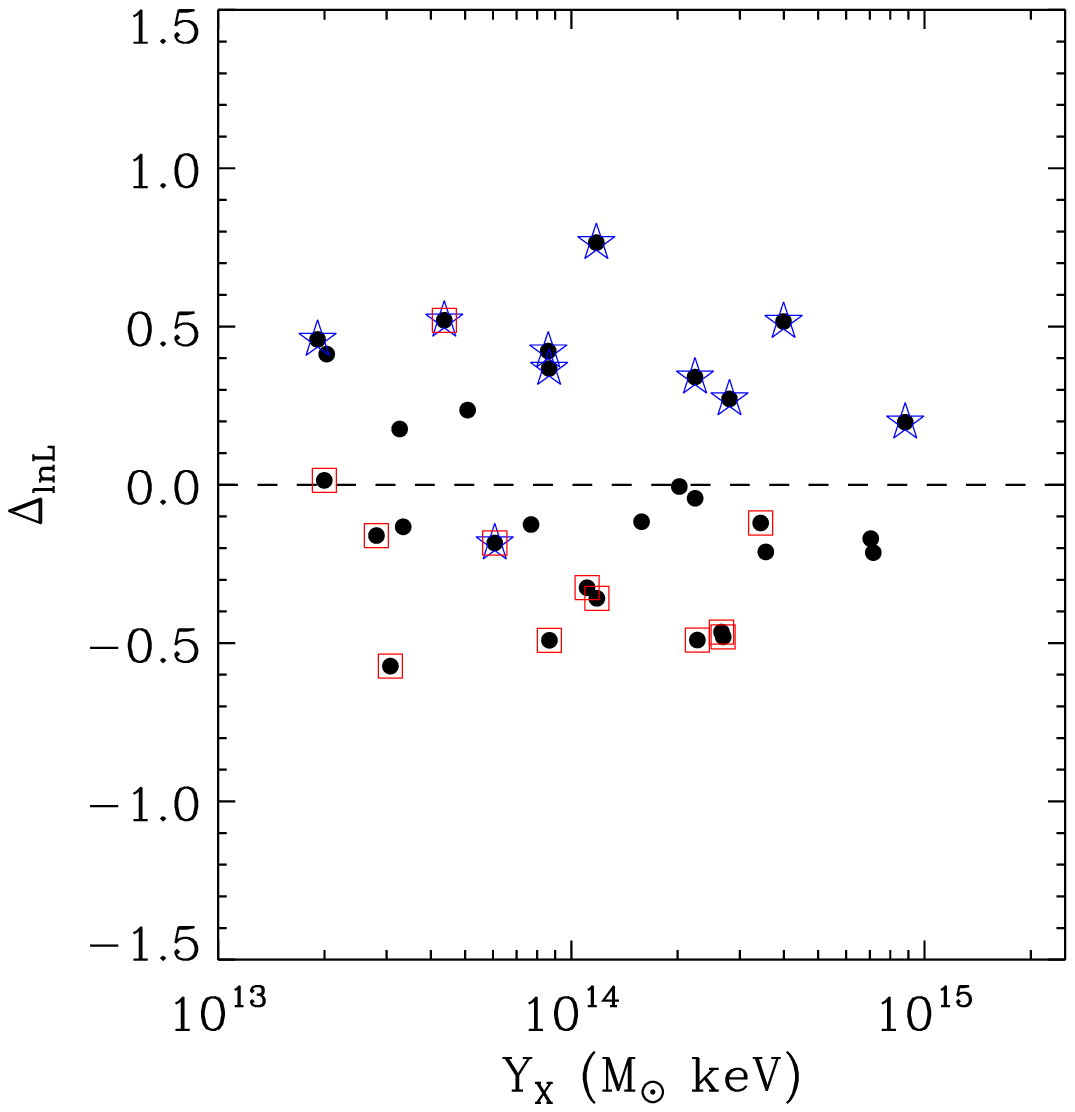}
\caption{{\footnotesize {\it Left}: $L_1$--$Y_X$ relation for the
    \rexcess\ sample, with luminosity derived from all
    emission interior to $\Rv$. The error bars are smaller than the
    points in many cases. The best fitting power law relation
    derived from the BCES orthogonal fitting method is 
    overplotted as a solid line (the BCES (Y$|$X) results are very similar); the shaded region corresponds to the $1\sigma$ uncertainty on the fit. The dashed line is the fit derived
    by \citet{maughan07} from observations of 115 galaxy clusters in
    the {\it Chandra} archive. The agreement is excellent. {\em Centre:} Histogram of the log space residuals about the best fitting orthogonal BCES relation.  {\em
      Right:} Log space residuals of the different subsets. Blue stars: cooling core clusters; red squares: morphologically disturbed systems.}}\label{fig:YxLxraw} 
   \end{figure*}

\begin{figure*}[]
\centering
\includegraphics[width=0.67\textwidth]{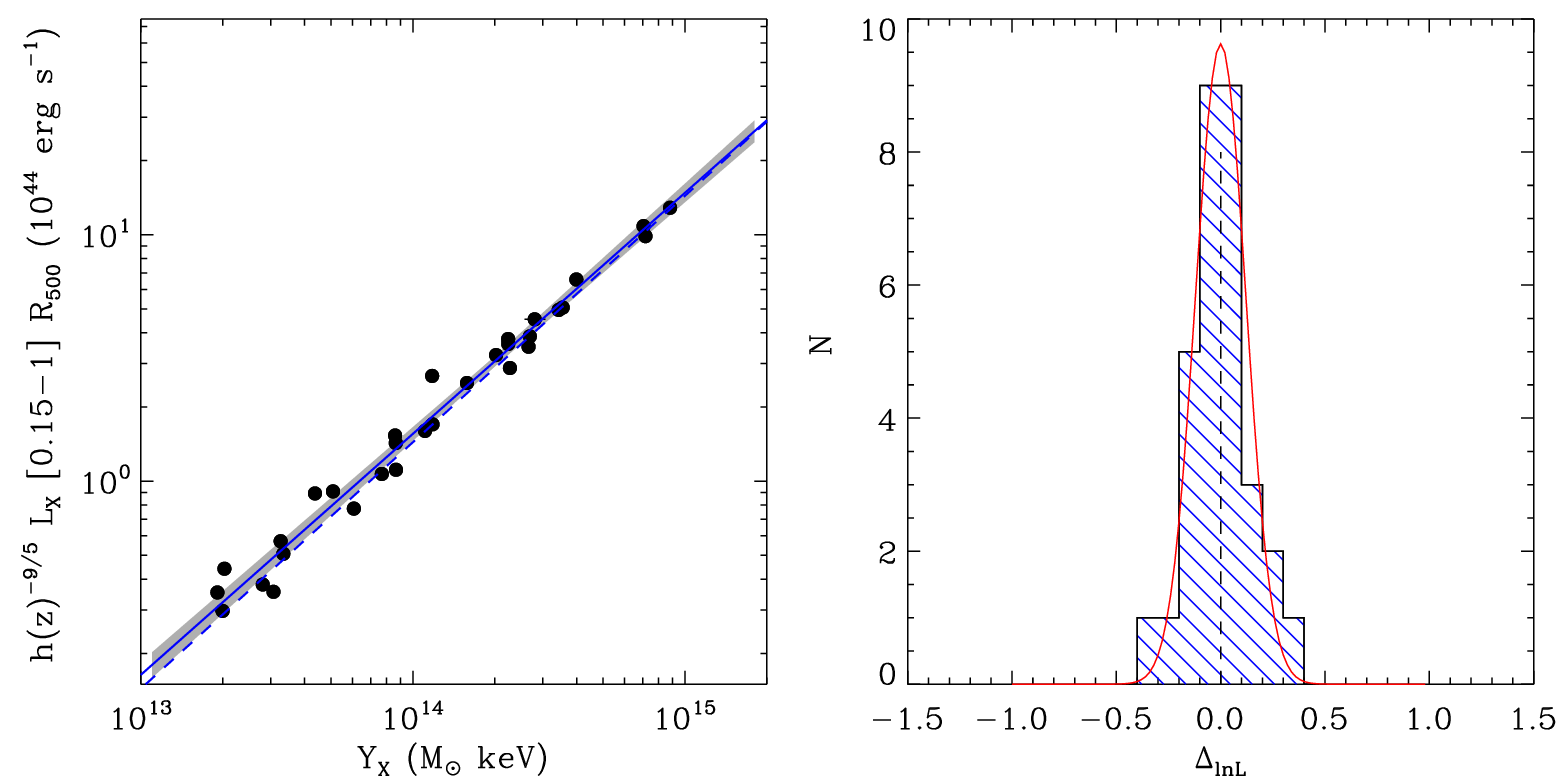}
\includegraphics[width=0.32\textwidth]{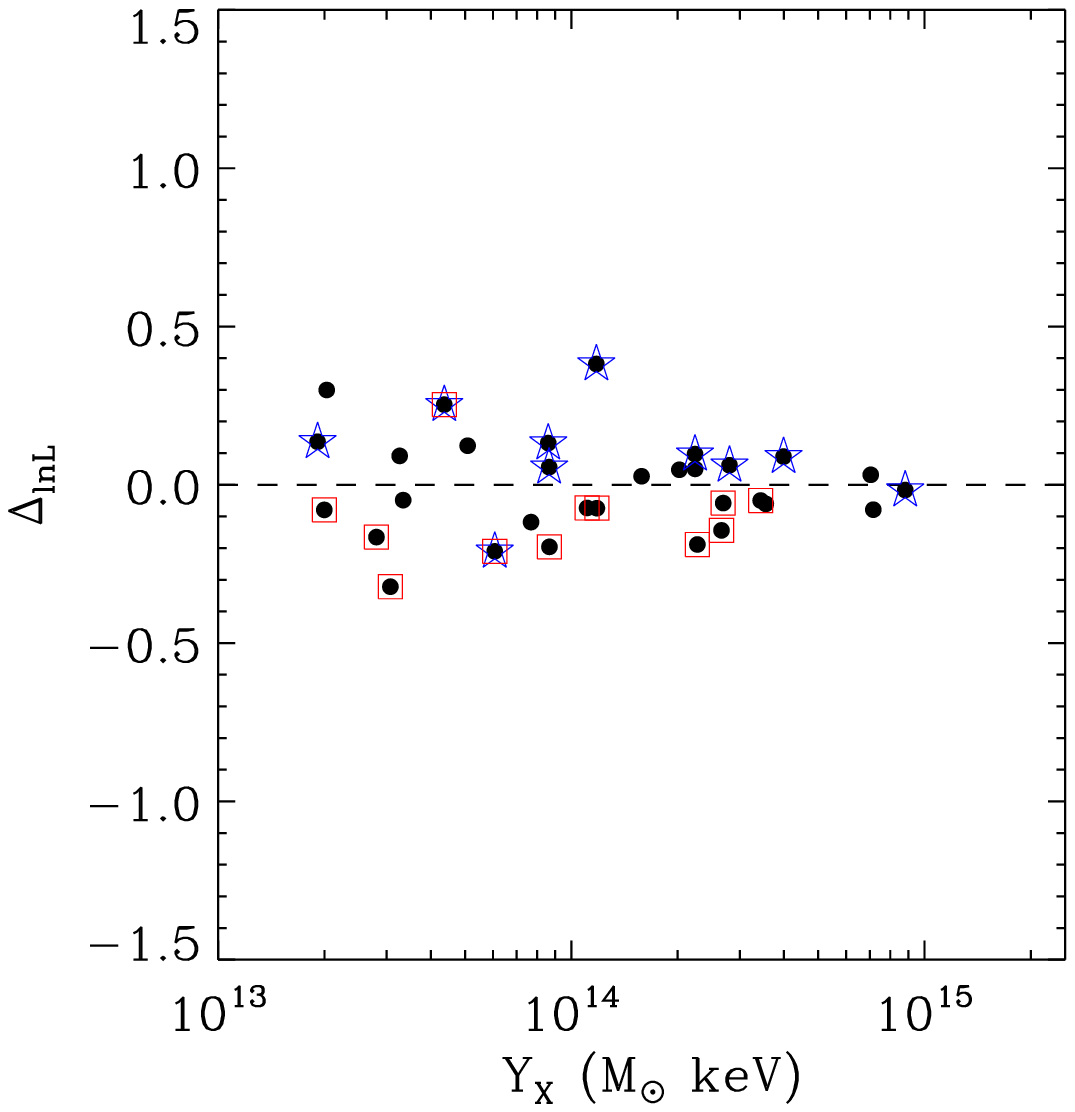}
\caption{{\footnotesize {\it Left}: $L_2$--$Y_X$ relation for the
    \rexcess\ sample, with luminosity derived from emission in the 
    $0.15\,\Rv < R < \Rv$ aperture. The best fitting power law relation
    derived from the BCES orthogonal fitting method, which takes into account errors in
    both coordinates and intrinsic scatter in the data, is 
    overplotted as a solid line (the BCES (Y$|$X) results are very similar); the shaded region corresponds to the $1\sigma$ uncertainty on the fit. The dashed line is the fit derived
    by \citet{maughan07} from observations of 115 galaxy clusters in
    the {\it Chandra} archive. The agreement is again excellent. {\em Centre:} 
    Histogram of the log space residuals about the best fitting orthogonal BCES relation. The solid curve is a Gaussian with $\sigma_{\ln{L}}=0.16$, corresponding to the scatter about the relation.{\em Right:} Log 
    space residuals of the different subsets. Blue stars: cooling core clusters, red squares: morphologically disturbed systems.}}\label{fig:YxLxcorr} 
   \end{figure*}

It is interesting to investigate the factors driving the very large
scatter about the $L_1$--$T_1$ relation.
The right hand panel of Figure~\ref{fig:LxTraw} shows the log space
deviations from the best fitting $L_1$--$T_1$
relations for the cooling core sample (blue stars), and for the
morphologically disturbed subsample (red squares). In both cases, the subsamples
clearly populate different regions of the residual space: cool core
systems are preferentially located above the main relation, while
morphologically disturbed systems lie below it.
The best fitting power law relations to the individual subsamples
are listed in Table~\ref{tab:lxrel}. 
Cool core clusters have a statistically identical slope to that of the
non-cool core systems, and to that of the sample as a whole. 
The cool core subsample has a higher normalisation than the non-cool
core subsample, significant at the $> 2 \sigma$ level, suggesting that
the primary effect of a cooling core is to move a given system
orthogonally from the standard relation. However, the logarithmic scatter about the cooling-core only relation ($\sigma_{\ln{L}} = 0.48\pm0.13$ for the BCES orthogonal fit) is higher than that about the non-cool core relation at slightly more than $1\sigma$, reflecting the wide variety of cooling core strengths in the present sample. The logarithmic scatter about the non-cool core
relation ($\sigma_{\ln{L}} = 0.29\pm0.07$, or $\sigma_{\rm log10} \sim 0.15$), is in good agreement with that found by \citet{ae99}, from a sample which contained only non-cool core systems.

A similar trend is seen when the clusters are divided according to the
morphology parameter. Firstly, it is clear from
Figure~\ref{fig:LxTraw} that the 
disturbed clusters preferentially populate the lower envelope of the
$L_1$--$T_1$ relation. The slopes of the relations are statistically
identical for both subsamples, and in agreement with that of the
entire sample, but the normalisation of the relaxed sample is higher
at the $1 \sigma$ level than that of the unrelaxed sample. This is partly due to the predominance of cool core systems in
the relaxed subsample, although disturbed cool core systems do
exist. The logarithmic scatter about the relations is very similar, at $\sigma_{\ln{L}} \sim 0.65$, although they are not well constrained.

The clear segregation of the cooling core clusters from the rest of
the population (Fig.~\ref{fig:LxTraw}), together with the small
relative segregation of dynamically disturbed systems and the
structural similarity at large radius (Fig.~\ref{fig:tci0}, right hand
panel), suggest
that simply excluding the central region should tighten the luminosity
scaling relations. Figure~\ref{fig:LxTcorr} shows the $L_2$--$T_2$ relation
derived from emission excluding the core region, where both
the luminosity and temperature are estimated in the $0.15\,\Rv < R <
\Rv$ aperture; best fitting slopes and normalisations are given in
Table~\ref{tab:lxrel}. 

The BCES (Y$|$X) slope, $2.78\pm0.13$ is
similar to the relation for that derived from all emission interior to
$\Rv$, and the slope of the BCES orthogonal fit, $2.94\pm0.15$ is slightly steeper but in good agreement within the $1\sigma$ uncertainties, as is the normalisation. This relation is in excellent agreement, both in terms of slope and normalisation, with those of \citet{mark98} and \citet{ae99}, which are overplotted in the same Figure. 

However, the logarithmic intrinsic scatter, $\sigma_{\ln{L}} = 0.27\pm0.06$ is smaller, as expected, by a factor of two. The
central panel of Figure~\ref{fig:LxTcorr} shows the histogram of
the log space residuals from the best fitting BCES orthogonal relation. The
overplotted curve is a Gaussian with raw $\sigma_{\ln{L}} =
0.27$, corresponding to the scatter about the relation. The KS test probability the residuals are drawn from a Gaussian distribution is 0.54 for the BCES orthogonal fit.

The scatter is clearly reduced on exclusion of the core regions. The relative effect of the change in luminosity and temperature in the reduction of scatter can be estimated simply by comparing the values estimated in the two apertures. We find $\langle T_1/T_2 \rangle = 1.02\pm0.07$ and $\langle L_1/L_2 \rangle =1.62\pm0.31$ for the full sample, indicating that the change in temperature is a very minor effect compared to the  change in luminosity. Unsurprisingly however, the change in temperature is negative for cool core systems ($\langle T_1/T_2 \rangle = 0.96\pm0.07$), while it is positive for non-cool core objects ($\langle T_1/T_2 \rangle = 1.05\pm0.05$).

Table~\ref{tab:lxrel} also lists the fits to the different
subsamples. Scatter decreases markedly (by approximately a factor of two) for the cool core subsample, as expected, but it also decreases somewhat ($\sim 15$ per cent) for the non cool core subsample.  Cool core clusters still tend to be found towards the upper envelope of the distribution, which is reflected in their
slightly higher normalisation compared to non-cool
core systems (although this is not significant). The slopes are stable however, and in agreement with
those found for the relation derived from all emission interior to
$\Rv$. In common with the full emission sample, morphologically
disturbed clusters tend to describe the lower envelope of the
distribution, having a slightly lower normalisation than the full sample, although this is not significant, and a similar slope.


\begin{figure*}[]
\includegraphics[width=0.32\textwidth]{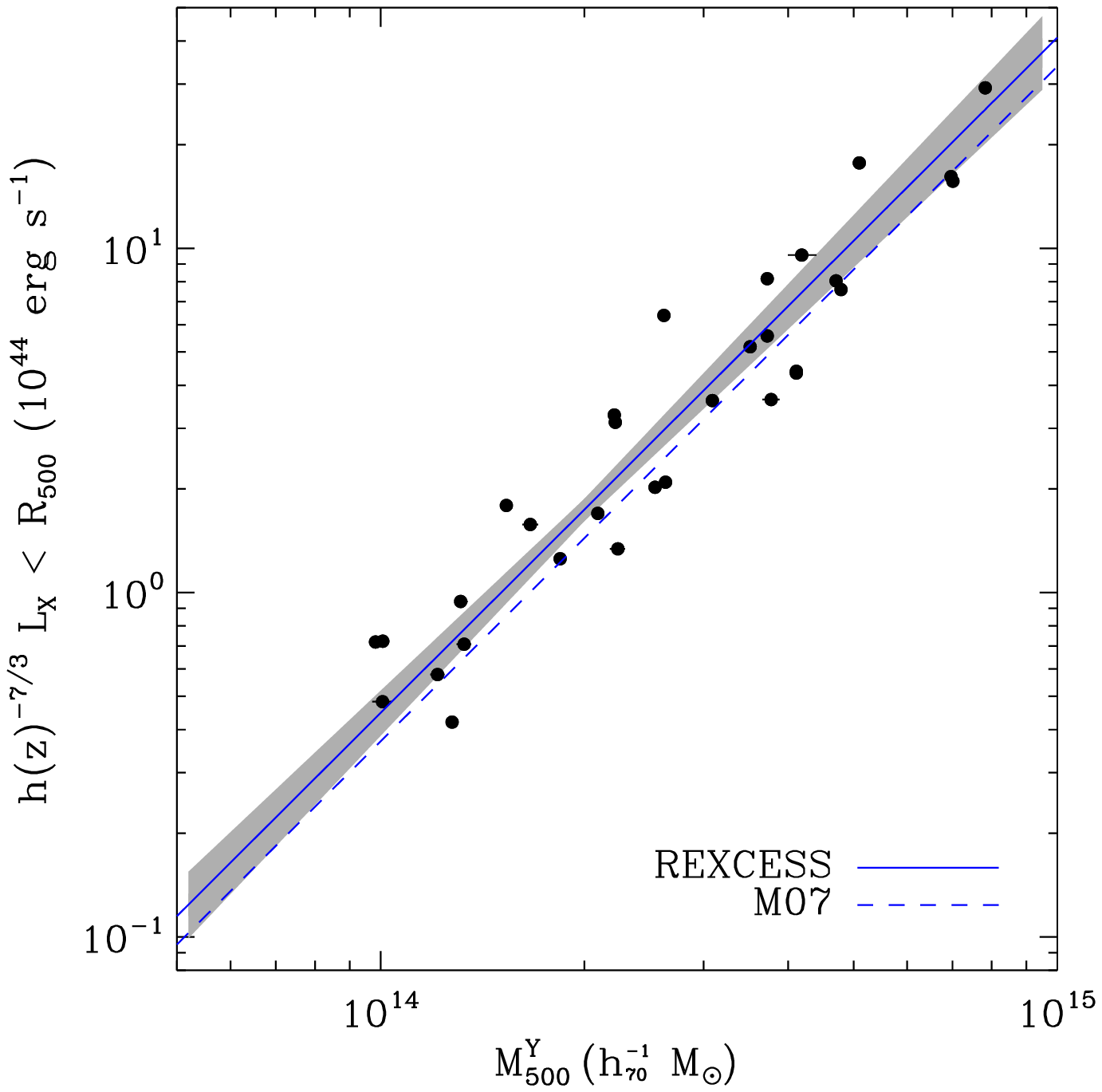}
\hfill
\includegraphics[width=0.32\textwidth]{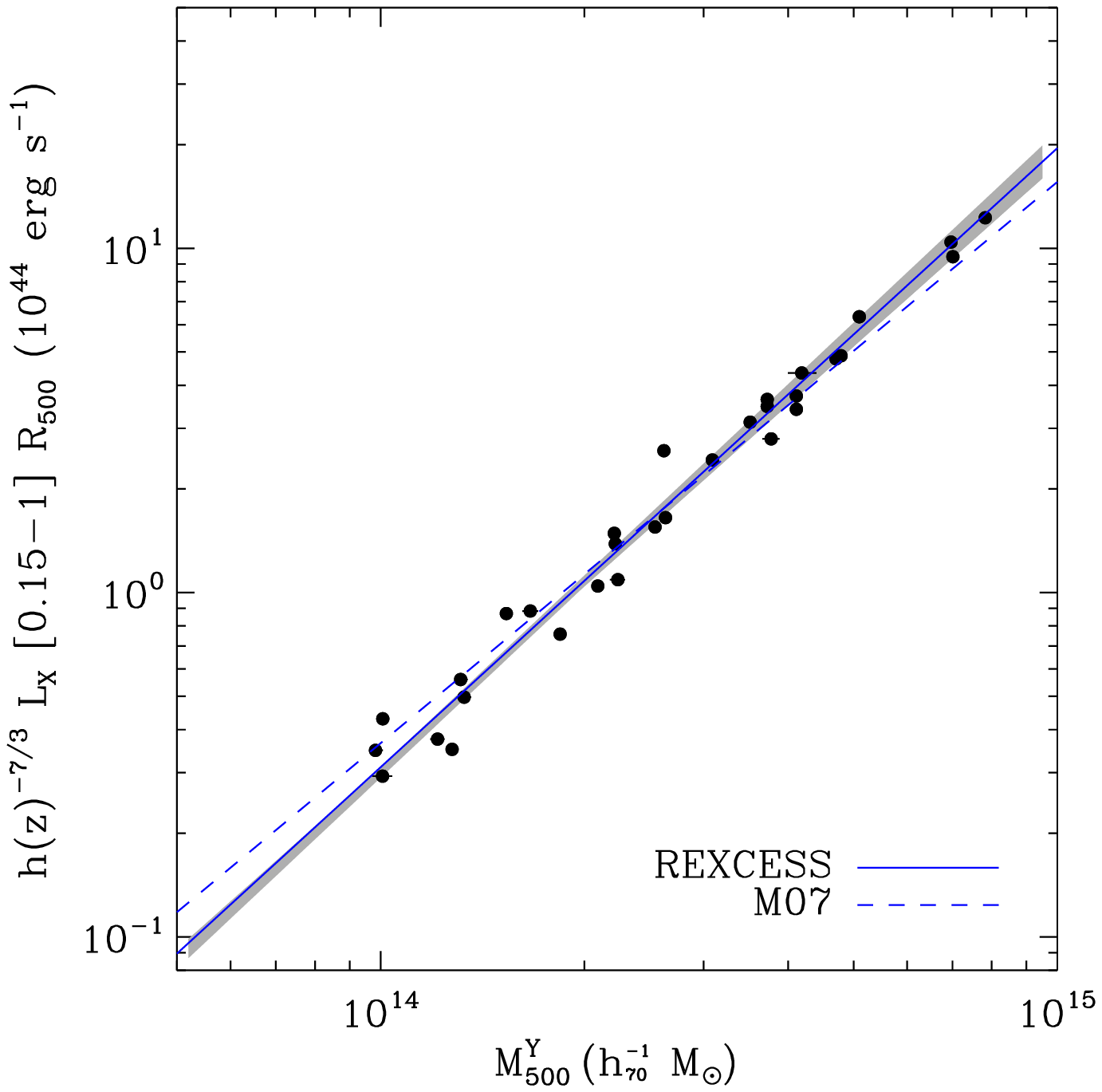}
\hfill
\includegraphics[width=0.32\textwidth]{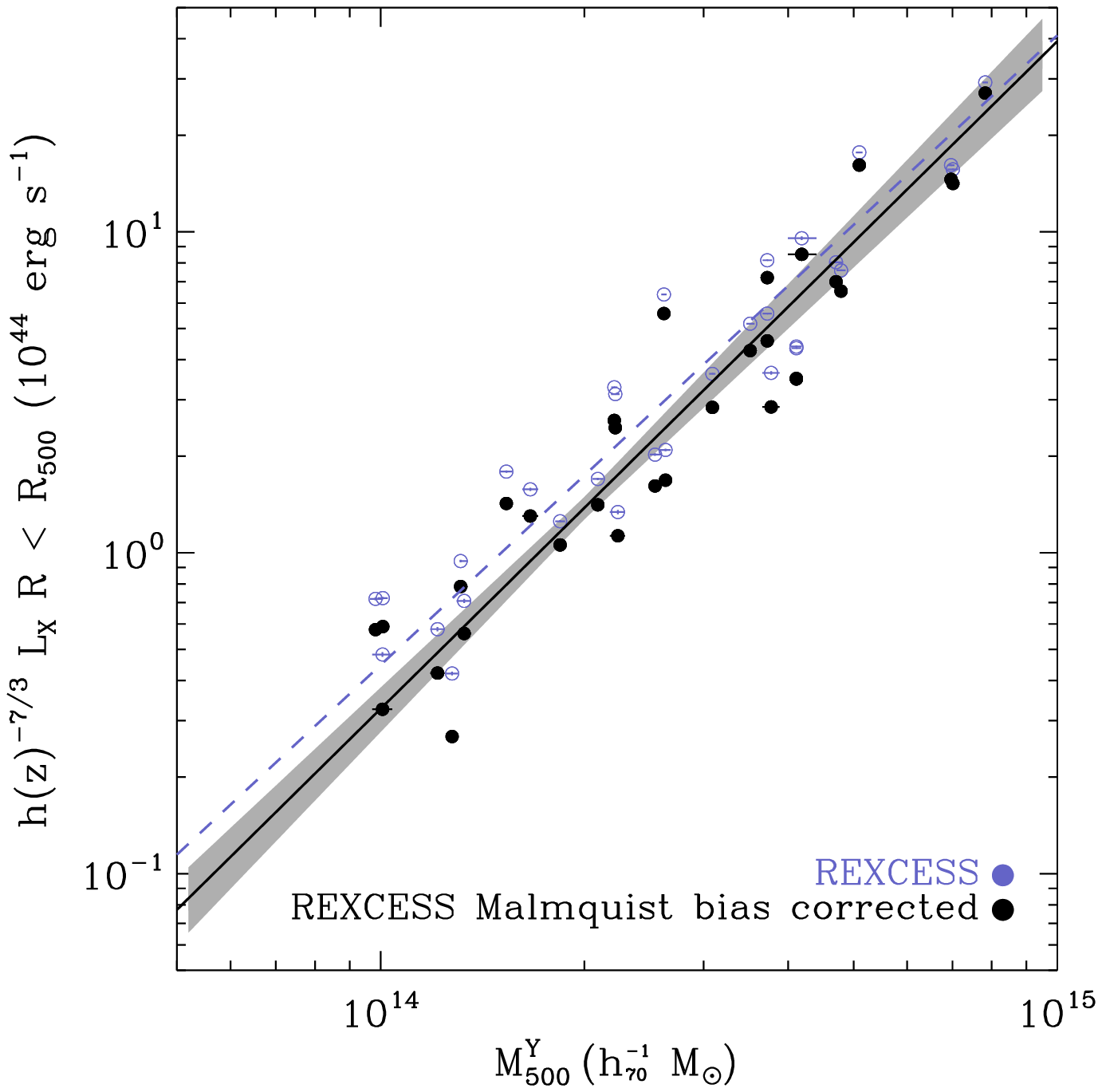}
\caption{{\footnotesize $L-M_{500}$ relation for the
    \rexcess\ sample, with the mass estimated from the
    $Y_X-M$ relation of \citet{app07}.  {\it Left:} Relation for all emission interior to $\Rv$. {\it Centre:} Relation for emission in the $[0.15-1]\,\Rv$ aperture. The best fitting power law relation
    derived from the orthogonal BCES fit method is 
    overplotted as a solid line. The dashed line is the fit derived
    by \citet{maughan07} from observations of 115 galaxy clusters in
    the {\it Chandra} archive. {\it Right:} Relation corrected for Malmquist bias as described in Appendix~\ref{sec:surveylx}.}}\label{fig:LxMcorr} 
   \end{figure*}

\subsection{The $L-Y_X$ relation}

$Y_X$ is an interesting quantity because simulations suggest that deviations in $M_{\rm gas,500}$ and $T$ for a given system are anti-correlated with respect to the self-similar expectations, leading to a reduction in scatter (although thus far this is empirically untested). That $Y_X$ is the X-ray analogue of the integrated SZ Comptonisation parameter $Y_{SZ}$ makes the calibration of its relationship with the X-ray luminosity of prime importance for the interpretation of data from the upcoming all-sky surveys from the {\it Planck} and {\it eROSITA} satellites.

The $L_1$--$Y_X$ relation, where the luminosity is derived from all emission interior to $\Rv$, is
shown in Figure~\ref{fig:YxLxraw}; the best fitting power law
values are given in Table~\ref{tab:lxrel}.  Because of the smaller scatter in these data, both BCES fitting
methods give consistent results. Our relation is in
good agreement with that of the {\it Chandra} archive study of
115 galaxy clusters by \citet{maughan07}: the slope
$B=1.10\pm0.04$ is consistent with our BCES orthogonal value, $\alpha=1.04\pm0.06$,
and the normalisation at $Y_X=2\times10^{14}\,M_\odot\keV$ is only 14 per cent lower than ours. The intrinsic logarithmic scatter is $\sigma_{\ln{L}} = 0.38\pm0.06$, considerably less than about the \LxT\ relation (note that since $Y_X$ is calculated using the temperature estimated in the $[0.15-1]\,\Rv$ aperture, this will tend to damp scatter somewhat). However, the residual histogram about the best fitting orthogonal BCES relation, plotted in the central panel of the same Figure, has a KS probability of only 0.09 of being compatible with a Gaussian distribution.

The residual distribution for the different subsamples mirrors that of
the $L_1$--$T_1$ relation: cool core clusters lie preferentially above and
morphologically disturbed systems lie preferentially below. This fact
is reflected in the different normalisations found when fitting the
different subsamples: the cool core systems have the highest normalisation and the disturbed systems the lowest. However, the slope
of the relation, when fitted to different subsamples, is remarkably
stable at 0.96-1.06, and the slopes of all subsamples are statistically
indistinguishable.

Figure~\ref{fig:YxLxcorr} shows the $L_2$--$Y_X$ relation, determined with
the core emission excluded. Once again there is excellent agreement in
both slope and normalisation between our relation and that of
\citet{maughan07}. The relation is very tight: the intrinsic logarithmic scatter is only $\sigma_{\ln{L}} = 0.16\pm0.04$, and the KS probability that the distribution of residuals is compatible with a Gaussian is 0.93. For the different subsamples, Table~\ref{tab:lxrel} shows that the slopes are remarkably similar, ranging from 0.94 to 0.99, and the power law normalisations for the best fitting models are segregated in a similar manner to the luminosity
temperature relation, although with much reduced significance. 


\subsection{The $L-\Mv$ relation}

It is interesting to make a first examination of the slope and normalisation of the \LxM\ relation for the present sample. Since we do not have independent measures of the mass, we use the $M_{500}-Y_X$ relation of \citet{app07} to estimate the masses of the clusters in the sample. For the purposes of this initial investigation, we ignore the impact of the intrinsic scatter about the  $M_{500}-Y_X$ relation because it is at present not sufficiently well quantified; X-ray calibrations are necessarily available for relaxed cluster samples only, and weak lensing calibrations are at present lacking sufficient dynamic range in mass. The present approach allows us to verify the slope and normalisation of the relation under the given assumptions, to check the coherence of the slopes, and to compare with previous work using similar approaches. 

The measured $L-M$ relations are summarised in
Table~\ref{tab:lxrel} and the relations obtained for bolometric $L$
measured in both apertures are plotted in Figure~\ref{fig:LxMcorr}. The slopes of the relations, $\sim 1.8$, are consistent with the $L$-$T$ and $M_{500}-Y_X$ relations, as expected. Comparing our measurements of the slope and normalisation with those of \citet{maughan07}, we find excellent agreement in slope for the
relation derived from all emission interior to $\Rv$, athough our normalisation is somewhat higher (by $< 20$ per cent). When the core
emission is excluded, the slope of \citeauthor{maughan07}'s
relation ($1.63\pm0.08$) is somewhat shallower than our BCES orthogonal measurement ($1.80\pm0.05$), at the $\sim 2\sigma$ level. However, the normalisations are in excellent agreement. 

In the right hand panel of Figure~\ref{fig:LxMcorr} we compare the raw $L-M$ relation with that corrected for the effect of Malmquist bias. The correction procedure, and the relations for the $[0.1-2.4]$ keV and $[0.5-2]$ keV bands, plus comparison with the results of \citet{vikh08}, are given in Appendix~\ref{sec:surveylx}.
The correction has the effect of steepening the relation slightly due to the under-representation of low-luminosity clusters on the \rexcess\ sample.

The scatter is, by definition, identical to that about the $L-Y_X$ relation, and
is in excellent agreement with that found by \citet{vikh08} from a similar analysis of a larger flux-limited sample of nearby clusters. 
Note that if $L \propto M^\gamma$, then a first order estimate of the scatter in mass is $\sigma_{\ln{M}} \sim \sigma_{\ln{L}}/\gamma$. However, this will only be true if $\sigma_{\ln{L}}$ is measured at fixed $M$ for a complete sample. Using the measurement of $\sigma_{\ln{L}}$ at fixed $T$ or $Y_X$ introduces covariance of $T$ and $Y_X$ into the relations, which would modify the first order scatter estimate. Nevertheless, the scale of this first order estimate of the scatter in mass is $\sigma_{\ln{M}} \sim 0.20-0.37$ for the full aperture and only $\sigma_{\ln{M}} \sim 0.09-0.16$ for core extracted quantities}.


\subsection{Relations including a third parameter}

The presence of a cool core is clearly the factor which contributes
most to the scattering of a given cluster about the best fitting 
relation. Figure~\ref{fig:tci0} shows that
the central density $n_{e,0}$ is a very reliable indicator of cool core
strength. Following \citet{ohara06}, it thus follows that $n_{e,0}$ may be taken into account as a third parameter in the scaling relations. 

Fitting a scaling relation of the form:

\begin{equation}
h(z)^n L = C (A/A_0)^\alpha (n_e)^\beta
\end{equation}

\noindent where $h(z)$ is the Hubble constant normalised to its present day value and $n$ was fixed to the expected scaling with $z$, and solving for $\alpha, \beta$ and the normalisation $C$, allows us to investigate the influence of central density $n_{e,0}$ on the scaling relations. For each relation the  fit was undertaken using standard linear regression in the log-log plane. We determine the best fitting values and associated $1\sigma$ uncertainties via 1000 bootstrap resamplings of the observed data set, and the raw scatter was estimated using the error weighted orthogonal distances to the regression line.

The resulting best fitting relations are summarised in Table~\ref{tab:3par}. Dependencies on the main scaling parameter ($T$, $Y_X$ and $M$) are similar to those derived for the two-parameter fits to core-excluded quantities, as expected. The scatter is comparable to that derived from a two-parameter fit to core-excluded quantities for all relations.

Thus the technique of using the central gas density  appears to be a promising method for reducing scatter about the luminosity scaling relations. 

\begin{table}
{\tiny 
\begin{minipage}{\columnwidth}
\caption{{\footnotesize Best fitting parameters for the three parameter scaling relation fits. 
Data were fitted with a power law of the form $h(z)^{n} L = C \, (A / A_0)^\alpha (n_e)^\beta$, with $A_0 = 5$ keV, $2 \times 10^{14}\, M_\odot$ keV and $2 \times 10^{14}\, M_\odot$, and $n=-1$, $-9/5$ and $-7/3$ for $T$, $Y_X$ and $M$, respectively. }}\label{tab:3par}
\centering
\begin{tabular}{l l l l l }
\hline
\hline
\multicolumn{1}{l}{Relation} & \multicolumn{1}{l}{$C$} & \multicolumn{1}{l}{$\alpha$} & \multicolumn{1}{l}{$\beta$} & \multicolumn{1}{l}{$\sigma_{\rm ln\,L,intrinsic}$} \\

\hline
\\
$L_1$--$T_1$--$n_{e,0}$ & $27.45\pm1.45$ & $2.61\pm0.36$ & $0.36\pm0.10$ & $0.47\pm0.04$\\

$L_1$--$Y_X$--$n_{e,0}$ & $13.90\pm1.13$ & $0.99\pm0.04$ & $0.26\pm0.03$ & $0.22\pm0.02$ \\

$L_1$--$M$--$n_{e,0}$ & $4.84\pm1.14$ & $1.82\pm0.07$ & $0.26\pm0.03$ & \ldots \\

\\
\hline
\end{tabular}
\end{minipage}
}
$L_1/T_1$: luminosity/temperature interior to $\Rv$.
\end{table}


\section{Discussion}

\begin{figure*}[]
\sidecaption
\centering
\includegraphics[width=0.75\textwidth]{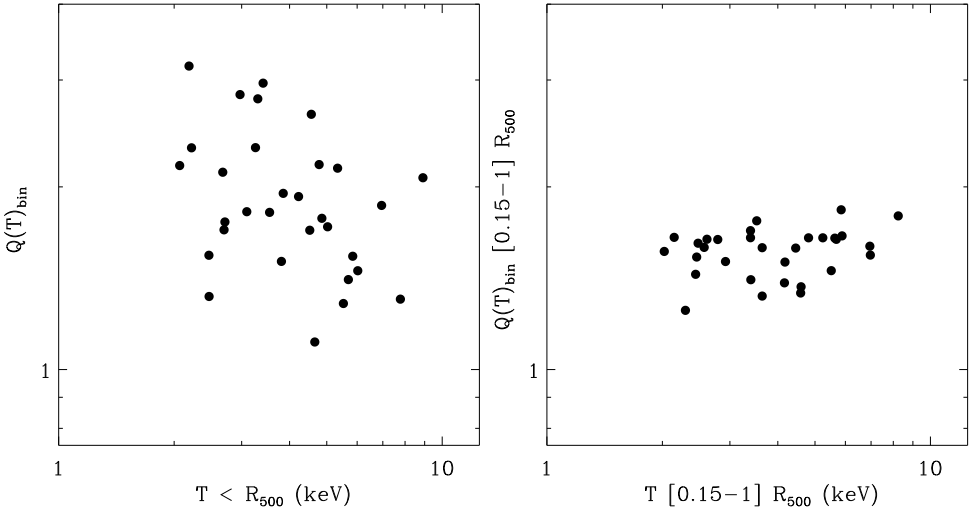}
\caption{{\footnotesize Structure factor $\hat{Q}(T)_{| \rm bin} = \langle \rho_{\rm gas}^2 \rangle / \langle \rho_{\rm gas}
\rangle^2$, estimated from the gas density profiles, versus temperature. {\it Left:} Quantities estimated from all emission interior to $\Rv$. {\it Right:} Quantities estimated in the $[0.15-1]\,\Rv$ aperture. There is no significant dependence of $\hat{Q}(T)_{| \rm bin}$ on temperature on a Kendall's $\tau$ test in either case.}}\label{fig:figQ} 
   \end{figure*}


\subsection{Scatter about the relations, and correcting for it}

The \rexcess\ data have allowed us to investigate the sources and
magnitude of the scatter about the various relations using a data set
which should be representative of any X-ray selected sample of
clusters. 

For all emission interior to $\Rv$, morphologically disturbed systems
tend to lie below the best fitting 
relation to the entire whole sample; however, this is mainly due to
the effect of cool core systems in the full sample, which tend to
increase its normalisation. A fairer test is to compare the subsamples
when the core emission is excluded: in this case, the normalisations
differ in all cases by less than $1.5\sigma$, although morphologically
disturbed systems still have the lowest normalisation.
In common with \citet{ohara06}, we find
that clusters with greater morphological substructure do not exhibit
more scatter about scaling relations than clusters with less
substructure. This is mostly a consequence of the fact that morphologically regular systems contain a preponderance of cool core clusters. This result would also suggest that the main effect of merging is to
move systems along the relation rather than orthogonal to
it. In this context, we note that numerical simulations predict
quasi-simultaneous boosting in temperature and luminosity at certain
epochs after a merging event, which would indeed tend to move clusters along
the relation. 

In common with most previous investigations, we find that
the vast majority of the scatter in all relations is due to the presence of
cool cores, which lie systematically above the best fitting relation
with an offset that appears to be related to the strength of the cool
core.  A fit to the cool core systems differs from a fit to the
whole sample only by a normalisation factor. However, the scatter
about the best fitting cool core subset relation is nearly twice that
about the non cool core subset relation, reflecting the different cool
core strengths. 

Excluding the central emission leads to a significant reduction
of the scatter about all relations. For example, the scatter about the
\LxT\ relation decreases by a factor of two on exclusion of the
core, and the reduction in scatter is similar for the \LxYx\
relation. Correcting for the presence of a cool core by assuming a power law dependence of central density $n_{e,0}$ with luminosity affords an alternative method to reduce scatter. The reduction in scatter obtained by the use of $n_{e,0}$ is of the same order as that obtained from simple core exclusion. We note that core exclusion may be difficult in the case of distant clusters or those detected with very low signal to noise, and in these circumstances it may be preferable to use the central density or surface brightness to reduce scatter about the scaling relations. 


\subsection{Slope of the relations}

The X-ray luminosity of a cluster can be written \citep{ae99}:

\begin{equation}
L(T) = f_{\rm gas}^2(T) [M(T) \Lambda(T)] \hat{Q}(T)
\end{equation}

where $\Lambda(T)$ is the cooling function. $\hat{Q}(T)$, introduced by \citet{ae99}, is equal
to $\langle \rho_{\rm gas}^2 \rangle / \langle \rho_{\rm gas}
\rangle^2$, with the angle brackets denoting an intrinsic volume average. $\hat{Q} (T)$ is thus a dimensionless structure factor which depends only on the
spatial distribution of the gas density (e.g., clumpiness at small scale, shape at large scale, etc). With the set of
additional assumptions (i) pure bremsstrahlung 
emission $[\Lambda(T) \propto T^{1/2}]$; (ii) virial equilibrium $[M
\propto T^{3/2}]$; identical internal cluster structure $[\hat{Q}(T) =
C_1]$; constant gas mass fraction $[f_{\rm gas}(T) = C_2]$, we arrive
at the standard self-similar expectation for the
luminosity-temperature relation, $L \propto T^2$. 
The self-similar \LxYx\ relation can be obtained from combination of
the gas mass-luminosity and luminosity-temperature relations to give
$L \propto Y_X^{4/5}$. Combining the self-similar mass-temperature
and luminosity-temperature relations leads 
to a dependence of luminosity with mass of $L \propto M^{4/3}$.

In common with most previous work on the subject, we find that the
slope of the \LxT\ relation of the \rexcess\ sample is steeper
than the prediction from the expectations of self-similar collapse
models. The steeper slope is found consistently in all subsamples, and
in all cases the statistical precision of the data allow us to rule
out the self-similar predictions. We find similar results for the
\LxYx\ relation, where the observed slope of $\sim 1.0$ is
significantly steeper than the self-similar expectation of 0.8, and
for the \LxM\ relation, where the observed slope of 1.8 is steeper
than the expected value of 1.3. These facts imply that one or more of
the assumptions listed above does not hold for the real cluster
population.

The assumption of pure bremsstrahlung is not strictly valid since line
emission becomes increasingly important as the temperature decreases,
having the effect of flattening the relation as lower temperature
systems are boosted in luminosity. While the lower temperature limit
of the \rexcess\ sample, 2 keV, should suffice to minimise these
effects, systematic differences in the metallicity between objects may
serve to change the temperature dependence of the X-ray emission from
the expected value of $T^{1/2}$. We tested this using the measured
temperatures and abundances of the \rexcess\ sample, finding a best fitting power law relation of 0.5, in full agreement with the expected dependence.

The assumption of virial equilibrium leads to the expected relation $M
\propto T^{3/2}$ between total mass and temperature. A topic of vigorous debate in previous years, several recent
investigations of the X-ray mass-temperature relation have shown percent-level agreement in normalisation and that
the slope is not greatly different from the self-similar
expectation \citep{app05,vikh06}, although it may be slightly steeper \citep{app05,sun08}.

The question of structural regularity has also received quantitative tests
in recent years. For instance, there is now converging evidence that the total mass density profiles of galaxy clusters and groups scale
quasi-self-similarly with a mass dependence that is in good agreement
with predictions from numerical simulations
\citep[eg.,][]{pap05,vikh06,gasta07}. On the mass scales we are
considering here, the variation of the total mass density
concentration with mass is in fact consistent with zero
\citep{pap05,vikh06}. This would imply that a variation of cluster
dark matter structure with mass cannot be responsible for the
steepening of the \LxT\ relation, at least in the mass range covered by the present data.

However, the baryonic components of clusters are subject to somewhat
different physics, and there are indications that the large scale ICM density
structure is temperature/mass dependent. The clear
correlation of the slope of the gas density profile measured in the
radial range $[0.3-0.8]\,\Rv$ with temperature seen in the \rexcess\
analysis of \citet{croston08} is one example.
At the same time \citeauthor{croston08} have shown that the
temperature dependence of the relative dispersion of scaled gas
density profiles has all but disappeared at $0.7\,\Rv$, suggesting
that clusters become increasingly structurally similar at larger
radii. This is borne out in the present data when we examine the more powerfully diagnostic structure factor $\hat{Q}(T)_{| \rm bin} = \langle \rho_{\rm gas}^2 \rangle / \langle \rho_{\rm gas} \rangle^2$, where the average is taken over the radial gas density profile. This quantity is identical to that presented in \citet{ae99} except that we use a fully deconvolved, deprojected gas density profiles rather than $\beta$-model fits.  
$\hat{Q}(T)_{| \rm bin}$ effectively probes the variation of the large scale shape of the gas density with temperature, that is to say, variation of the gas concentration with mass. However, since $\rho_{\rm gas} (r)$ is derived from spherically symmetric deprojection of the surface brightness profile, by construction $\rho_{\rm gas} (r)$  is $\sqrt{\langle \rho_{\rm gas}^2 \rangle}$, where $ \langle \rho_{\rm gas}^2 \rangle$ is the average within each radial shell. Thus we emphasise that $\hat{Q}(T)_{| \rm bin}$ is only a partial estimator of $\hat{Q}(T)$ which does not probe more subtle effects such as variations of the gas clumpiness with mass, or substructuring  at small scale. 

\begin{figure*}[]
\centering
\includegraphics[width=0.47\textwidth]{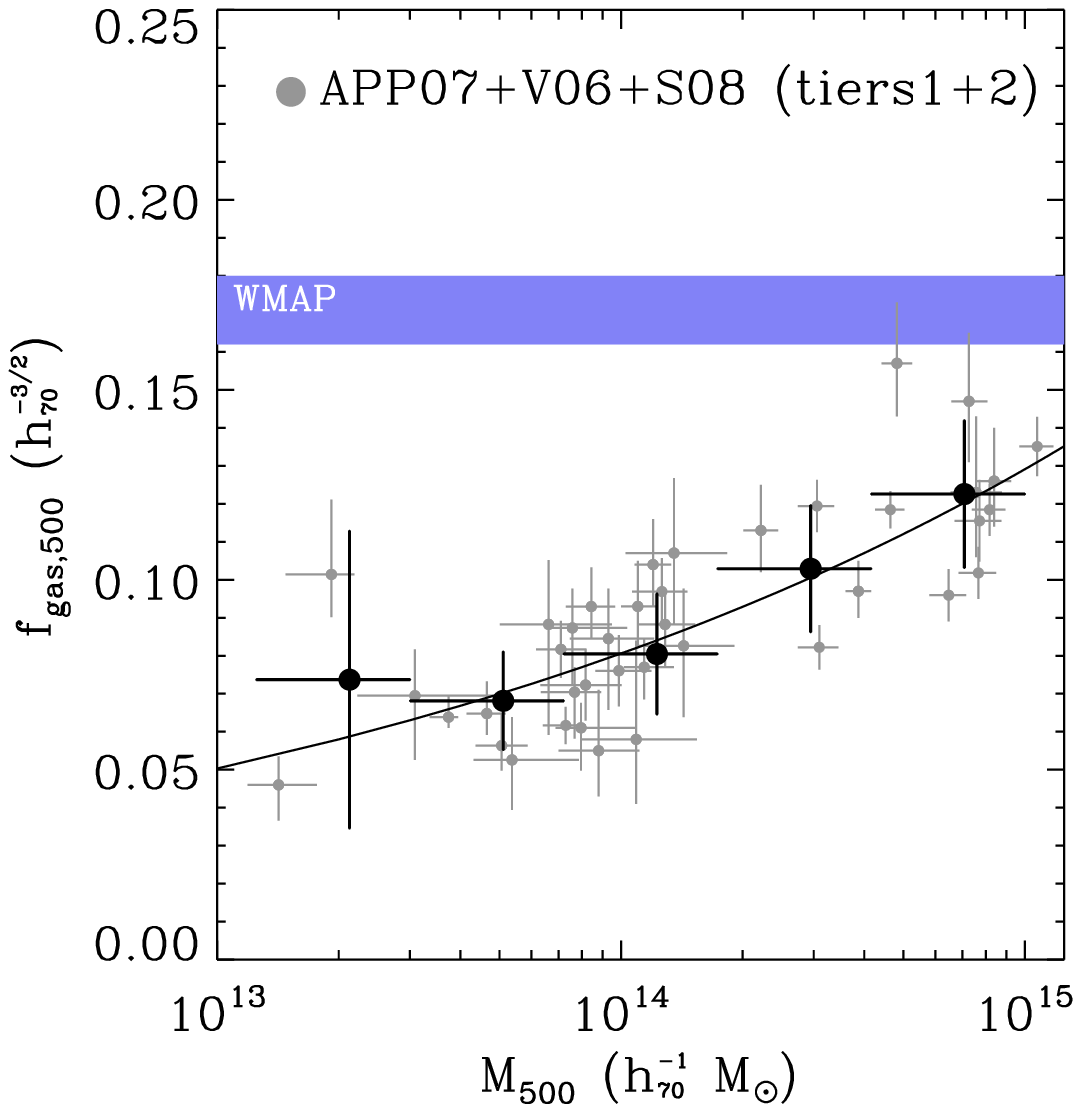}
\hfill
\includegraphics[width=0.47\textwidth]{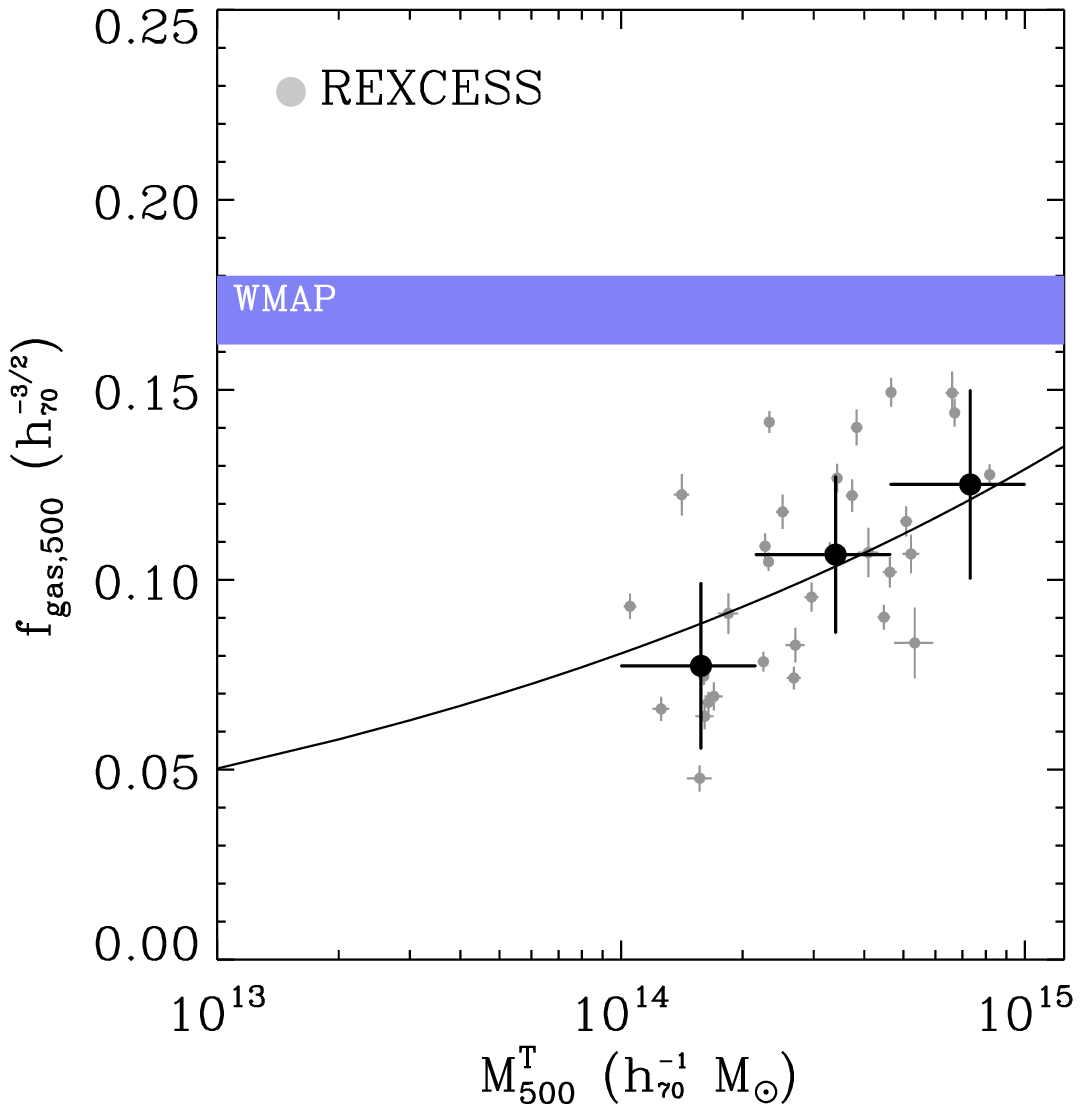}

\caption{{\footnotesize Gas mass fraction vs mass. {\it Left panel\/:} Trend of gas mass fraction versus mass derived from X-ray measurements of 41 groups and clusters with high quality hydrostatic mass estimates \citep{app07,vikh06,sun08}. Grey points are actual measurements; black points are mean values in logarithmic mass bins. The solid line is the orthogonal BCES fit to the unbinned data, $f_{\rm gas,500} \propto M^{0.2}$. {\it Right panel\/:} Approximate gas mass fraction versus mass measurements for the \rexcess\
    sample, where the masses have been estimated from the $M_{500}-T$ relation of
    \citet{app05}. Black points again show the mean trend for three logarithmic mass bins. The solid line is the same as in the left panel. The band illustrates the WMAP 5-Year baryon fraction constraints \citep{dunk09}.}}\label{fig:fgas} 
   \end{figure*}

In Figure~\ref{fig:figQ} we plot $\hat{Q}(T)_{|\rm bin}$ versus system temperature for both apertures considered in the present work. The evidence for a
correlation between $\hat{Q}(T)_{|\rm bin}$ and system temperature is very weak on a Kendall's $\tau$ test, being 
significant at only $\sim 8$ per cent for the full aperture and $\sim 15$ per cent for the $[0.15-1]\,\Rv$ aperture. Interestingly, the standard deviation of $\hat{Q}(T)_{| \rm bin}$ in the $[0.15-1]\,\Rv$ aperture, $\sigma_{\hat{Q}(T)} = 0.14$, gives an observational limit of the variation of cluster structure outside the core regions at fixed temperature. That it is only on the order of 15 per cent argues for a cluster population of remarkable structural similarity. It is thus unlikely that a systematic dependence of cluster structure on temperature/mass can be a major cause of steepening of the \LxT\ relation of the present sample. However, the effect of a systematic dependence of gas structure at smaller scale with temperature/mass remains an open issue. More detailed assessment of the gas clumpiness requires combining X-ray data with high quality SZ data. 

A more likely explanation is if the gas mass-temperature relation
$M_{\rm gas}-T$ deviates from predictions (or equivalently, if the gas
mass fraction, $f_{\rm gas} = M_{\rm g,500}/\Mv$ varies with mass), as was discussed in previous work based on {\it ROSAT\/} data \citep[e.g.,][]{na01,mme99}. Recent
results on the relaxed cluster sample of \citet{app07} and on the
\rexcess\ sample itself \citep{croston08} have shown that $M_{\rm
g,500} \propsim T^{2}$, in agreement with previous work, thus implying a steeper dependence than the self-similar prediction of
$M_{\rm gas} \propto T^{3/2}$. Such a dependence of $M_{\rm gas}$ on
$T$ would imply $f_{\rm gas} \propto M^{1/3}$ for a self-similar $M
\propto T^{3/2}$ relation. 

The left hand panel of Figure~\ref{fig:fgas} shows the $f_{\rm
gas,500}-M_{500}$ relation for 41 systems ranging from $10^{13}$ to $10^{15}\,M_\odot$ for which gas mass fraction estimates derived from hydrostatic mass measurements are available \citep{app07,vikh06,sun08}. The trend of mean gas fraction with mass is evident to the eye. To better illustrate this, we have divided the data into five approximately equal logarithmic total mass bins and calculated the mean and standard deviation of the gas mass fraction measurements in each bin.  A BCES orthogonal fit to the combined (not binned) data set in log-log space yields the relation:

\begin{eqnarray*}
h(z)^{3/2} \ln{f_{\rm gas,500}} & = &(-2.37\pm0.03) \\
& & + (0.21\pm0.03) \times \ln{(M / 2\times10^{14}\, M_\odot})
\end{eqnarray*}

\noindent with $\sigma_{\rm lnf} = 0.12\pm0.03$ dispersion. The right hand panel of Figure~\ref{fig:fgas} shows the corresponding trend of $f_{\rm gas,500}$ with cluster mass for the \rexcess\ sample, where the data values have been estimated using the $M$--$T$ relation of \citet{app05}. Dividing the data into three mass bins and averaging the gas mass fraction measurements in each bin yields the thick solid points, which are in good agreement with the best fitting relation to the combined data from hydrostatic estimates discussed above. 

The trend in $f_{\rm gas}$ implies a decrease in gas content in
poorer systems relative to higher mass systems, a fact which 
manifests itself in an increase in ICM entropy and consequent suppression in luminosity at lower
masses. The two most likely physical possibilities are a variation with mass in the efficiency of conversion of baryons into stars
or {\it in situ} non-gravitational energy input from e.g., supernova feedback or AGN \citep*[e.g.][]{pss08,bmb08}, or a combination of the two. In a forthcoming paper we will use the entropy distributions to probe the source(s) and extent of the entropy redistribution.

Better constraints in the group regime are still required, especially in the light of the increased scatter apparently seen there (\citealt{op04}; but see \citealt{sun08}).


\section{Conclusions}

We have presented a detailed study of the luminosity scaling relations
of \rexcess, a galaxy cluster sample selected by X-ray luminosity in
such a way as to optimally sample the cluster X-ray luminosity
function. \rexcess\ contains objects of different dynamical states
with a range of core X-ray properties, allowing us to investigate
the effect of the presence of these systems on the scaling
relations. The homogeneous nature of the sample data, which have all
been observed with the same satellite to approximately the same depth, combined with an analysis approach based on extraction of relevant quantities within
scaled apertures, has been designed to minimise measurement scatter. 
We found the following results:

\begin{itemize}

\item The slope of the luminosity-temperature, luminosity-$Y_X$ and
  luminosity-mass relations are all steeper  at greater than 99 per cent confidence, than expected for
  self-similar gravitational collapse scenarios. 
  
\item  The dependence of the radially averaged structure
  factor $\hat{Q}(T)_{| \rm bin}$ on temperature, where both quantities have been
  estimated for the first time within $\Rv$, is not significant either when measured from all emission, or from core-excluded emission. This suggests that, contrary to previous results, structural variation cannot be
  a significant contributor to the steepening of the relations unless there is a very strong temperature/mass dependence of gas clumpiness. Furthermore, the scatter in $\hat{Q}(T)_{| \rm bin}$ measured in the $[0.15-1]\,\Rv$ aperture is $\sim 15$ per cent, illustrating the remarkable structural similarity of the present sample in the outer regions.

\item There is strong evidence for a decrease in gas mass content in poorer
  systems relative to higher mass systems, or in other words, a
  dependence of the gas mass fraction on total mass. This effect is clearly 
  seen in the $M_{\rm gas}-T$ relation of the
  present sample \citep{croston08} and has been seen in many previous samples. Using the
  total mass determined from published hydrostatic estimates of a combined sample of clusters and groups spanning $10^{13}$--$10^{15}\, M_\odot$, we find that the gas mass fraction depends on total mass such that $f_{\rm gas}
  \propto M^{0.21\pm 0.03}$. The trend in the \rexcess\ data, when total masses are estimated using the $M$--$T$ scaling relation, is similar.  This dependence is the dominant cause of the steepening of the X-ray luminosity scaling relations.

\item For the whole sample, the scatter of the X-ray luminosity-temperature relation derived from all emission interior to $\Rv$ is up to 75 per cent depending on the exact fitting method; the
  scatter is less than 40 per cent for the X-ray luminosity-$Y_X$ relation. The scatter is not strongly compatible with a Gaussian distribution in either case, the distribution being characterised by a tail caused by the presence of cool core systems.
 
\item Cooling core systems, the one-third of the sample with the highest central density, describe the high luminosity envelope and contribute the majority of the variance to
  the relations. A fit only to the cool core systems suggests that
  they are offset from the relation by a simple normalisation
  factor. However, there is nearly two times more logarithmic scatter about
  the X-ray luminosity-temperature relation for cool core systems compared
  to that for non cool core systems. This suggests that there are
  large differences in the core structure even for cool core systems
  as a class. 

\item Systems exhibiting morphological substructure tend towards the
  lower luminosity envelope of the relations. Partly this is due to the increase
  in normalisation of the total sample due to the presence of cool core
  systems. The scatter about the X-ray luminosity-temperature relation
  of morphologically disturbed systems is 65 per cent, compared to a scatter of 48 per cent for cool core systems; within the uncertainties, the variance is in fact identical. Furthermore, the scatter in morphologically disturbed
  systems is identical to that for morphologically relaxed systems, due partly to the preponderance of cool core systems in the relaxed subsample.

\item Simple exclusion of the emission interior to $0.15\,\Rv$ results
  in a reduction of scatter in all relations. For the X-ray luminosity-
  temperature relation, the natural logarithmic scatter is 30 per cent, a
  reduction of more than a factor of two. Similarly significant 
  reductions are seen in  other relations. After exclusion of the core, the scatter in luminosity-temperature and luminosity-$Y_X$ relations is well described with a Gaussian distribution at $> 85$ per cent confidence. A reduction in scatter can also be achieved by considering
  the central gas density, $n_{e,0}$, as a third parameter in the
  scaling relations. The $L-T-n_{e,0}$ relation has a natural logarithmic
  scatter of 47 per cent; the $L-Y_X-n_{e,0}$ relation has a scatter of 22 per cent.

\item Using $Y_X$ as a mass proxy, a Malmquist bias corrected luminosity mass relation for \rexcess\ is steeper than the raw relation due to the under-representation, for a given mass, of low luminosity clusters in the sample. 

\end{itemize}

The behaviour of the observed luminosity scaling relations thus appears to be driven principally by a mass dependence of the total gas content. Plausible physical explanations for the dependence are a variation with mass in the efficiency of conversion of baryons into stars or {\it in situ} heating after accretion. Greater understanding of the source of the dependence will require deep observations of a similarly representative sample of group scale haloes, to measure accurate luminosities and probe the underlying physical causes; furthermore, precise calibration of the evolution of the scaling relations is needed, ideally with a similarly-selected distant cluster sample, to probe the effect over time. 

\begin{acknowledgements}

We thank the referee for a useful report which helped to improve the quality of the paper. During the course of this work, we had fruitful discussions with J. Ballet, A.E. Evrard, D. Pierini, E. Pointecouteau, T.J. Ponman and A. Vikhlinin. Pertinent comments on the manuscript were also received from S. Borgani, J.P. Henry, R. Kneissl, T.H. Reiprich, A.J.R. Sanderson, and G.M. Voit. GWP acknowledges support from DfG Transregio Programme TR33. The present work is
based on observations obtained with {\it XMM-Newton}, an ESA science
mission with instruments and contributions directly funded by ESA
Member States and the USA (NASA). The \xmm\ project is supported
in Germany by the Bundesministerium f\"ur Wirtschaft und
Technologie/Deutsches Zentrum f\"ur Luft- und Raumfahrt (BMWI/DLR, FKZ
50 OX 0001), the Max-Planck Society and the Heidenhain-Stiftung.
\end{acknowledgements}

\bibliographystyle{aa} 
\bibliography{gwpbib}

\appendix

\section{Cluster image gallery, sorted by cool core and $\langle w \rangle$ classification}\label{app:app1}

As an aid to visualisation of the sample, in this Appendix we present the images of each cluster that were used to calculate the centroid shift parameter $\langle w \rangle$. We remind the reader that $\langle w \rangle$ was evaluated with the central $0.1\,\Rv$ excised, to avoid biasing as a result of the highly peaked surface brightness of cool core systems. The images are derived from the three EPIC detectors and have been corrected for vignetting; in addition, point sources have been removed and replaced by Poisson noise sampled from counts in an annulus surrounding the excised source. 

\begin{figure*}[]

\includegraphics[scale=1.,angle=0,keepaspectratio,width=0.195\textwidth]{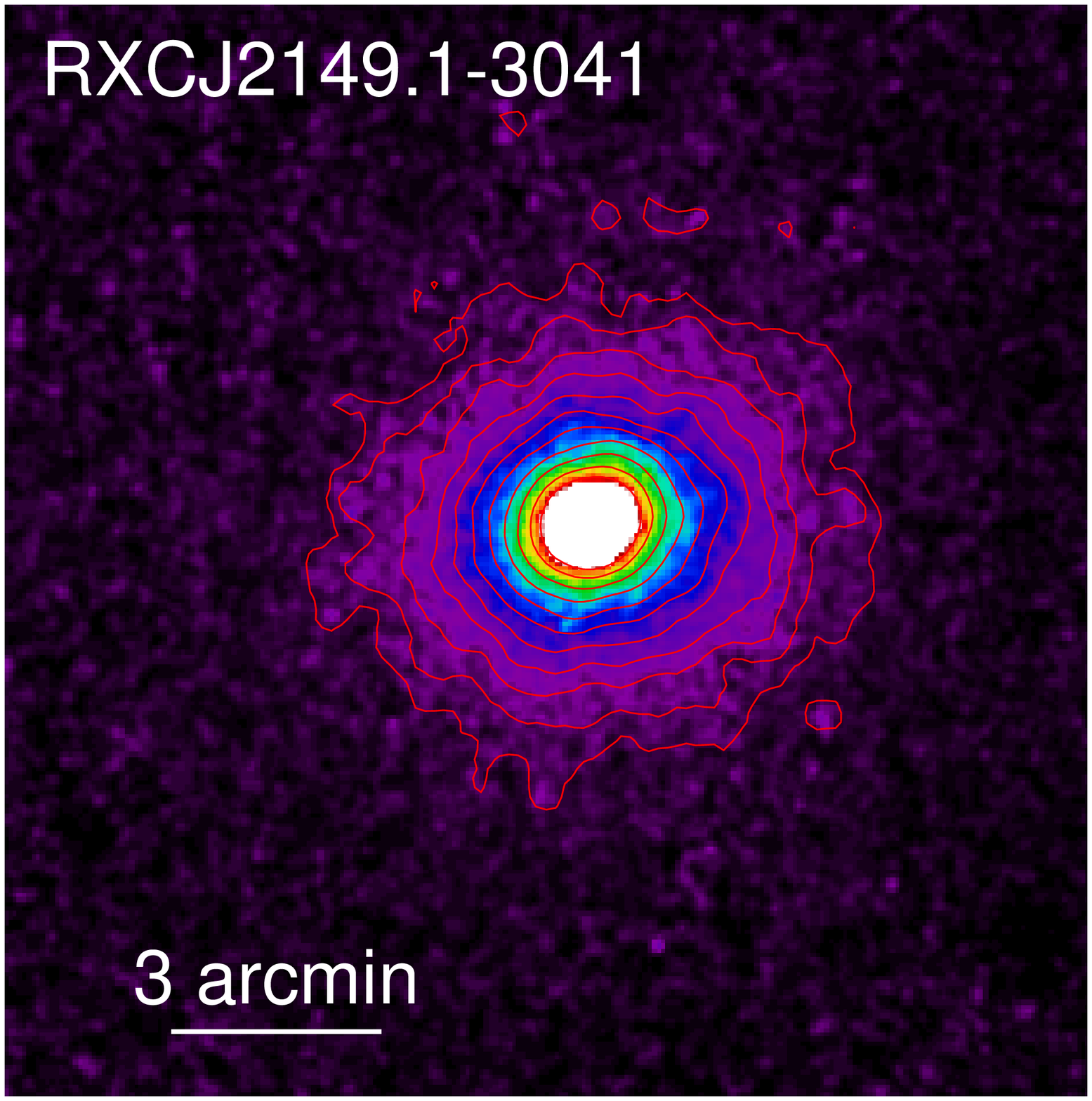}
\hfill
\includegraphics[scale=1.,angle=0,keepaspectratio,width=0.195\textwidth]{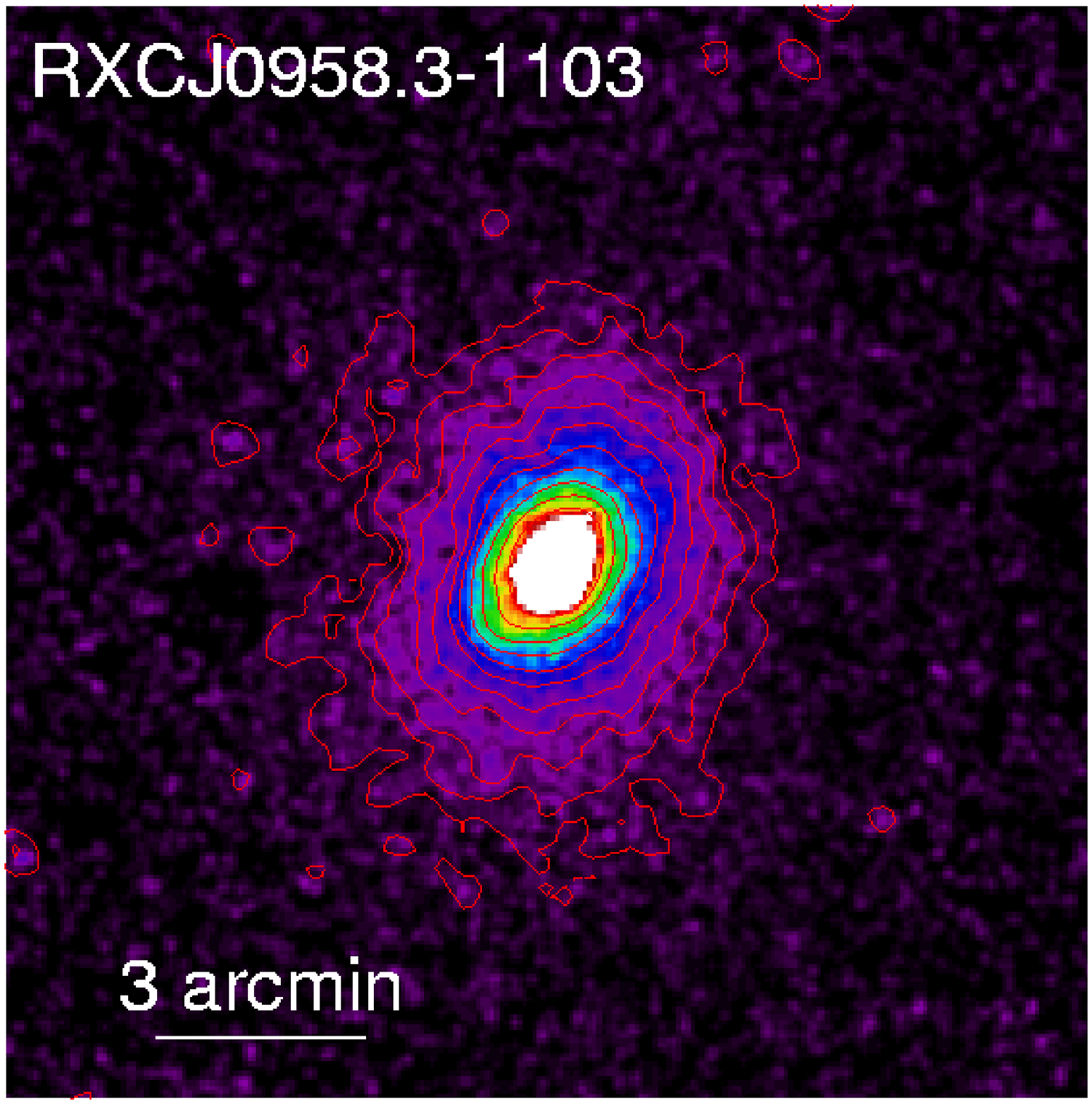}
\hfill
\includegraphics[scale=1.,angle=0,keepaspectratio,width=0.195\textwidth]{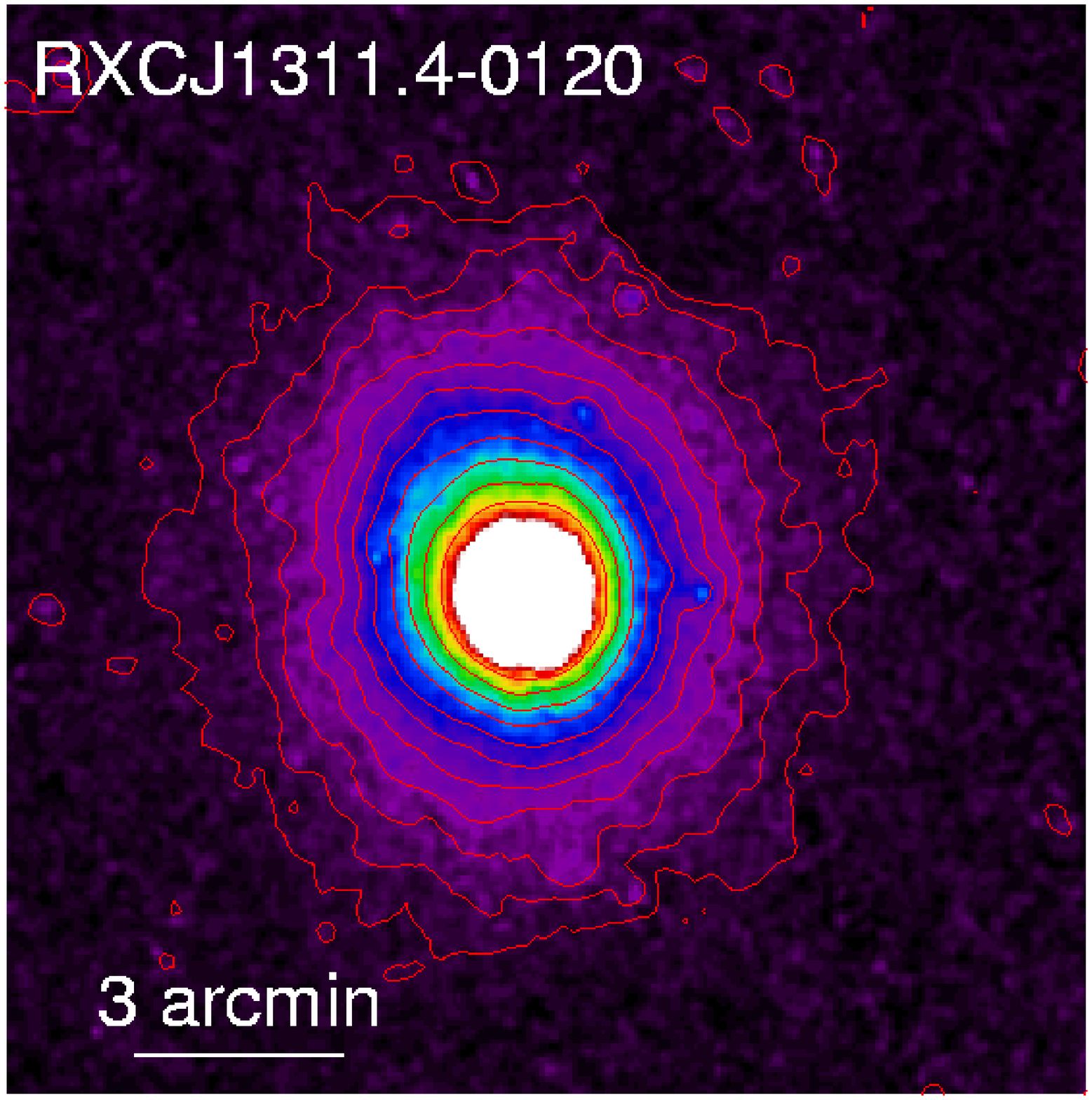}
\hfill
\includegraphics[scale=1.,angle=0,keepaspectratio,width=0.195\textwidth]{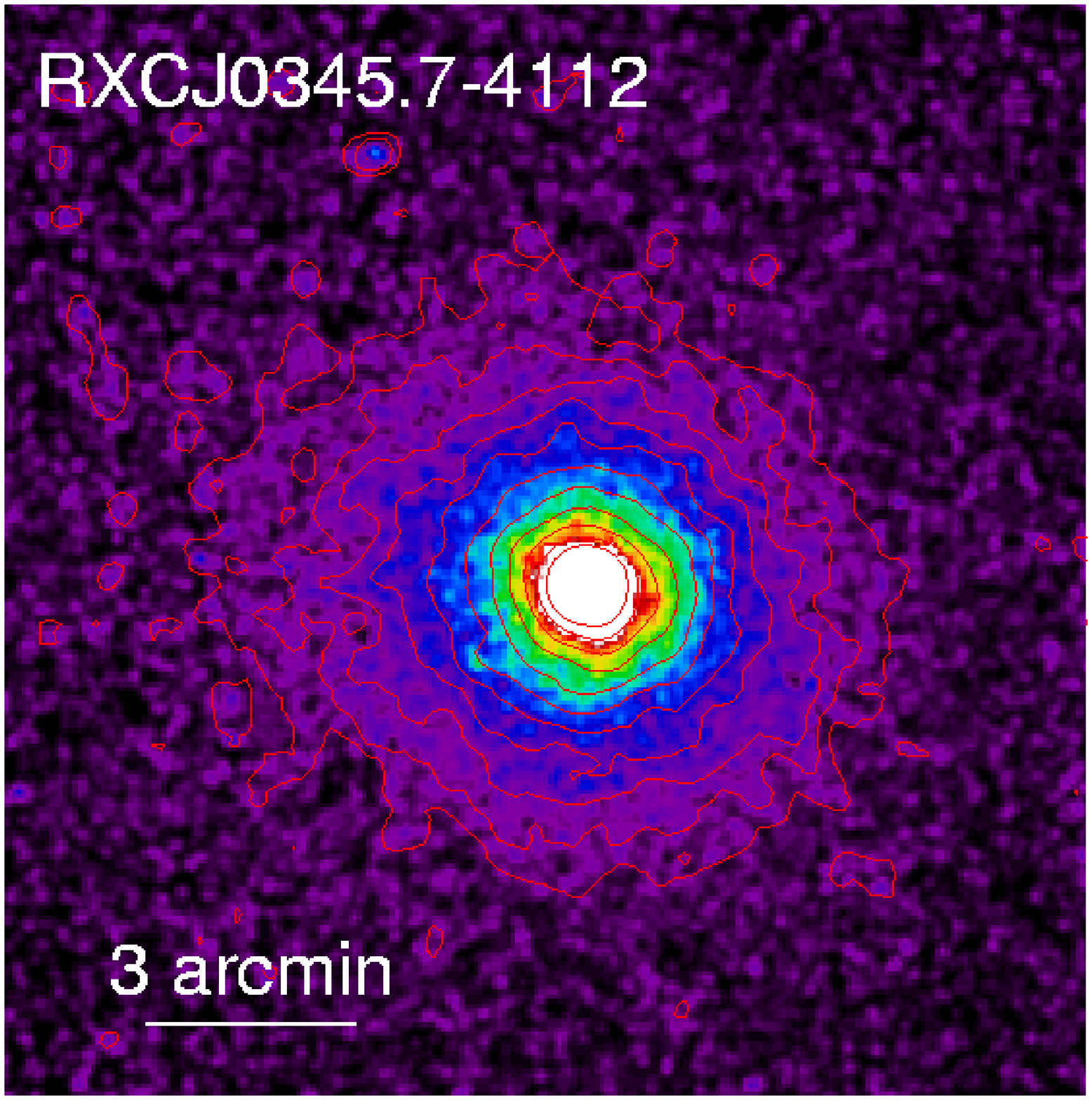}
\hfill
\includegraphics[scale=1.,angle=0,keepaspectratio,width=0.195\textwidth]{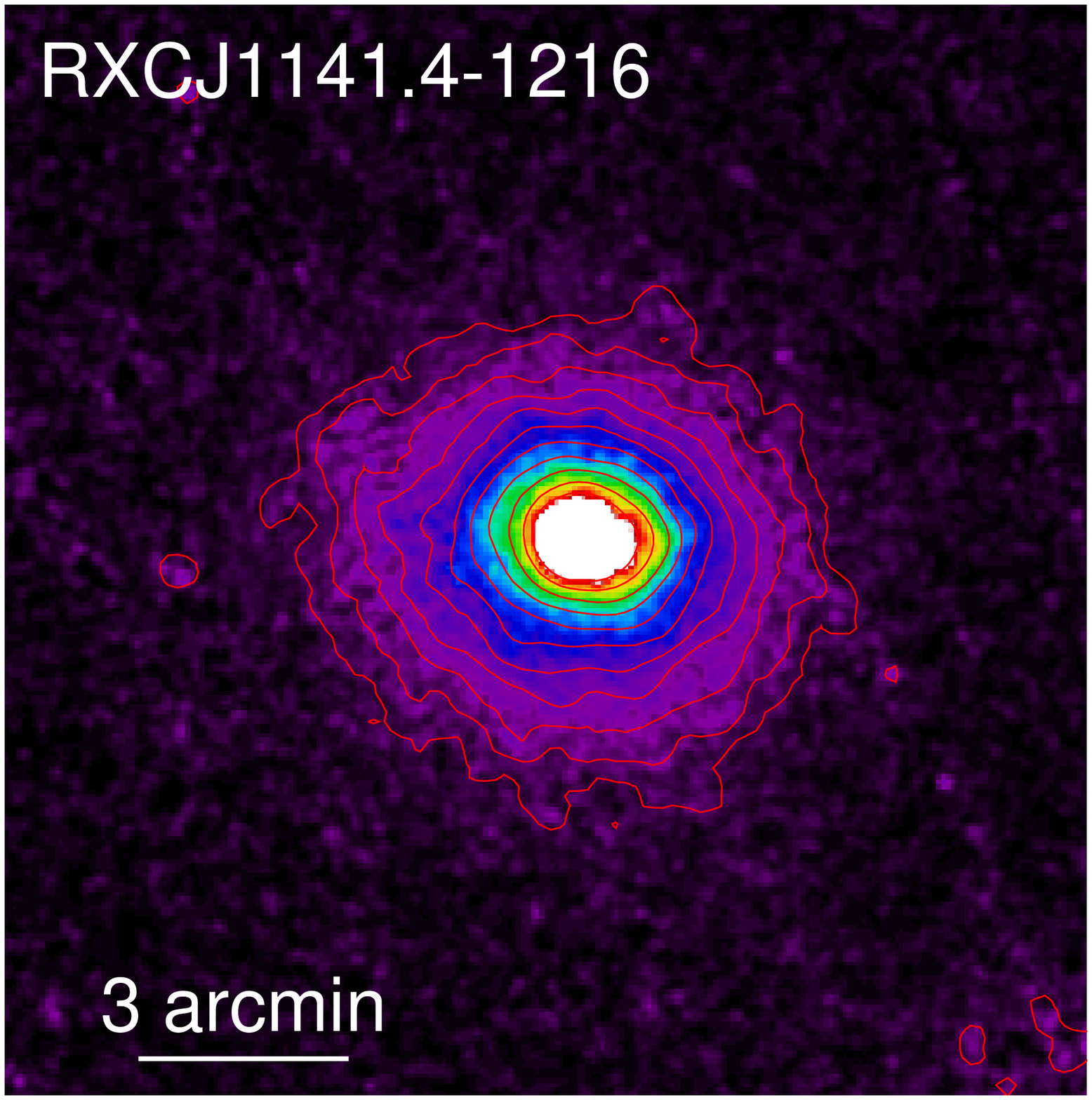}

\includegraphics[scale=1.,angle=0,keepaspectratio,width=0.195\textwidth]{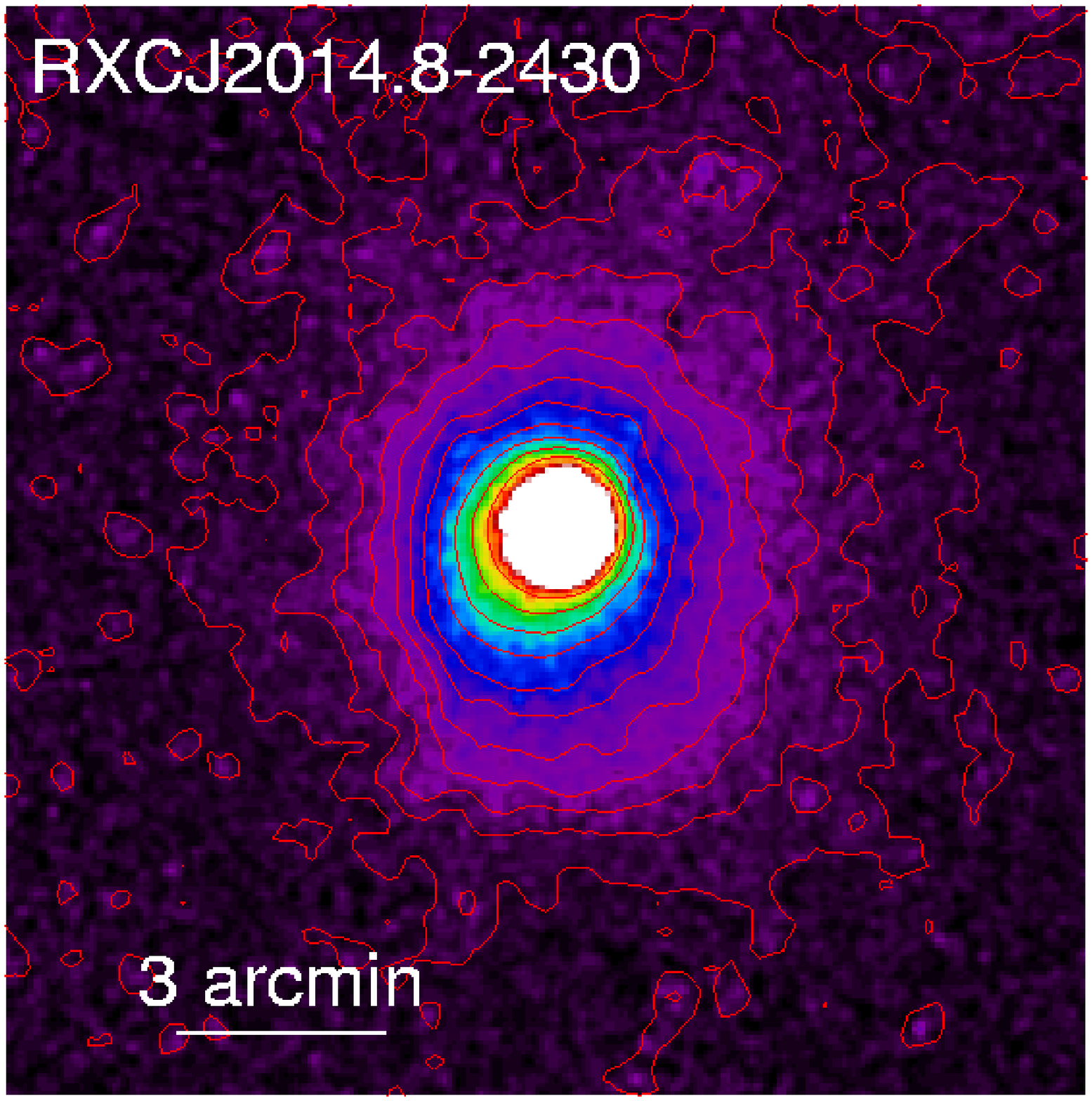}
\hfill
\includegraphics[scale=1.,angle=0,keepaspectratio,width=0.195\textwidth]{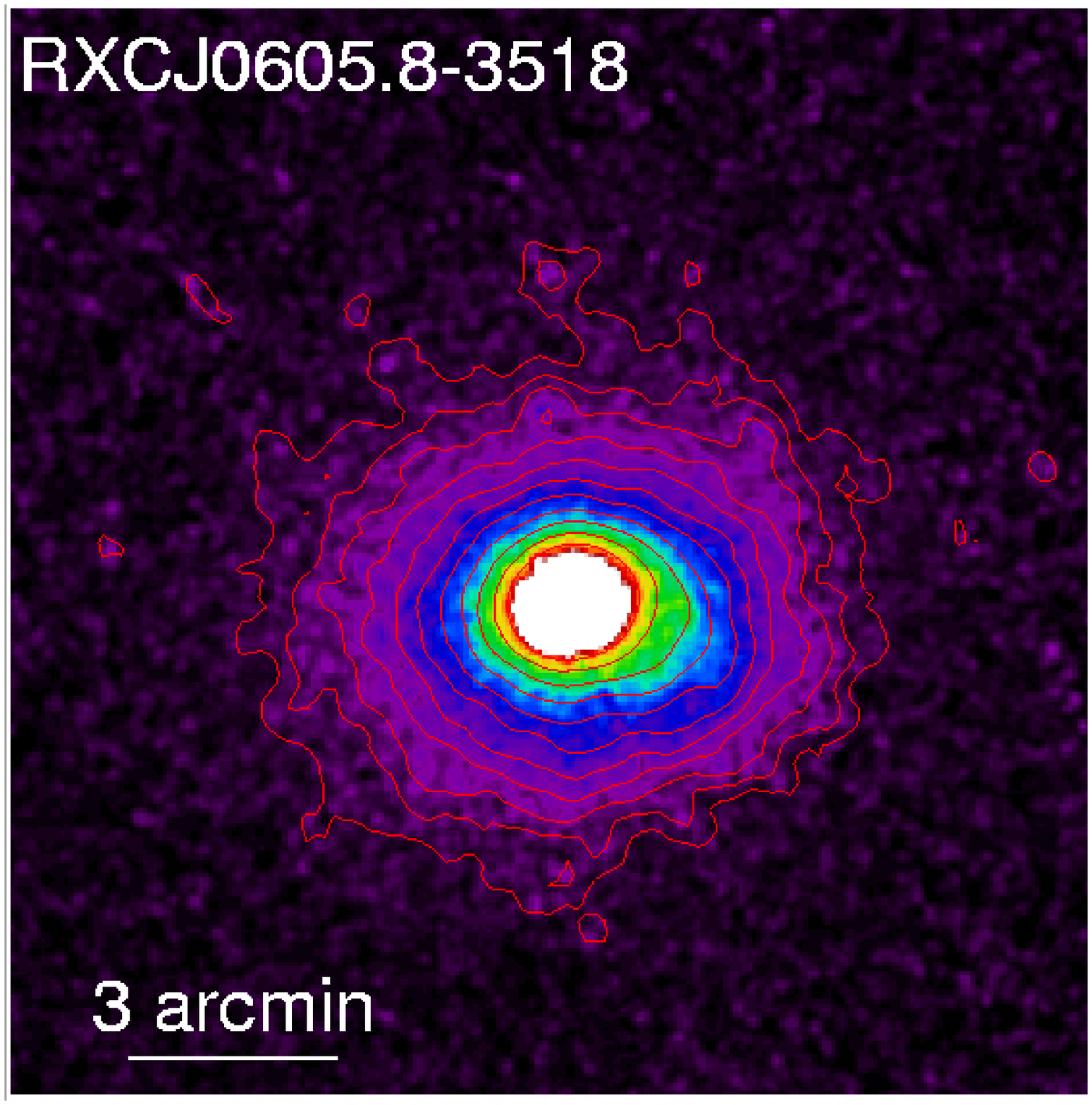}
\hfill
\includegraphics[scale=1.,angle=0,keepaspectratio,width=0.195\textwidth]{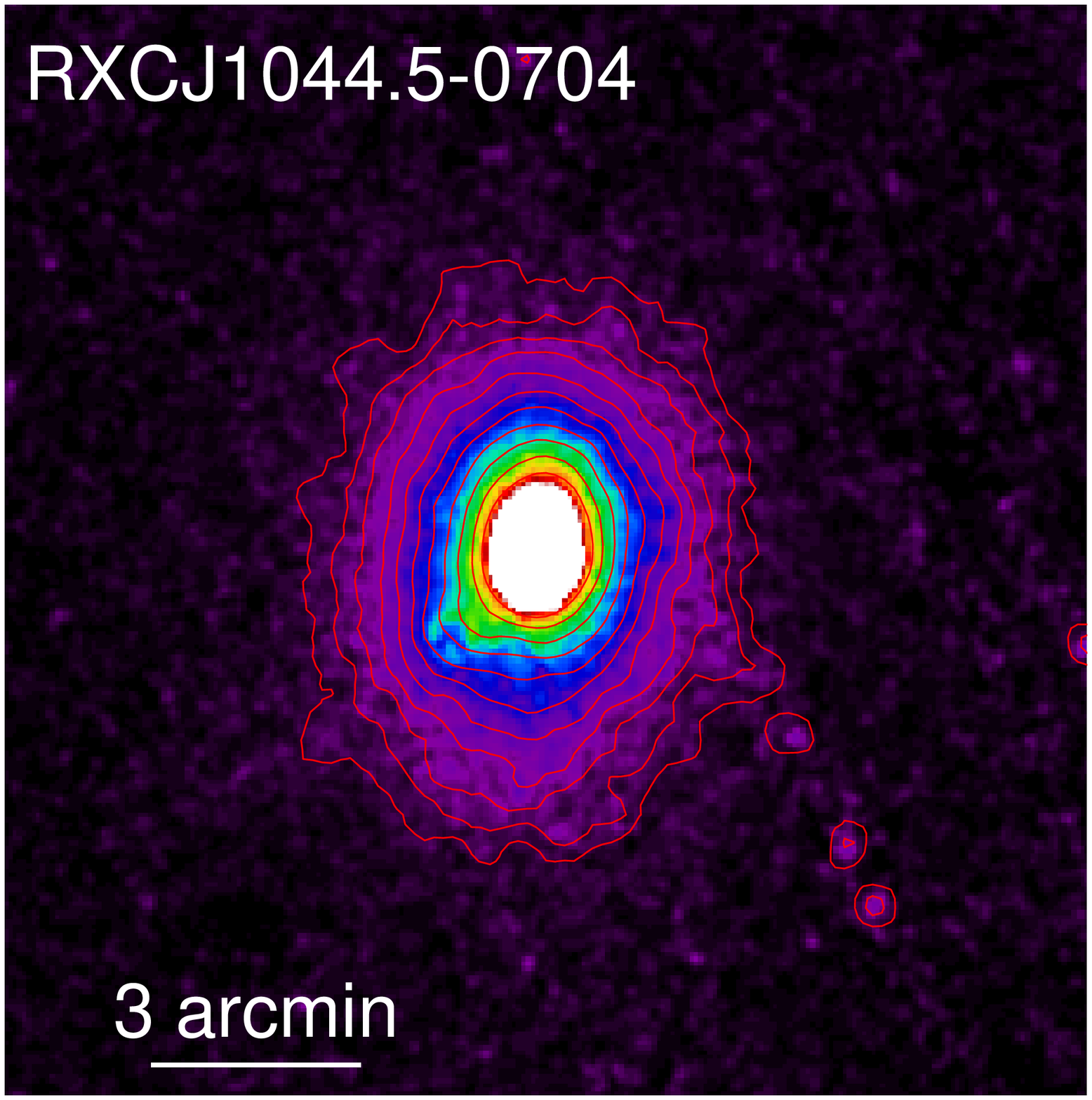}
\hfill
\includegraphics[scale=1.,angle=0,keepaspectratio,width=0.195\textwidth]{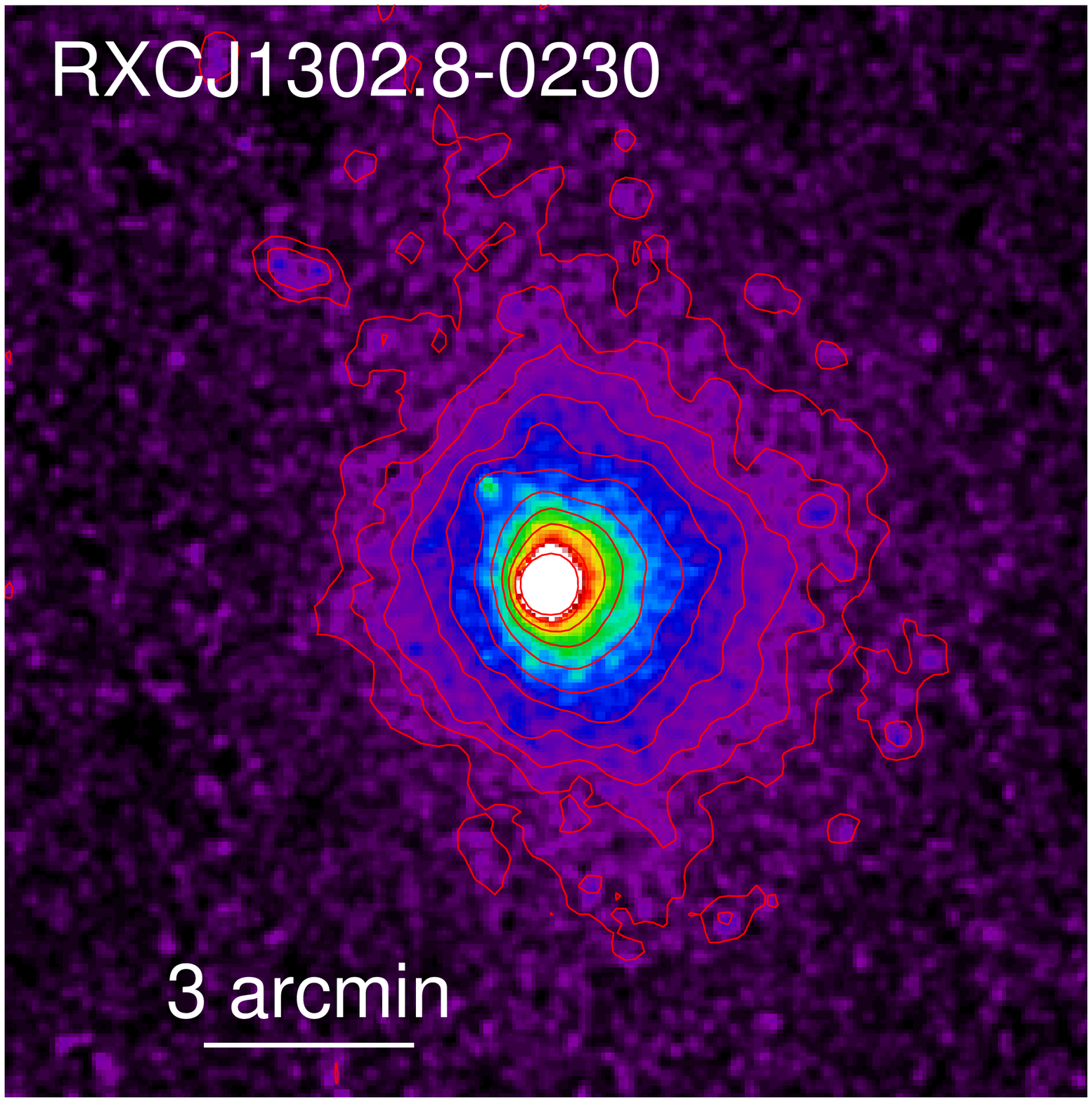}
\hfill
\includegraphics[scale=1.,angle=0,keepaspectratio,width=0.195\textwidth]{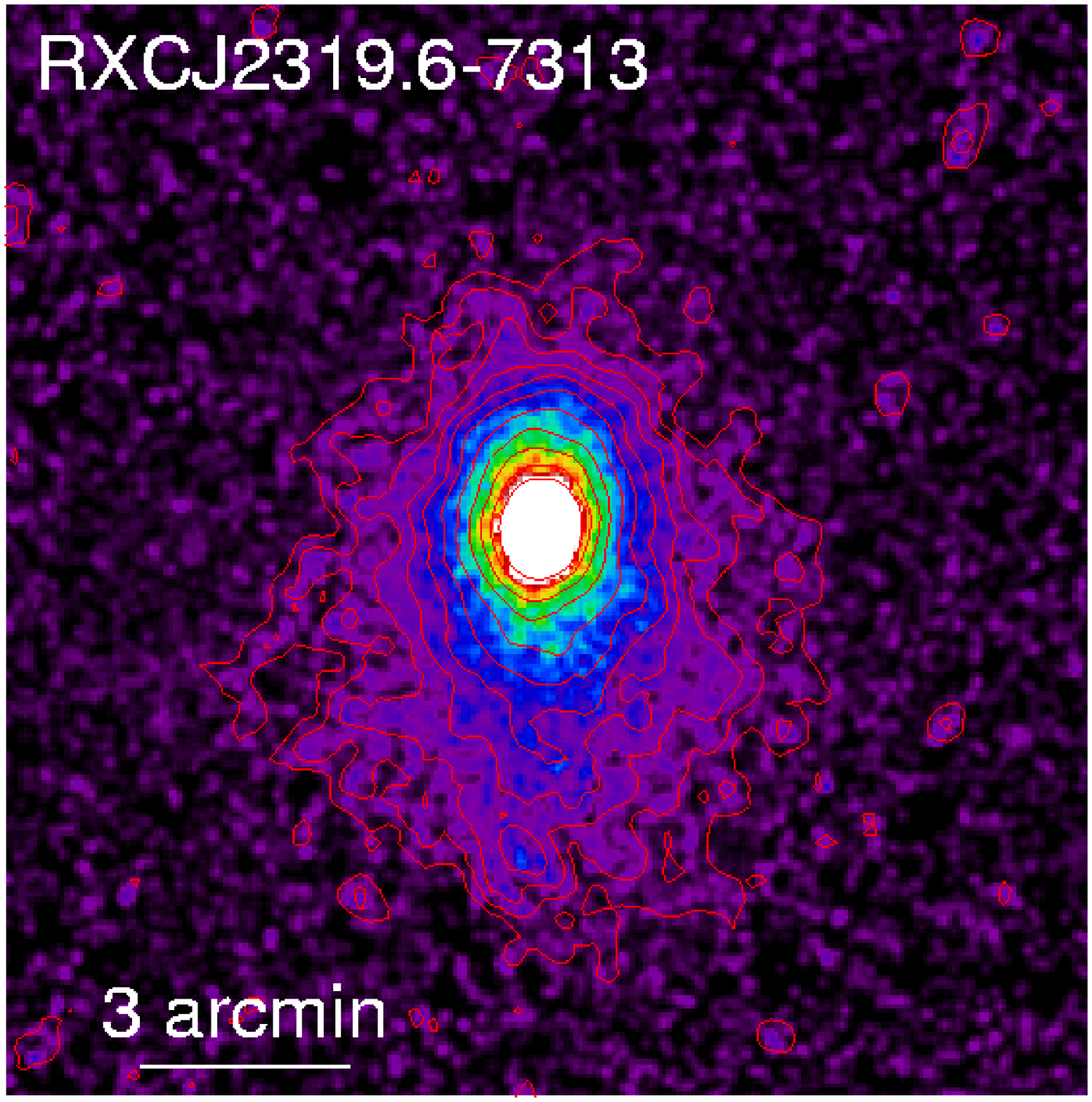}

\caption{{\footnotesize Cool core clusters, sorted from top left to bottom right in order of increasing centroid shift parameter, $\langle w \rangle$. The images are derived from the three EPIC detectors and have been corrected for vignetting. Point sources have been removed and replaced by Poisson noise sampled from counts in a surrounding annulus. Contours increase in steps of $\sqrt{2}$. The colour table of each panel is scaled by a factor of $L_X^{0.22}$ (see \citealt{boehringer07} for details). Note that RXC\,J1302 +0230 and RXC\,J2319 -7313 are classified as both cool core and morphologically disturbed. }}\label{fig:gallery1}   
\end{figure*}

\begin{figure*}[]

\includegraphics[scale=1.,angle=0,keepaspectratio,width=0.195\textwidth]{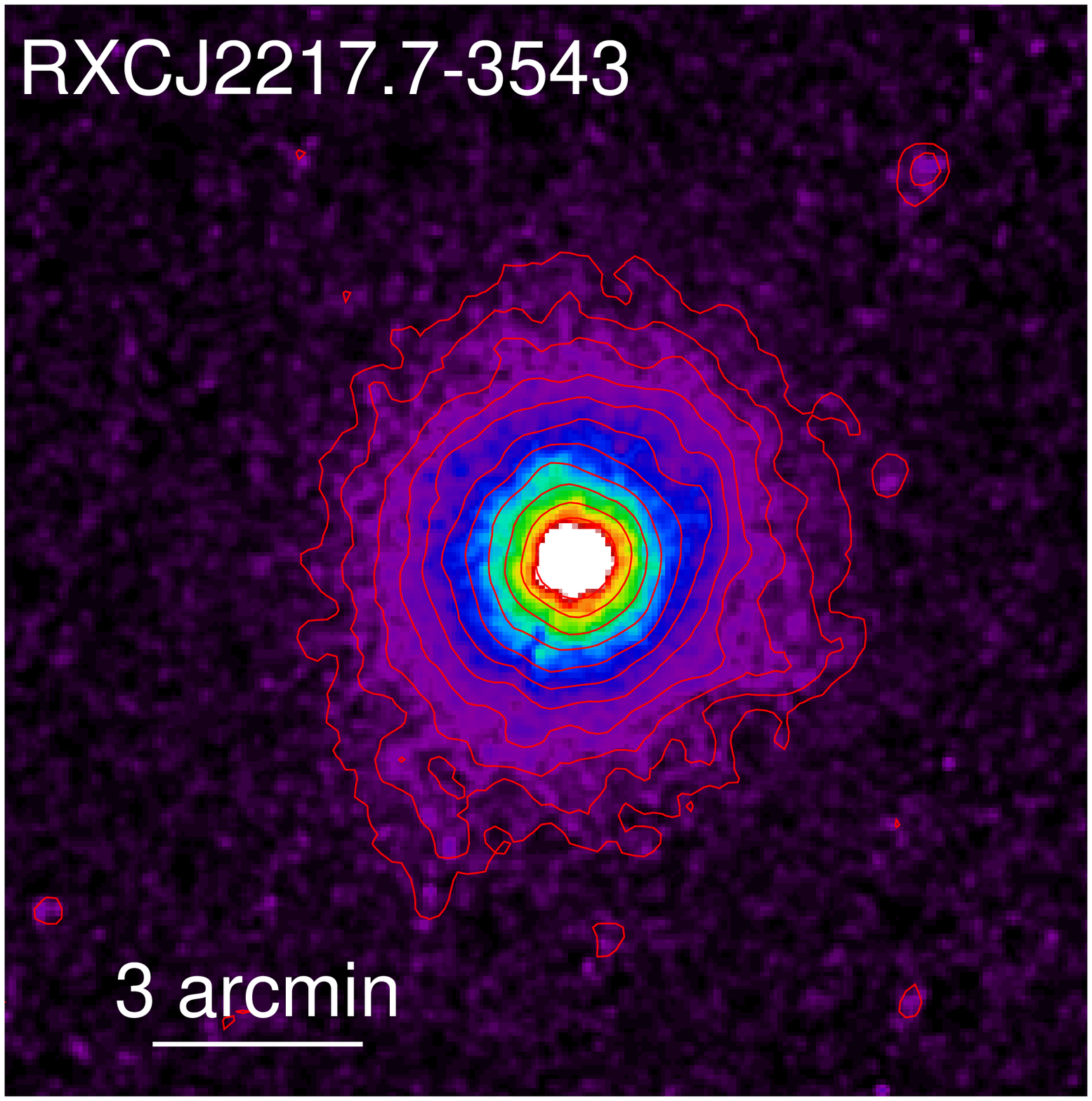}
\hfill
\includegraphics[scale=1.,angle=0,keepaspectratio,width=0.195\textwidth]{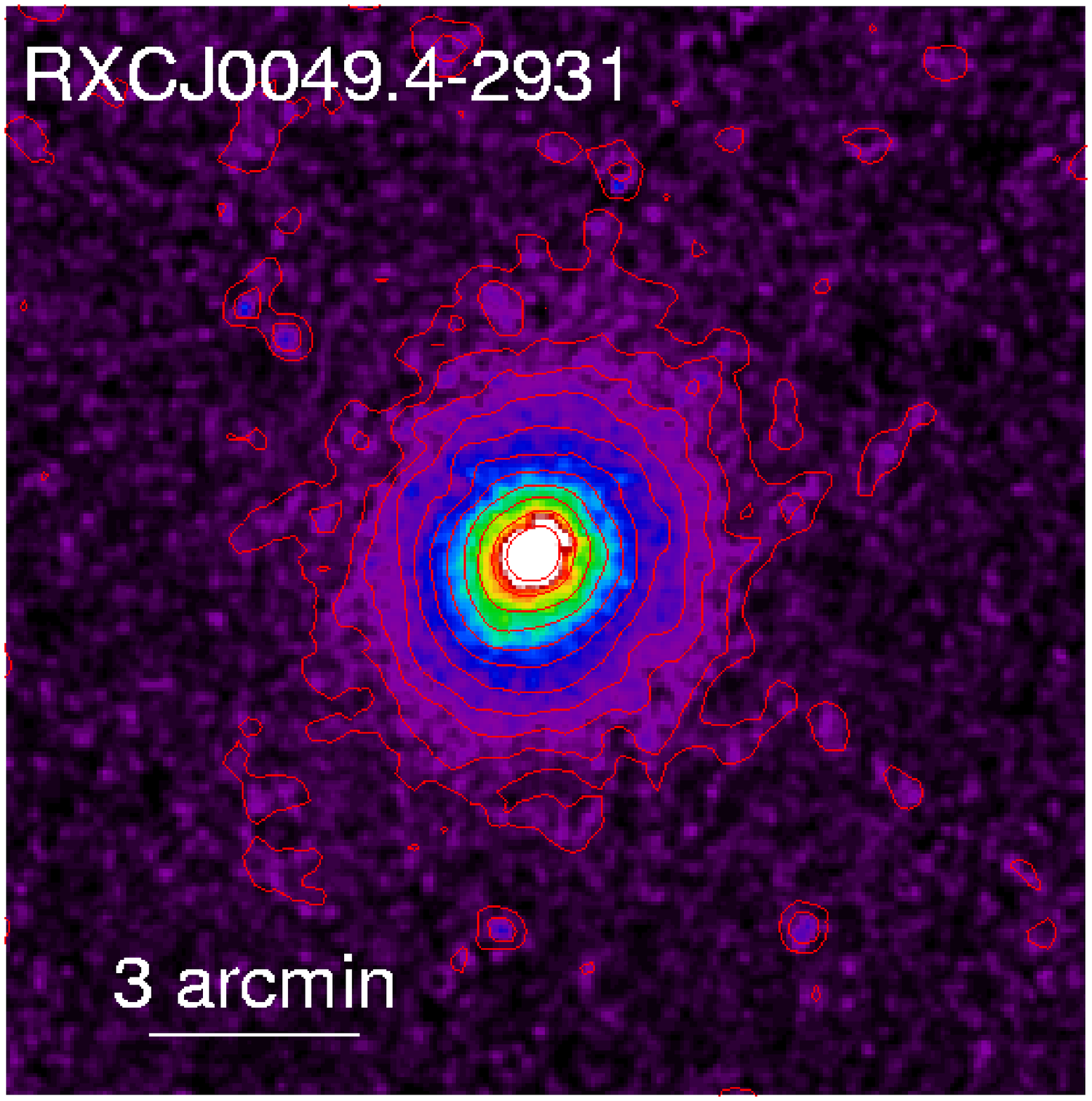}
\hfill
\includegraphics[scale=1.,angle=0,keepaspectratio,width=0.195\textwidth]{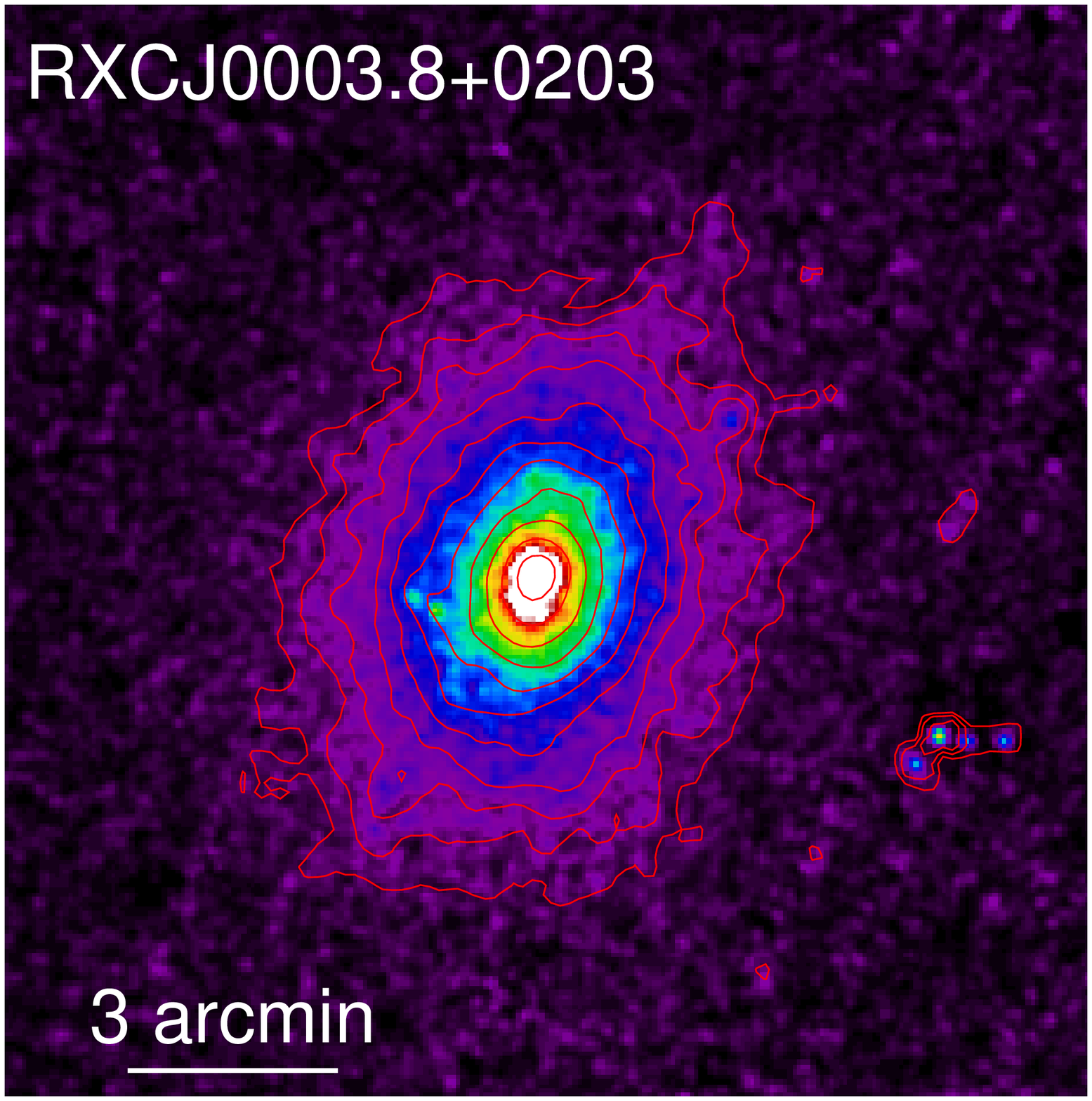}
\hfill
\includegraphics[scale=1.,angle=0,keepaspectratio,width=0.195\textwidth]{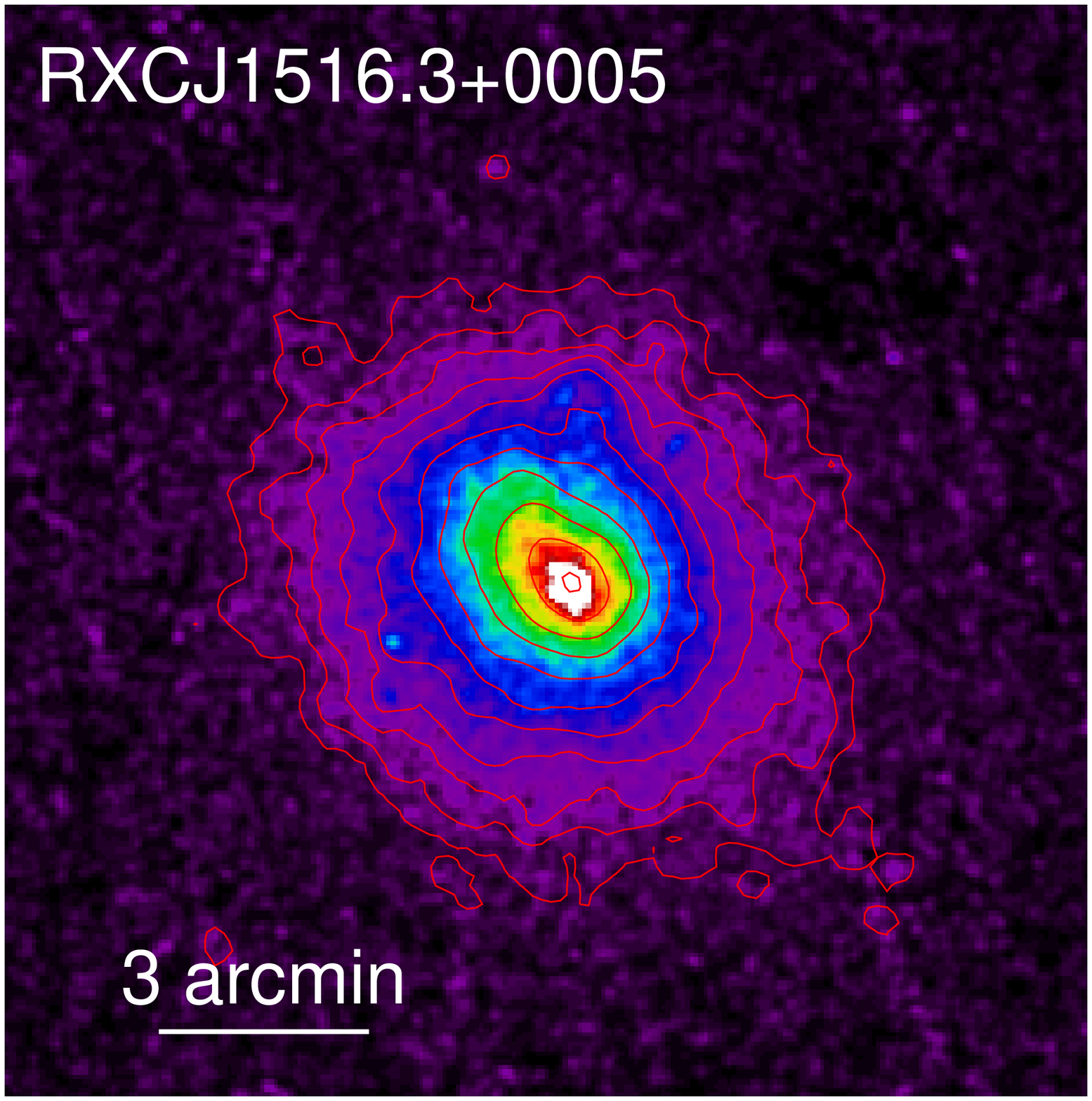}
\hfill
\includegraphics[scale=1.,angle=0,keepaspectratio,width=0.195\textwidth]{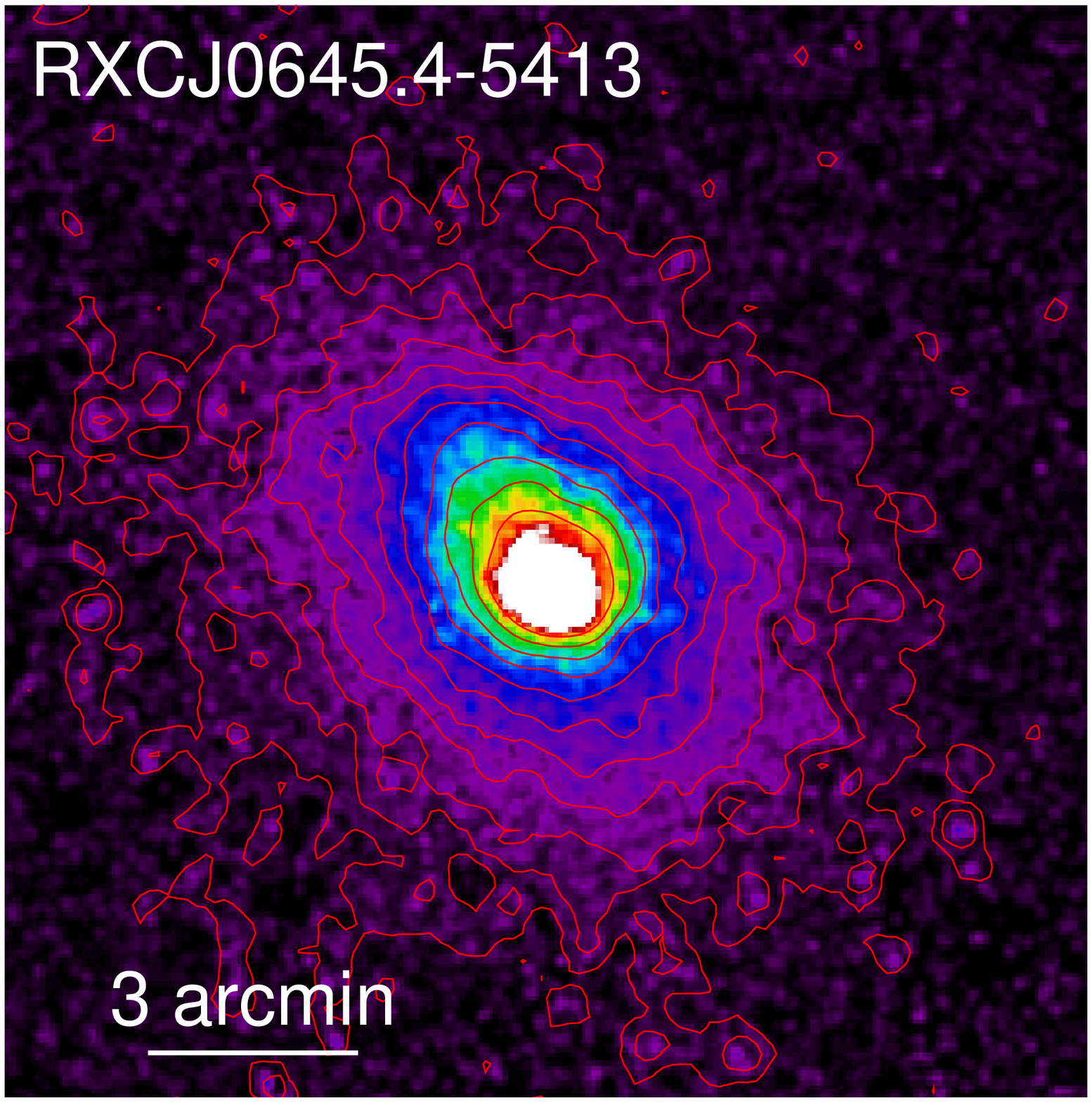}

\includegraphics[scale=1.,angle=0,keepaspectratio,width=0.195\textwidth]{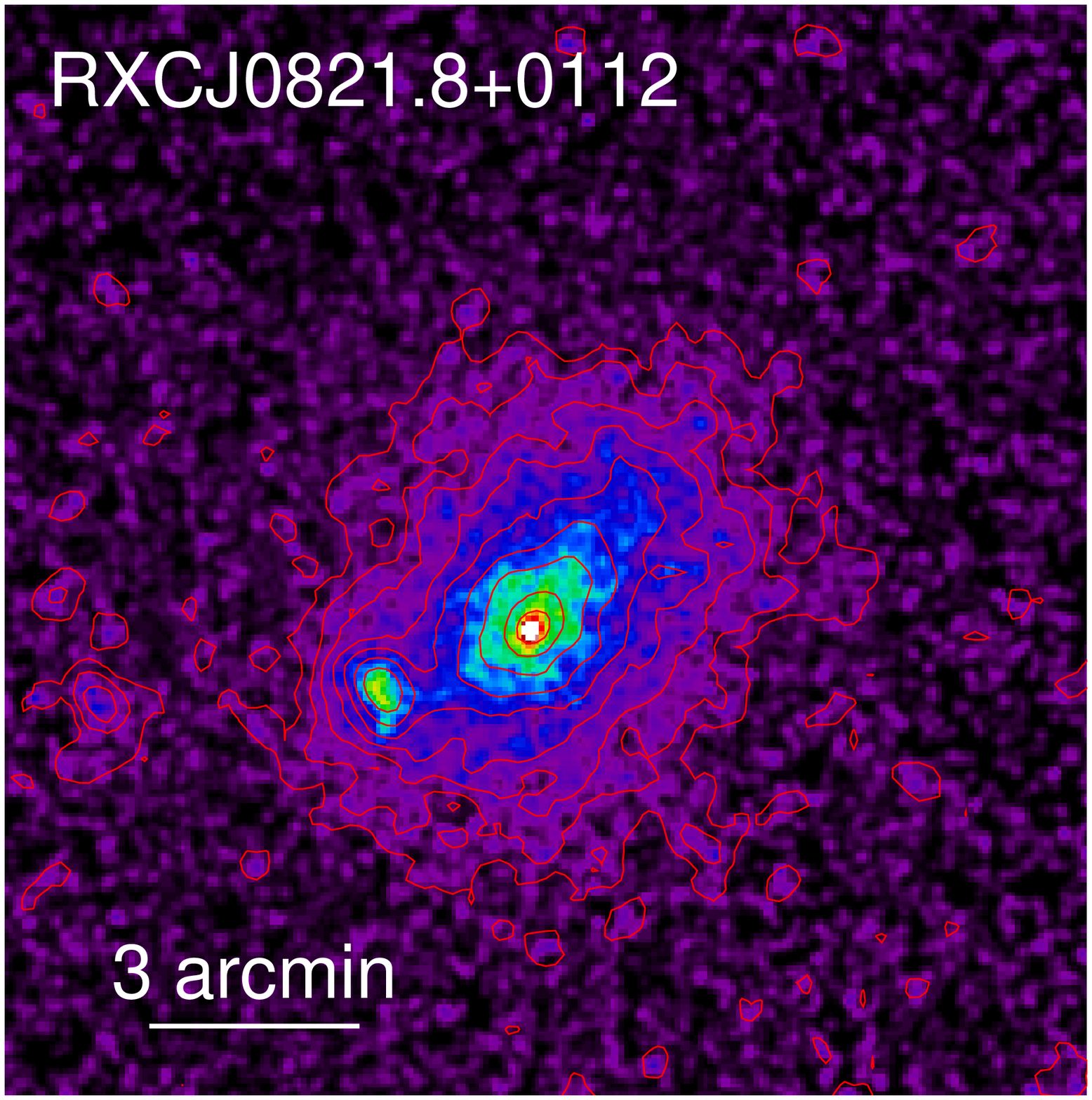}
\hfill
\includegraphics[scale=1.,angle=0,keepaspectratio,width=0.195\textwidth]{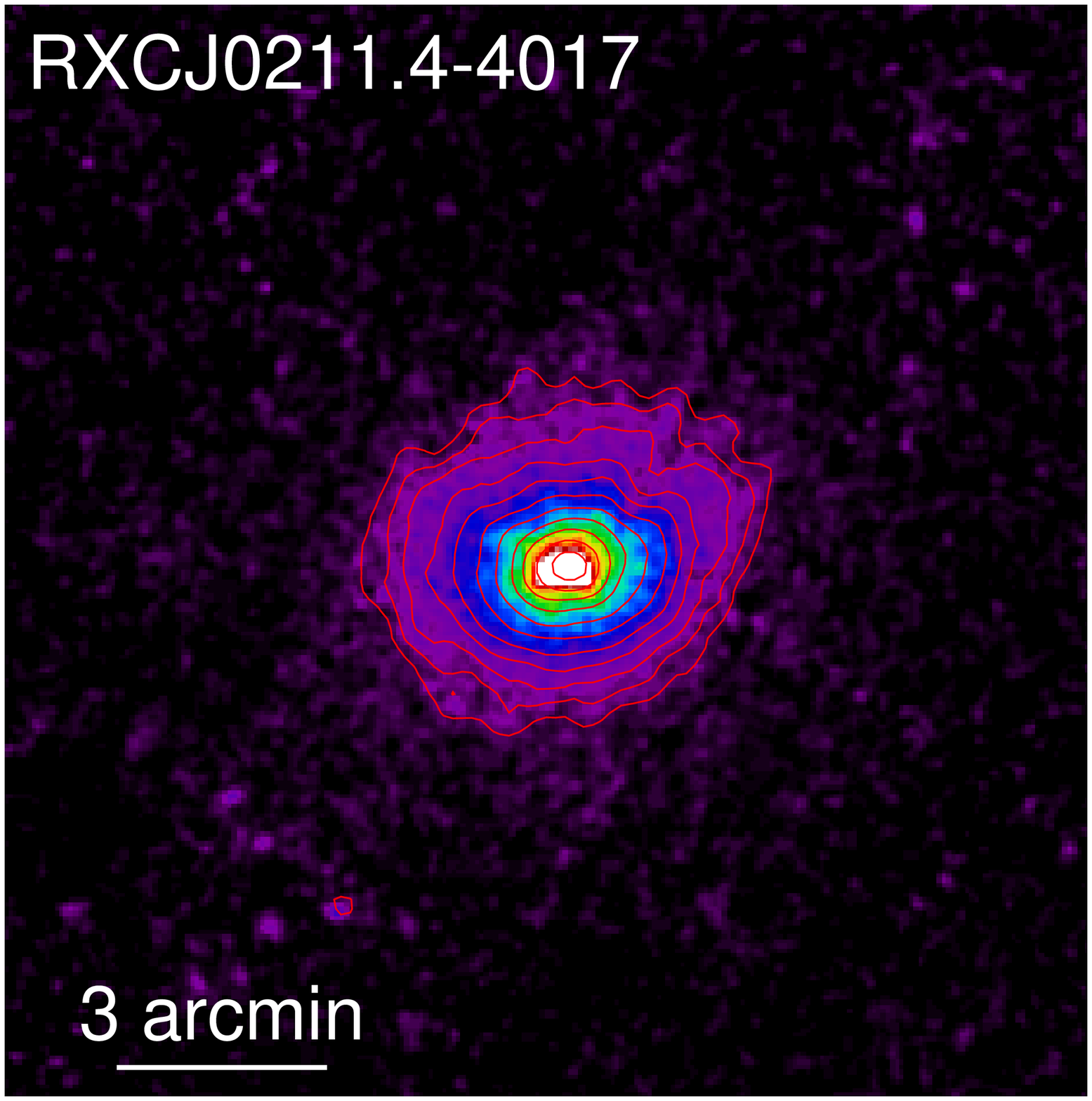}
\hfill
\includegraphics[scale=1.,angle=0,keepaspectratio,width=0.195\textwidth]{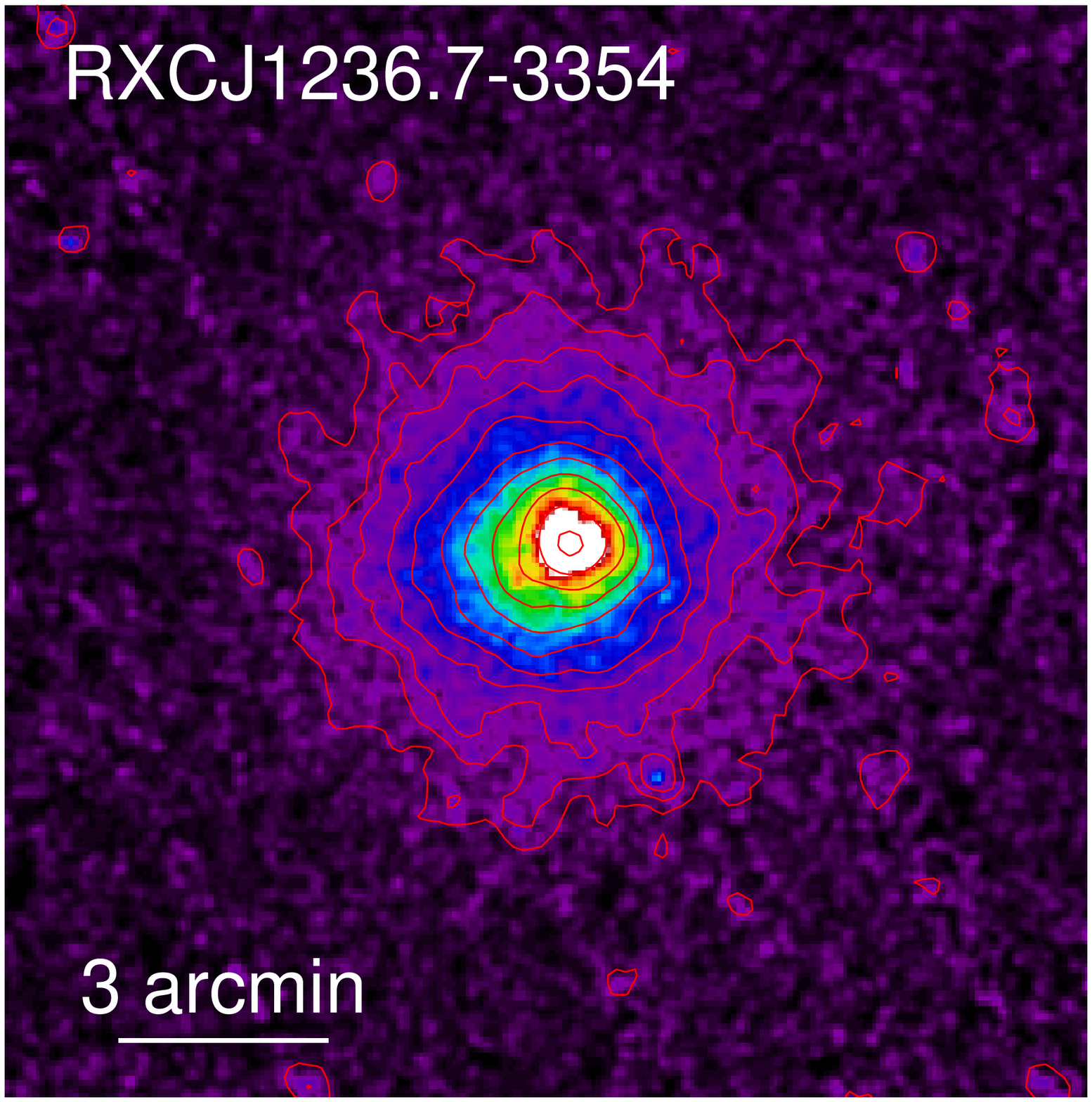}
\hfill
\includegraphics[scale=1.,angle=0,keepaspectratio,width=0.195\textwidth]{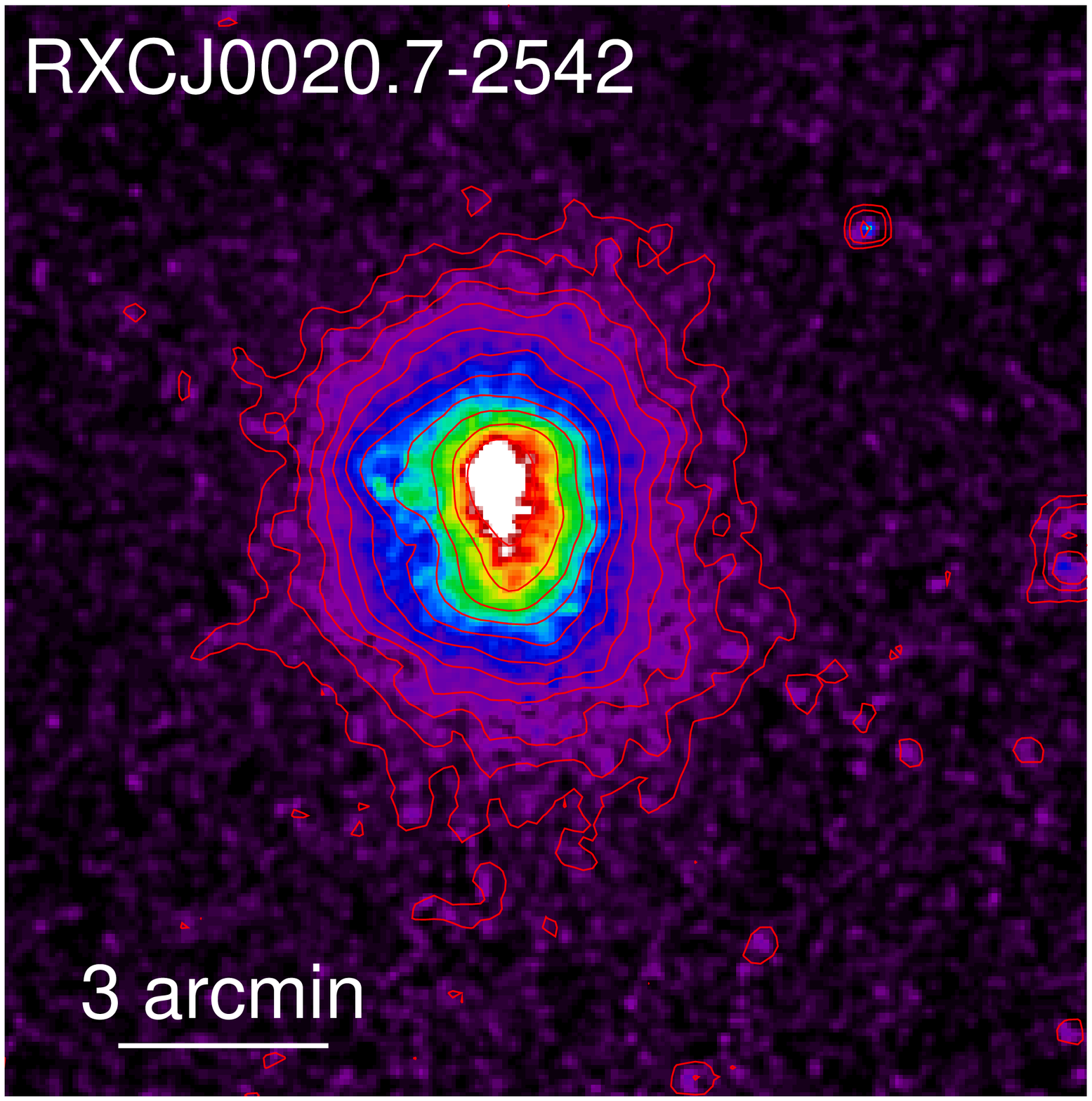}
\hfill
\includegraphics[scale=1.,angle=0,keepaspectratio,width=0.195\textwidth]{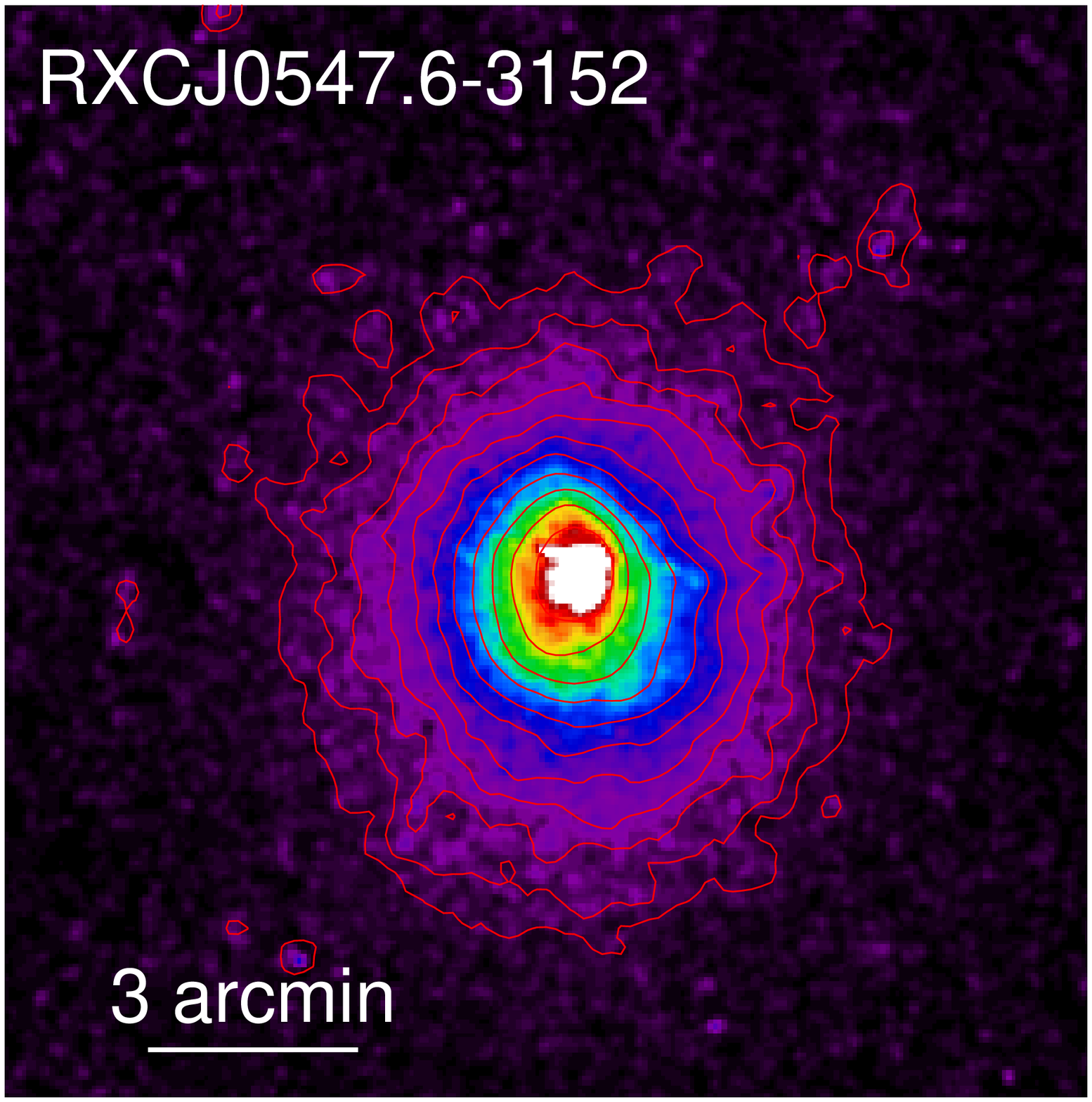}

\includegraphics[scale=1.,angle=0,keepaspectratio,width=0.195\textwidth]{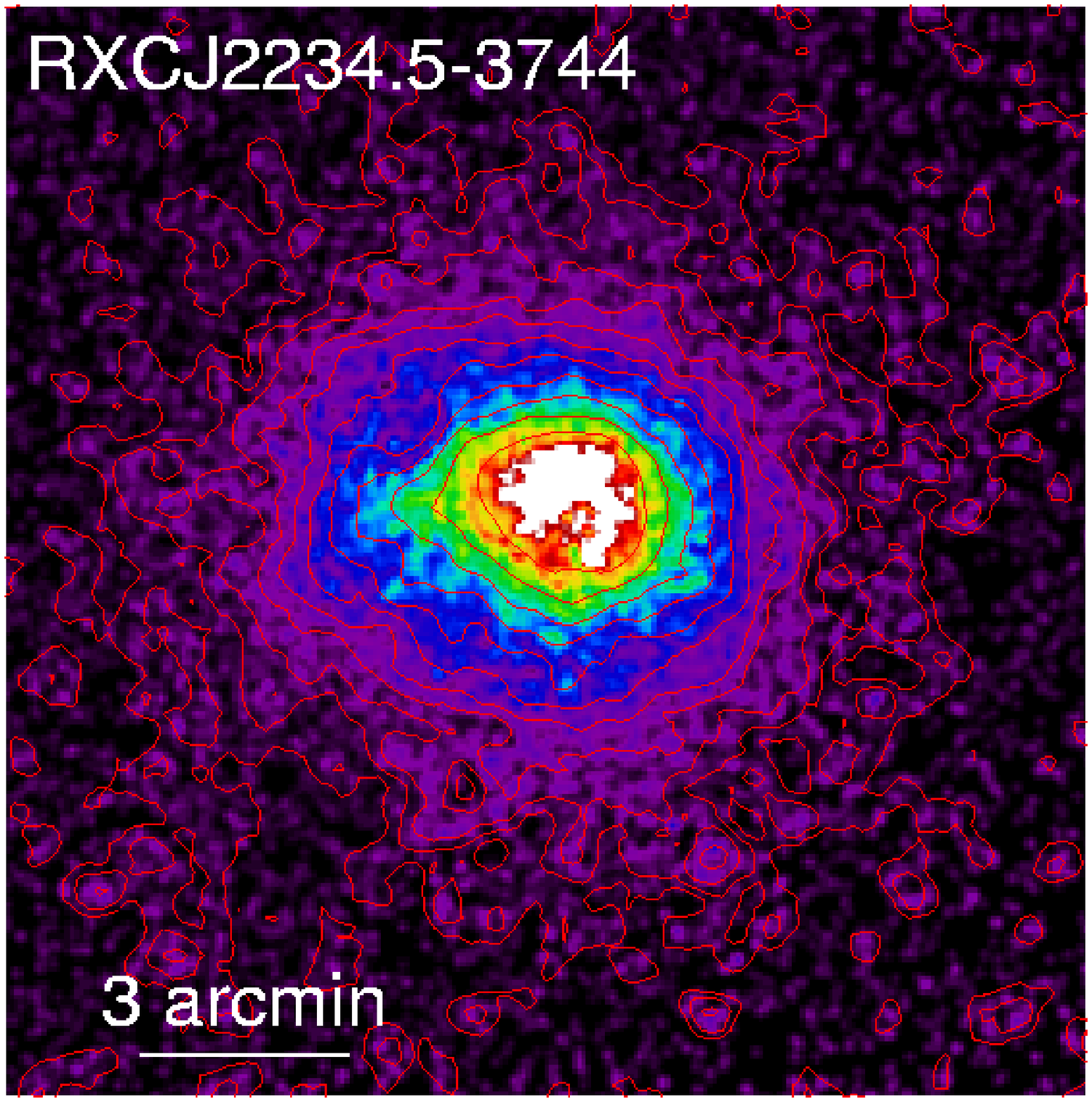}
\hfill
\includegraphics[scale=1.,angle=0,keepaspectratio,width=0.195\textwidth]{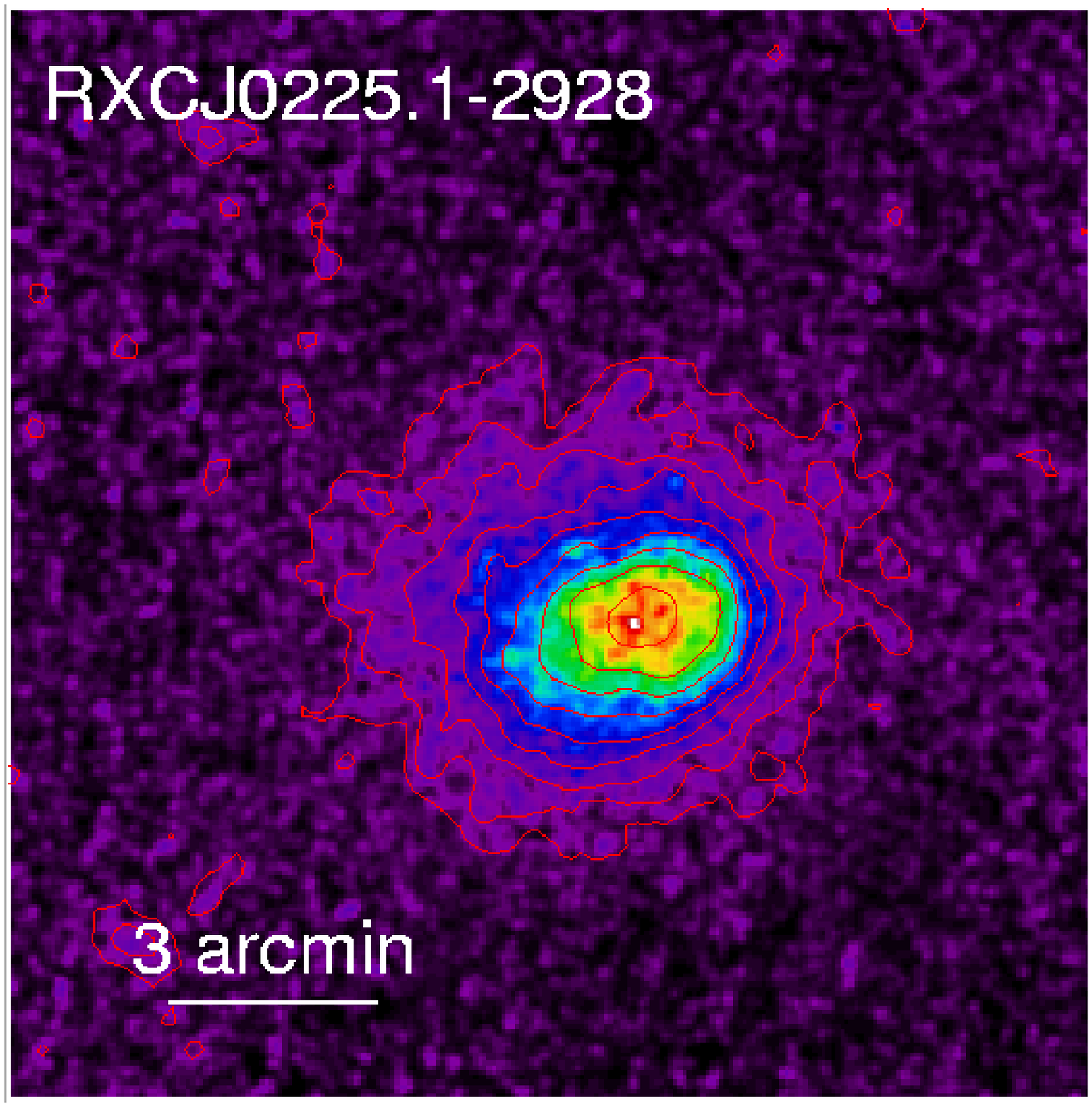}
\hfill
\includegraphics[scale=1.,angle=0,keepaspectratio,width=0.195\textwidth]{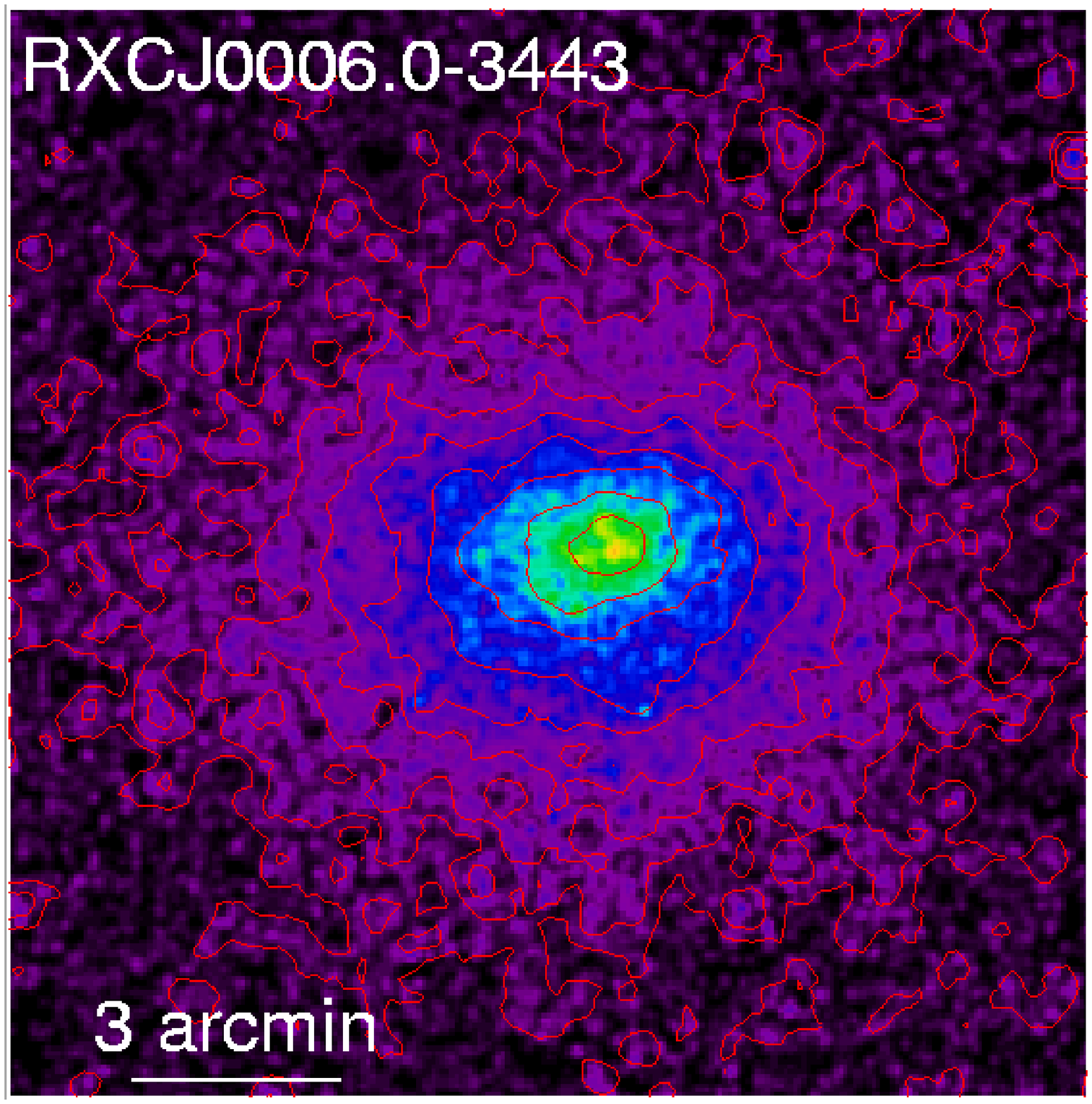}
\hfill
\includegraphics[scale=1.,angle=0,keepaspectratio,width=0.195\textwidth]{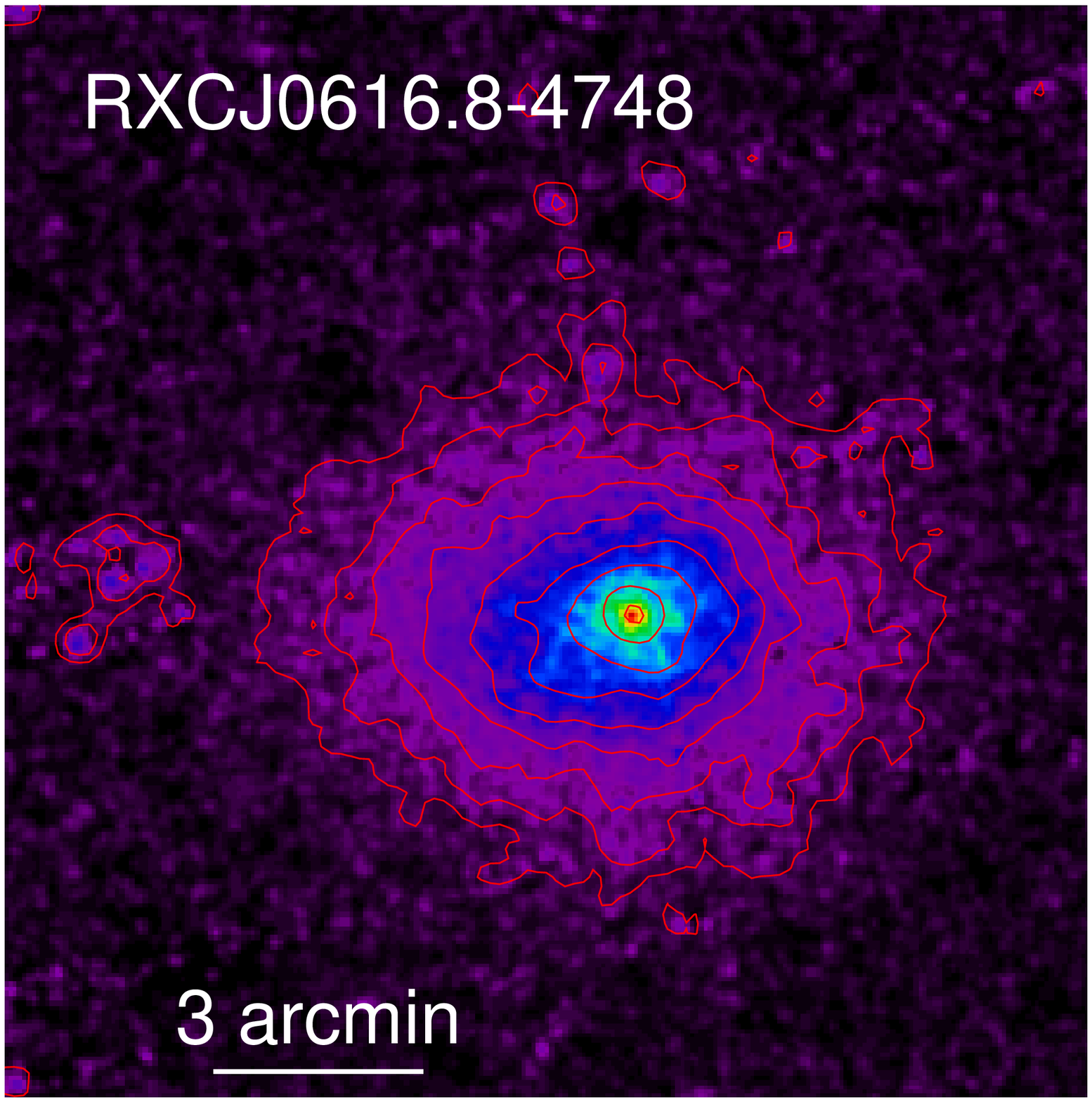}
\hfill
\includegraphics[scale=1.,angle=0,keepaspectratio,width=0.195\textwidth]{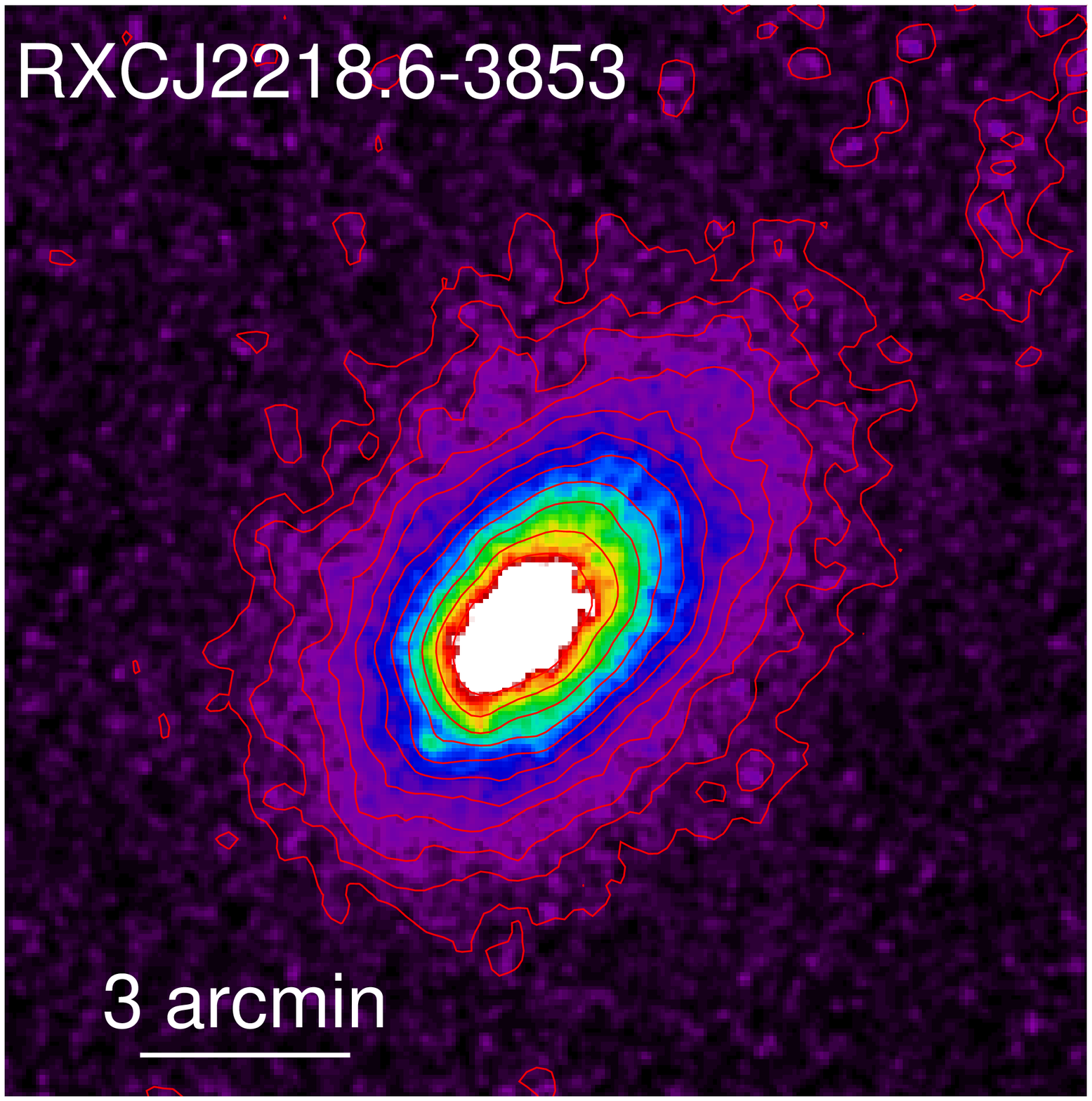}

\includegraphics[scale=1.,angle=0,keepaspectratio,width=0.195\textwidth]{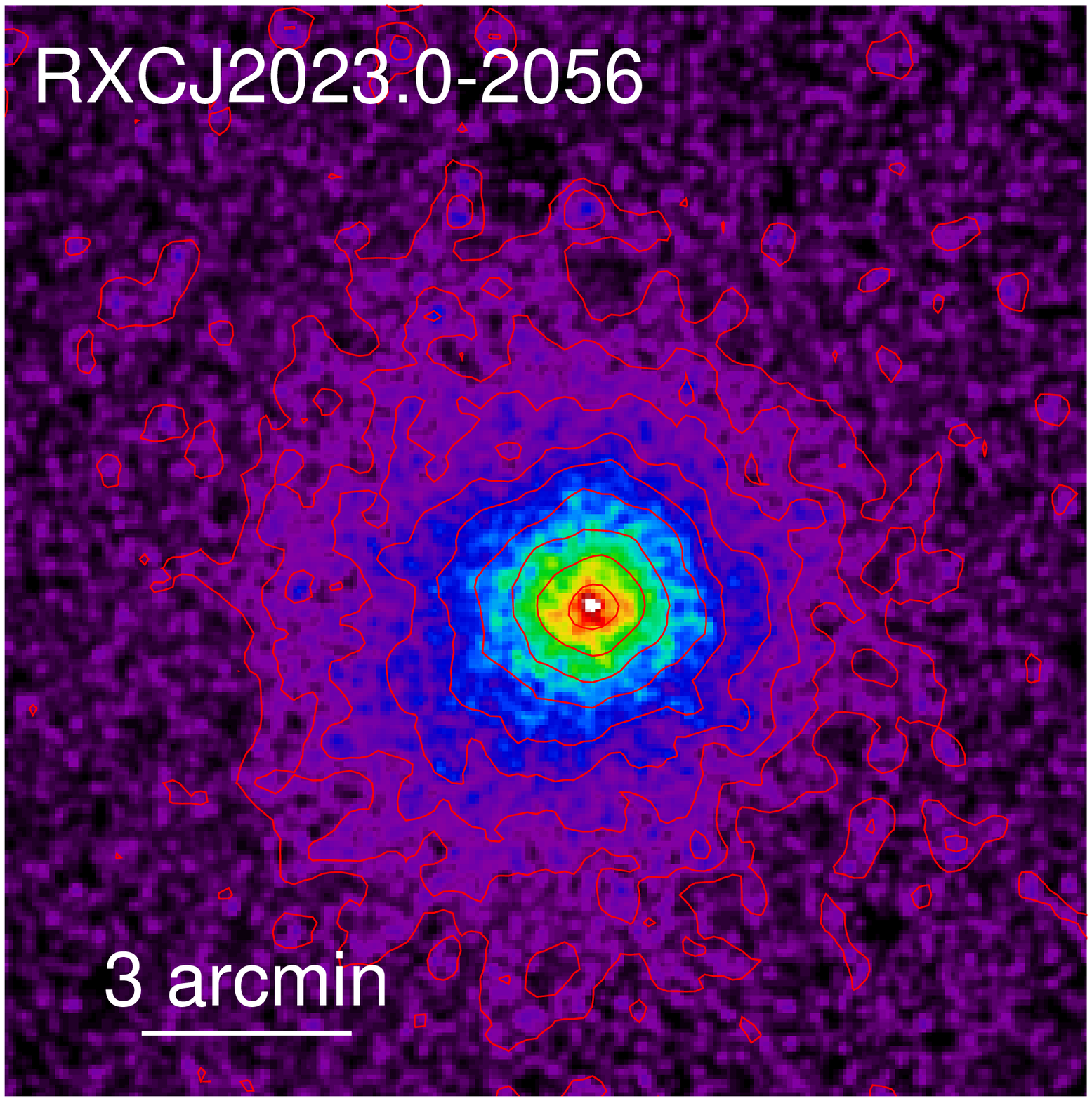}
\hfill
\includegraphics[scale=1.,angle=0,keepaspectratio,width=0.195\textwidth]{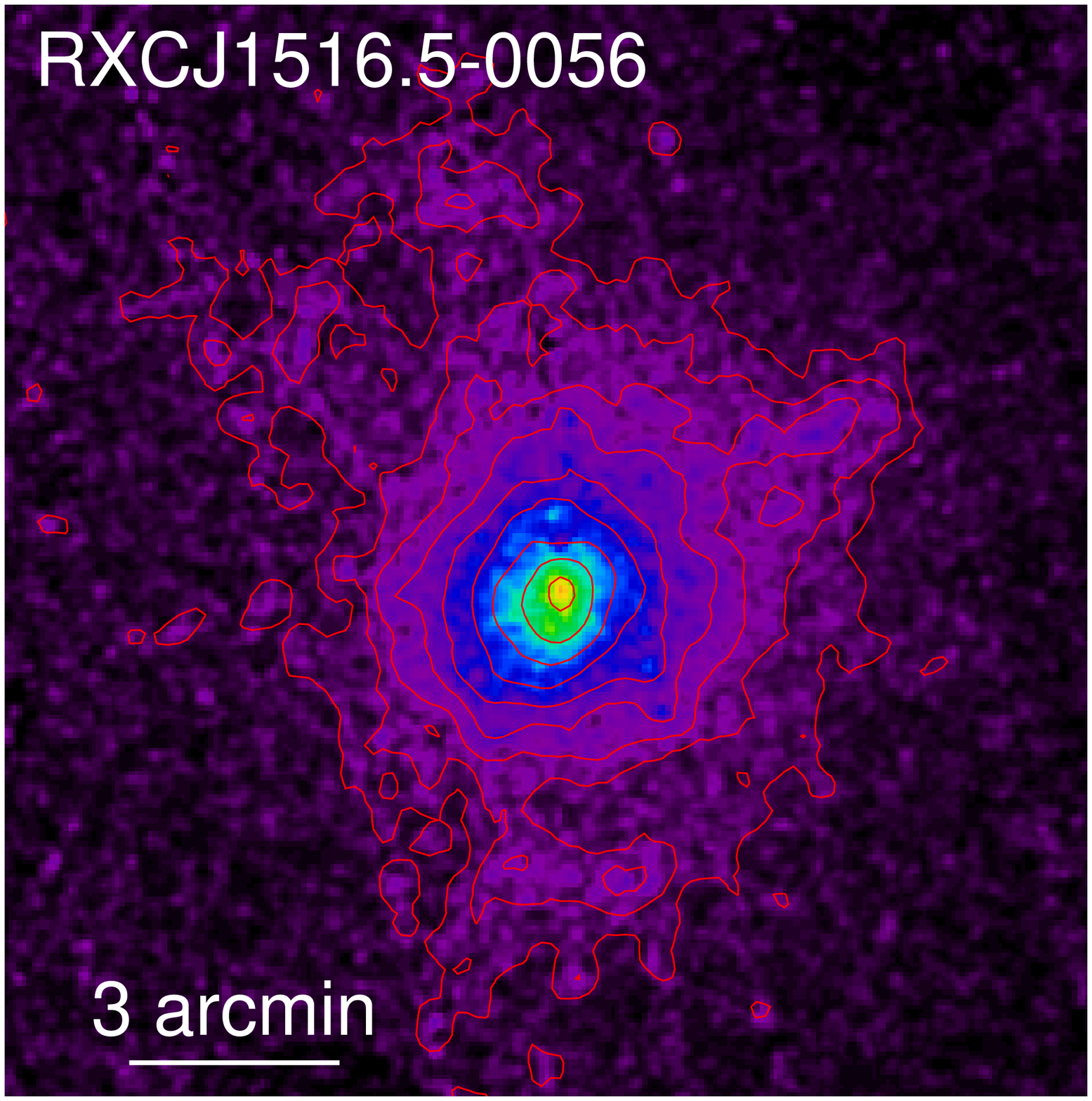}
\hfill
\includegraphics[scale=1.,angle=0,keepaspectratio,width=0.195\textwidth]{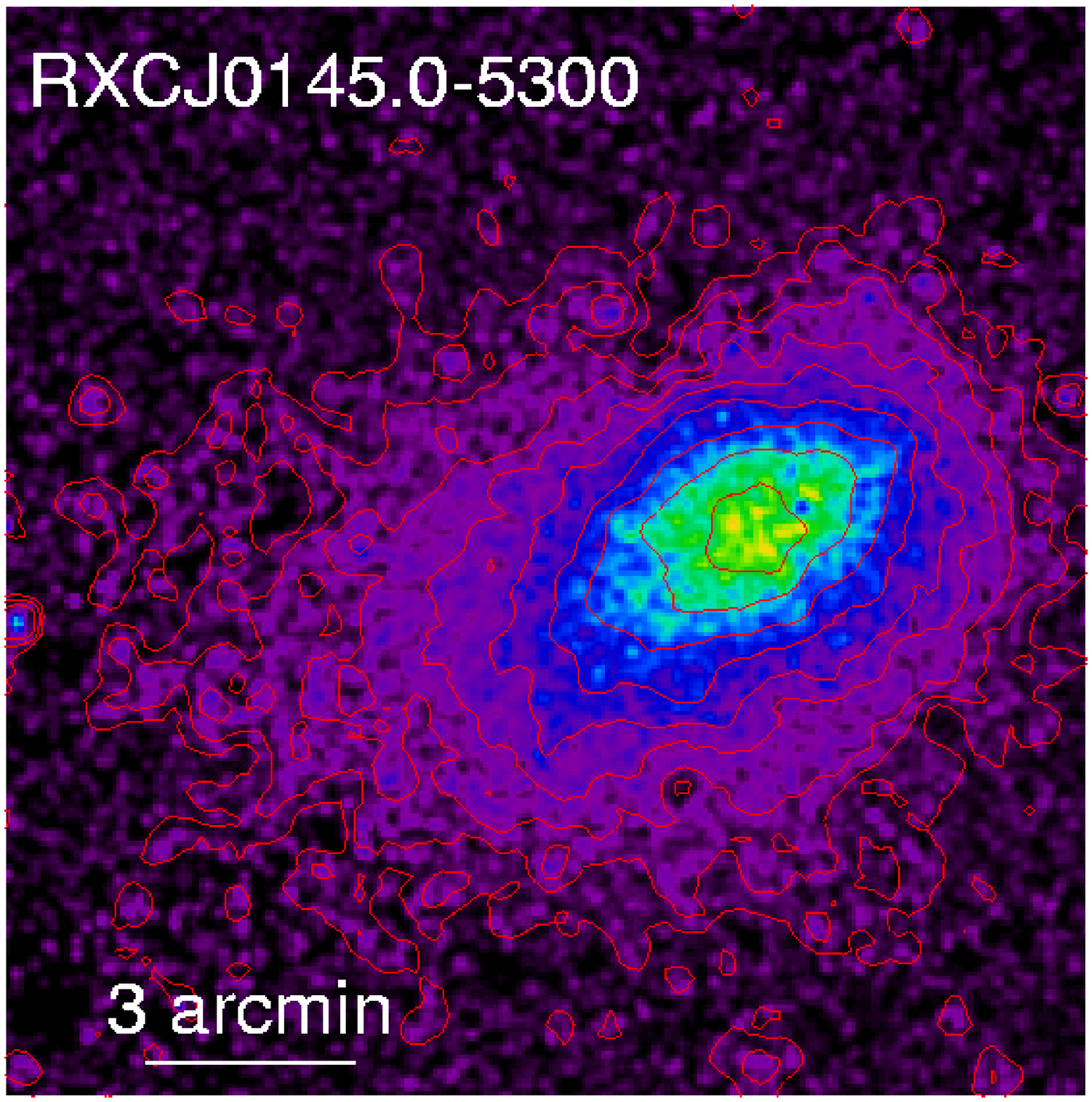}
\hfill
\includegraphics[scale=1.,angle=0,keepaspectratio,width=0.195\textwidth]{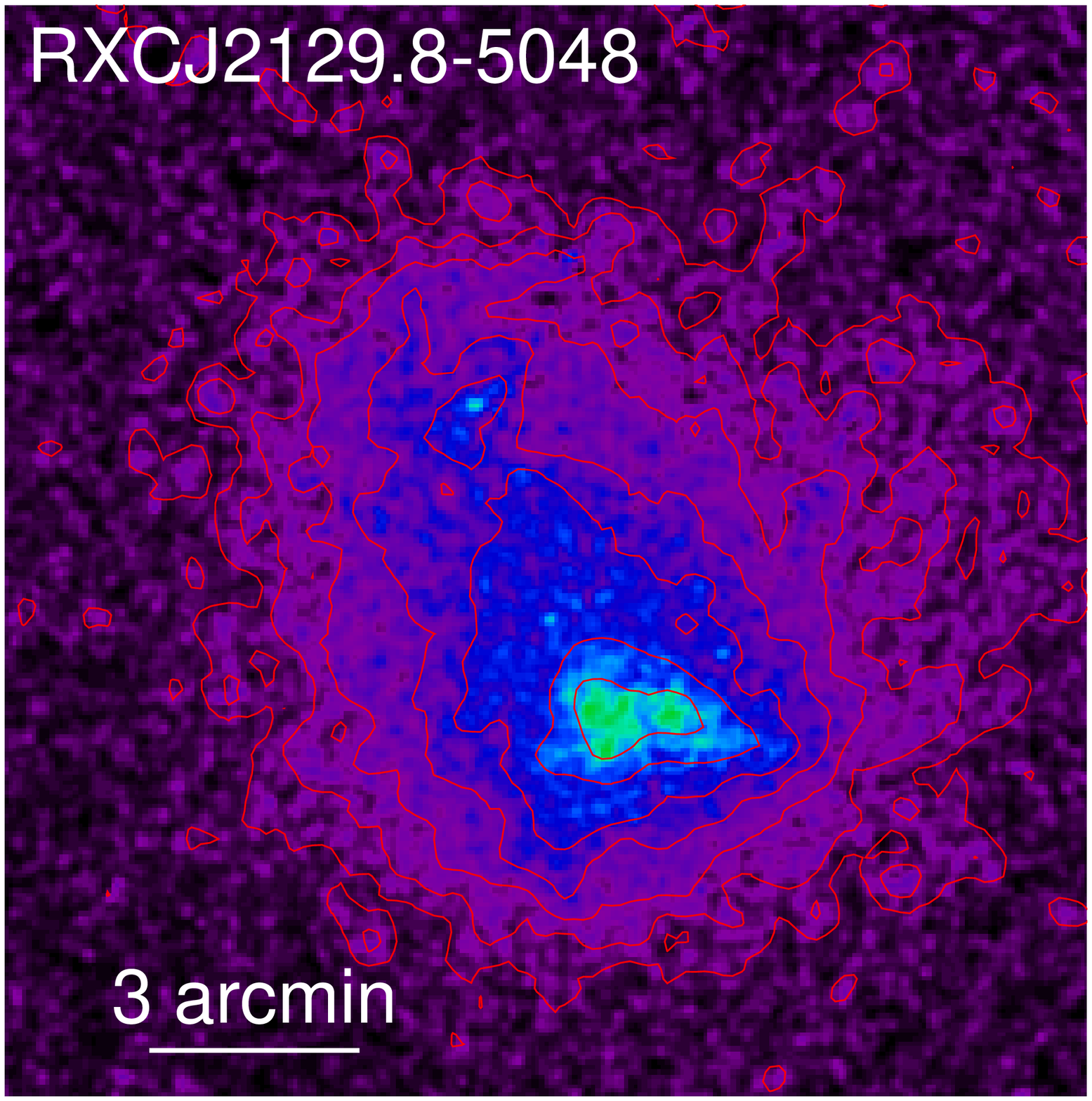}
\hfill
\includegraphics[scale=1.,angle=0,keepaspectratio,width=0.195\textwidth]{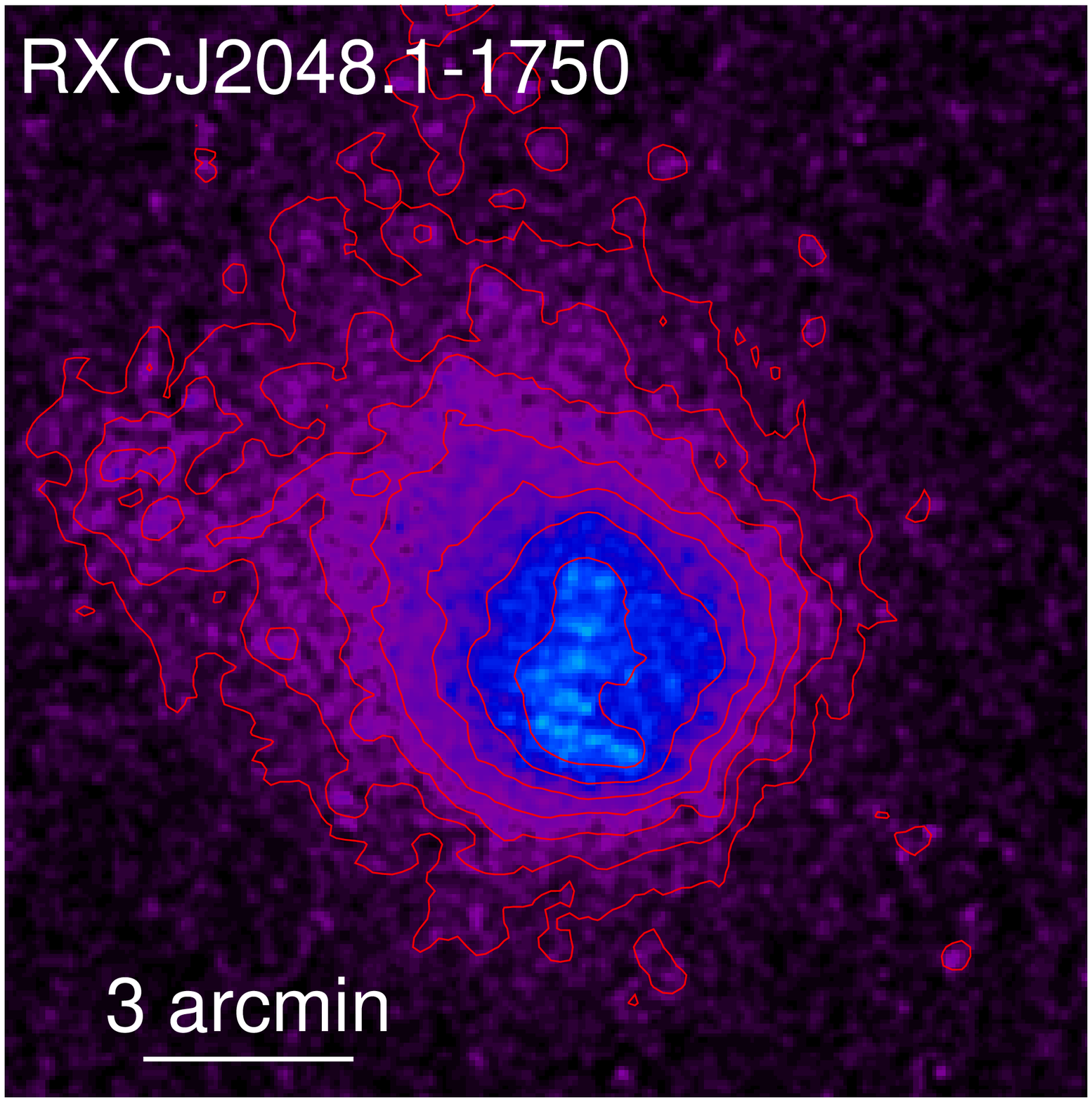}

\includegraphics[scale=1.,angle=0,keepaspectratio,width=0.195\textwidth]{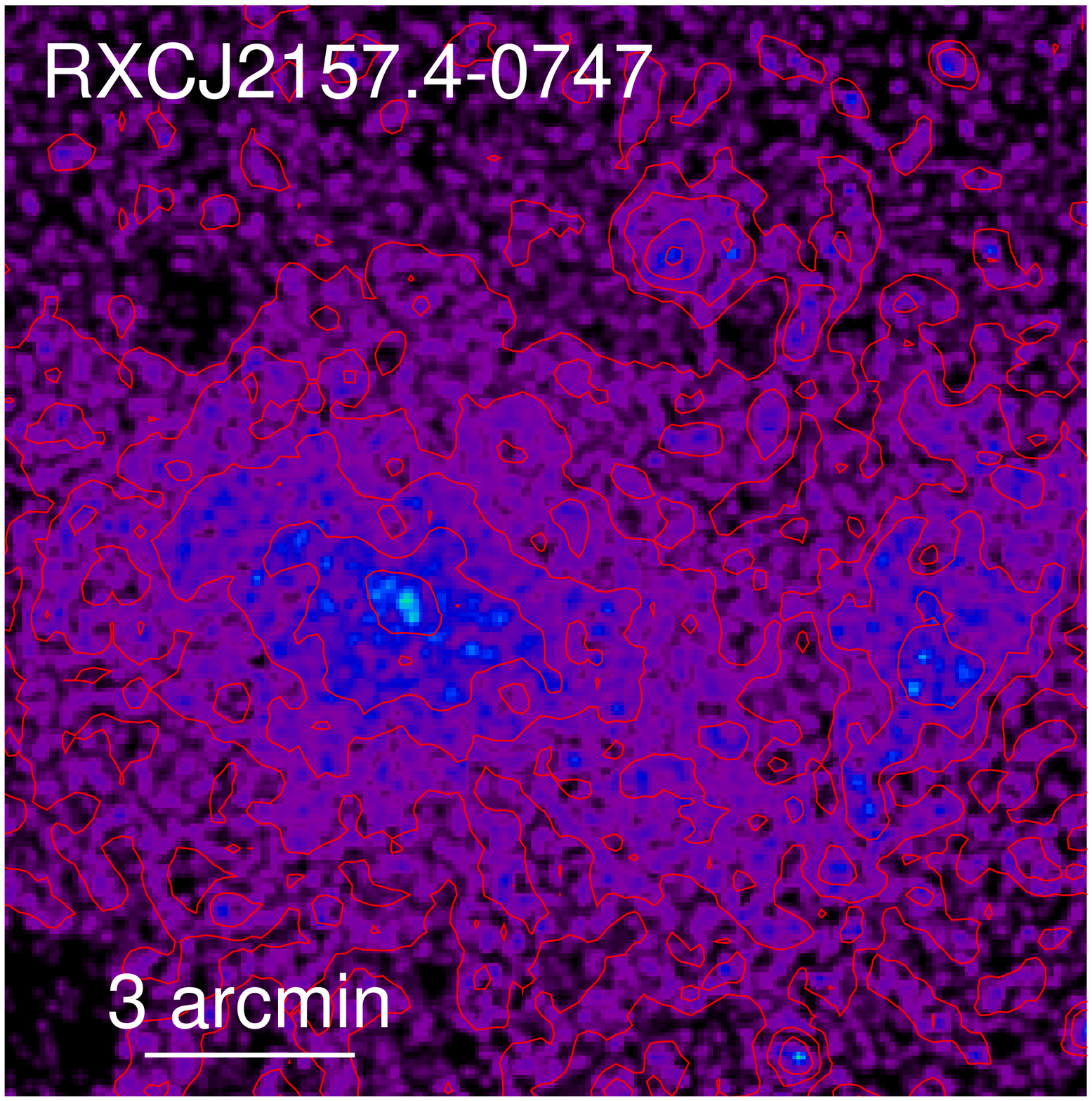}
\hfill

\caption{{\footnotesize Non-cool core clusters, sorted from top left to bottom right in order of increasing centroid shift parameter, $\langle w \rangle$.}}\label{fig:gallery2}   
\end{figure*}

Figures~\ref{fig:gallery1} and~\ref{fig:gallery2} show the resulting images. Contours increase in steps of $\sqrt{2}$. As in the original \rexcess\ paper \citep{boehringer07}, the colour table of each panel is scaled by a factor of $L [0.1-2.4]^{0.22}$, derived from the theoretical relationships between surface brightness and radius ($S_X \propto R$), radius and mass ($R \propto M^{1/3}$) and luminosity and mass ($L \propto M^{4/3}$). 

Clusters are divided into cool core and non cool core subsamples and then arranged in order of increasing centroid shift parameter $\langle w \rangle$.  Cool core systems generally appear more morphologically undisturbed; however, the increase in morphological complexity is always evident at higher values of $\langle w \rangle$. Note that two clusters are classified as both cool core and morphologically disturbed: RXC\,J1302 +0230 and RXC\,J2319 -7313. Figure~\ref{fig:gallery1} shows that they exhibit both centrally peaked surface brightness and morphological complexity, as expected.

\section{Survey luminosities and associated scaling relations}
\label{sec:surveylx}

\subsection{Introduction}

Past (e.g., {\it ROSAT}) and future (e.g., {\it eROSITA}) X-ray survey satellites are primarily sensitive in the soft X-ray band. In this Appendix we give the soft band luminosities of the \rexcess\ sample, together with the scaling relations between luminosity and temperature and $Y_X$. All quantities are calculated as described in Section~\ref{sec:lumin}.

\subsection{Correction for selection bias}

Selection effects can bias the observed scaling relations if not 
all clusters are completely sampled in
a well defined test volume. The classical problem is the selection
bias of an X-ray flux-limited sample, where the more luminous clusters
are sampled from a larger volume than the less luminous clusters. Thus for
any cluster property that is correlated with luminosity and features a
finite scatter, the more luminous clusters 
will always be over-represented with respect to the less luminous clusters. This
effect has previously been accounted for in flux-limited surveys in works by e.g., \citet[][]{ikebe02,stanek06,pacaud07,vikh08}.
In the case of \rexcess, the survey geometry is more complex, and has properties of both a flux- and a volume-limited sample. 

The basis for the calculation of the bias effect is the survey selection function,
which for our case specifies the size of the sampling volume as a function of 
X-ray luminosity. The selection function is given by three ingredients (i) the sky area covered by REFLEX ($\sim 4.24$ st) and the luminosity intervals and redshift shells defined for the \rexcess\ sample selection, (ii) the incompleteness of the sky coverage due to low exposure regions of the ROSAT All-Sky Survey (RASS), and (iii) the flux measurement
error which introduces a dispersion between the observed and true X-ray luminosity.
For the luminosity and redshift boundaries we use the definitions given in
Table 1 of \citet{boehringer07}. For those bins containing more clusters than 
the four clusters per bin selected for \rexcess\ we use an appropriate fractional weighting factor. The incompleteness is given by 
the sensitivity function in Table 8 of \citet{reflex}, illustrated in Fig. 23 of 
\citet{boehringer01}, which accounts for the
fact that not the complete redshift shell given in step (i) is sampled, but there
is an incomplete coverage due to lack of sensitivity in some areas with low 
exposure and/or high interstellar column density, $N_H$. An additional factor must be used in connection with the sensitivity function since \rexcess\ clusters were selected to have at least 30 counts
in the RASS;
this is taken into account as explained in B\"ohringer et al. (2001, 2004). For the first seven bins 
we used a further restriction in the \rexcess\ selection, $N_H \le 6 \times 10^{20}$ cm$^{-2}$, so for these bins we have recalculated the sky coverage for the
restricted sky region meeting the $N_H$ criterion. This and some other corrections
are small in our case (e.g., the part of the sky covered up to the nominal 
flux-limit for a minimal detection of 30 photons increases from 78\% to 79.5\%
for the low $N_H$ region), and will not significantly affect the final results here, but it is nevertheless important to have all these effects under control.

\begin{figure*}[]
\includegraphics[width=0.47\textwidth]{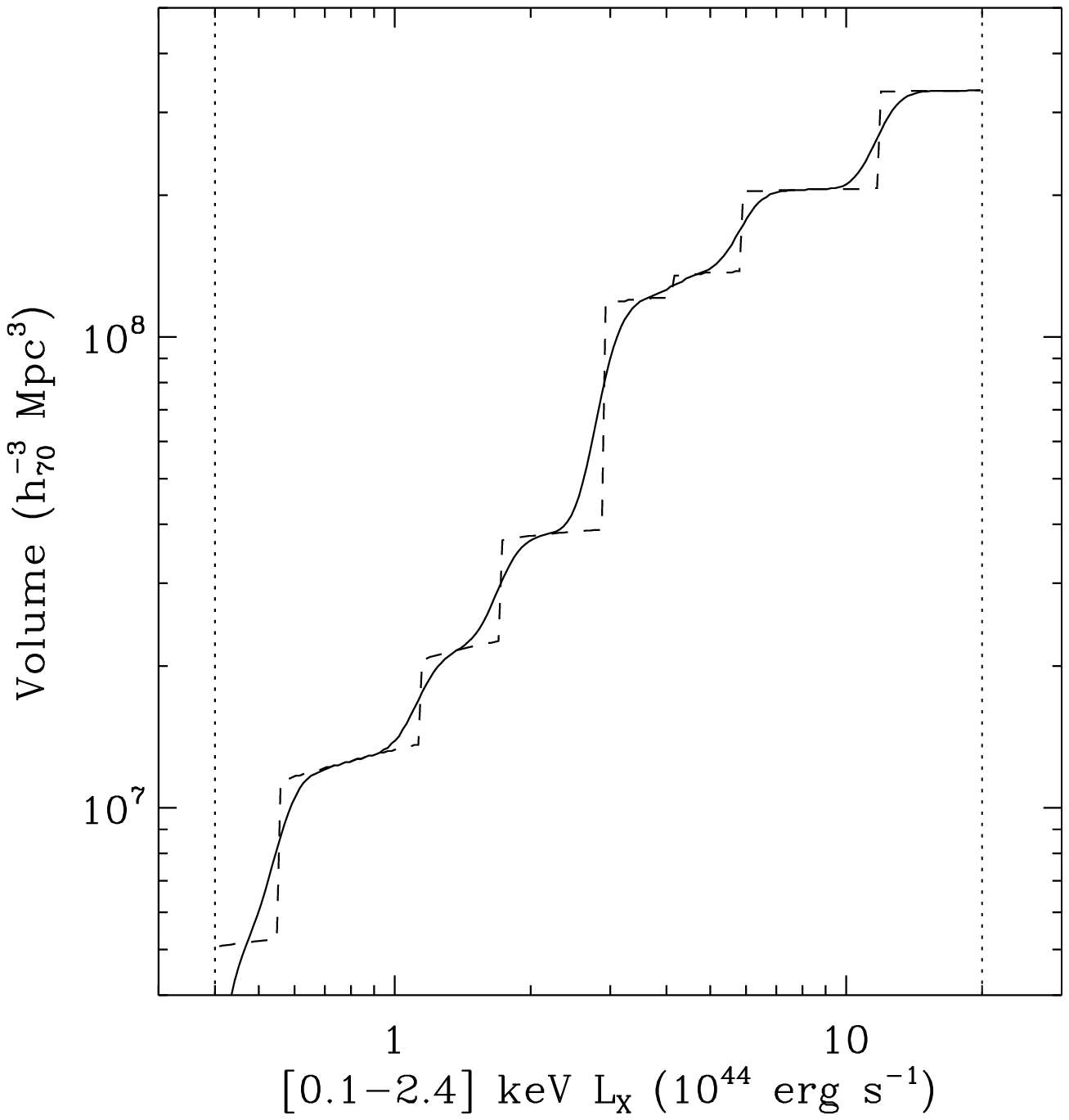}
\hfill
\includegraphics[width=0.47\textwidth]{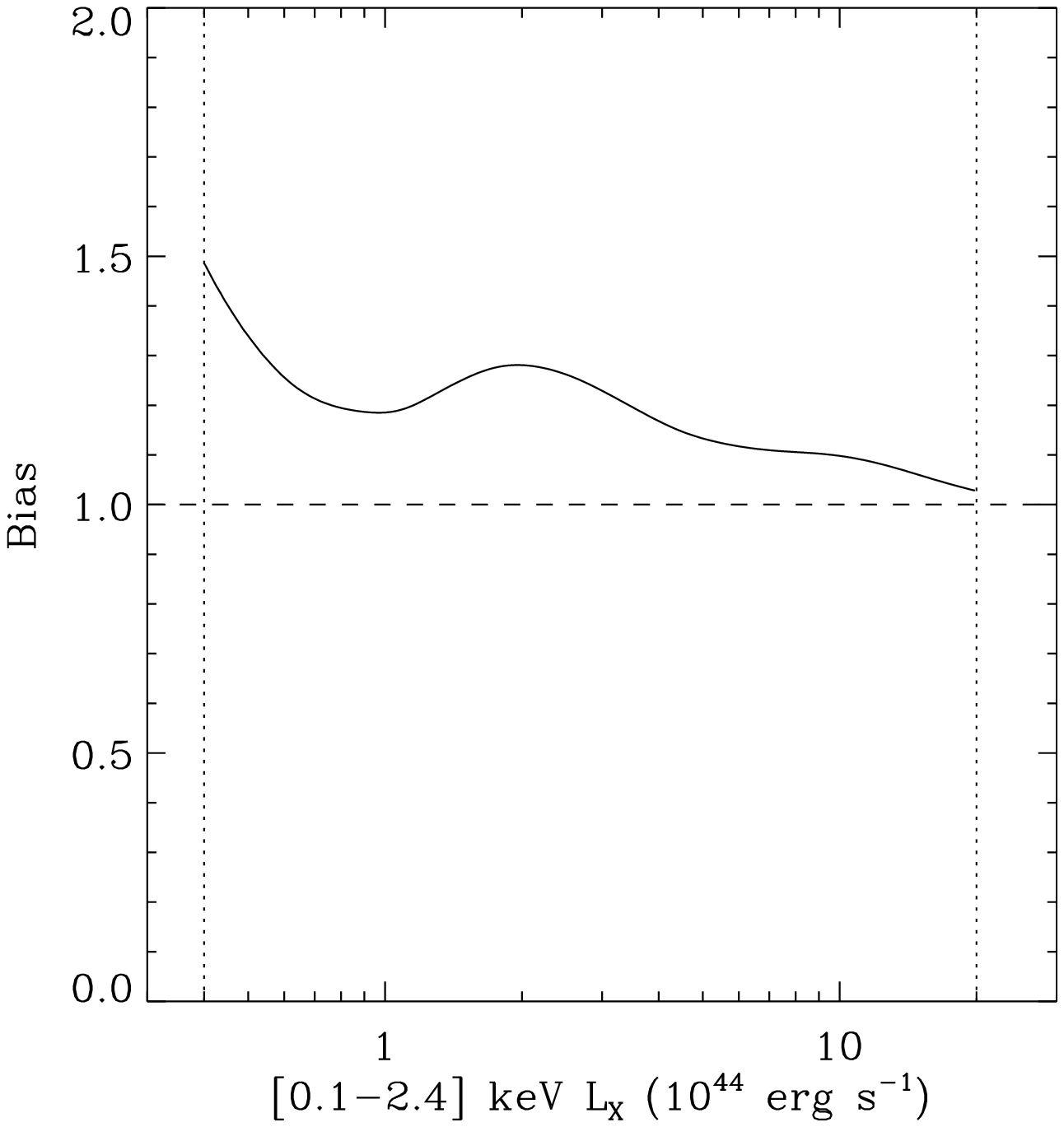}
\caption{{\footnotesize {\it Left panel}: Survey volume for the \rexcess\ sample. The dashed line is the raw survey volume; the solid line is the volume folded with an assumed measurement error of 10 per cent on the luminosity. {\it Right panel}: Malmquist bias for the \rexcess\ sample as a function of luminosity.  }}\label{fig:volbias} 
   \end{figure*}

The resulting selection volume
function shown in the left hand panel of Figure~\ref{fig:volbias} is a step function for the 9 luminosity bins of \rexcess, with the steps slightly
tilted due to correction (ii). Folding in the scatter between measured and
intrinsic luminosity, which is assumed to be $\sim 10\%$, smooths the step function. It is monotonically increasing with X-ray luminosity. 
It implies that luminous clusters are more represented than less luminous systems within \rexcess. This can then be straightforwardly included in a weighting factor for the clusters of different luminosity for a given mass. 
One approach is to give each cluster a weighting factor inverse to the selection volume ratio of the measured and nominal luminosity for given mass before fitting a scaling relation. Or, as we do here, we can follow a recipe analogous to that described by  Vikhlinin et al. (2008, Apendix A.2) where the mean bias for given mass is calculated via,

\begin{eqnarray*}
{\rm Bias}\ (\ln{L} |\ln{L_{0}}) & = & \langle \ln{L} \rangle  - \ln{L_{0}} \\
 & =  & {\int^{+\infty}_{-\infty}~ 
\Delta \ln{L}~  p(\ln{L} | \ln{L_{0}})~ V_{\rm sel}(\ln{L})~ d \ln{L}
\over \int^{+\infty}_{-\infty}~  p(\ln{L} | \ln{L_{0}})~ V_{\rm sel} (\ln{L})~ d \ln{L} }       
\end{eqnarray*}

\noindent where $L_{0}$ is the mean (zero scatter) $L$ for given mass, $\Delta \ln{L} = \ln{L} - \ln{L_{0}}$,
and $p(\ln{L} | \ln{L_{0}})$ characterizes the scatter of $L$ in the mass - luminosity relation, assumed to be lognormally distributed, and given by the observed scatter in the $L-M_Y$ relation (Table~\ref{tab:lxrel}). The bias factor is shown as a function of luminosity in the right hand panel of Figure~\ref{fig:volbias}. 

A final subtlety is that the luminosities used as the basis of the \rexcess\ selection, and thus for the bias calculation above, are those calculated as in the original REFLEX catalogue. These were iteratively calculated in the [0.1-2.4] keV band in the detection aperture and extrapolated to an assumed radius of $R_{200}$, and are thus not equivalent to the luminosities derived in this paper (see \citealt{reflex} for luminosity calculation details; the appropriate REFLEX luminosities for the \rexcess\ sample are given in Table~3 of \citealt{boehringer07}). 
We fit a linear relation in log-log space between the present luminosities and those from REFLEX, finding $L_{\rm REFLEX} = 1.15 \times  ([0.1-2.4]\, {\rm keV}\, L_{\rexcess})^{0.94}$.
The bias correction factor for the appropriate luminosity, ${\rm Bias}\ (\ln{L} | \ln{L_{0}})$, is then applied to each data point in the sample and the relation is refitted.

The Malmquist bias-corrected bolometric $L-M_Y$ relation is shown in the right hand panel of Figure~\ref{fig:LxMcorr}, and the corresponding fitted power law relation is given in Table~\ref{tab:lxrel}. The correction steepens the relation somewhat, due to the under-representation, at a given mass, of low luminosity clusters in the \rexcess\ sample. 

\subsection{Comparison to other results}

\begin{figure*}[]
\includegraphics[width=0.47\textwidth]{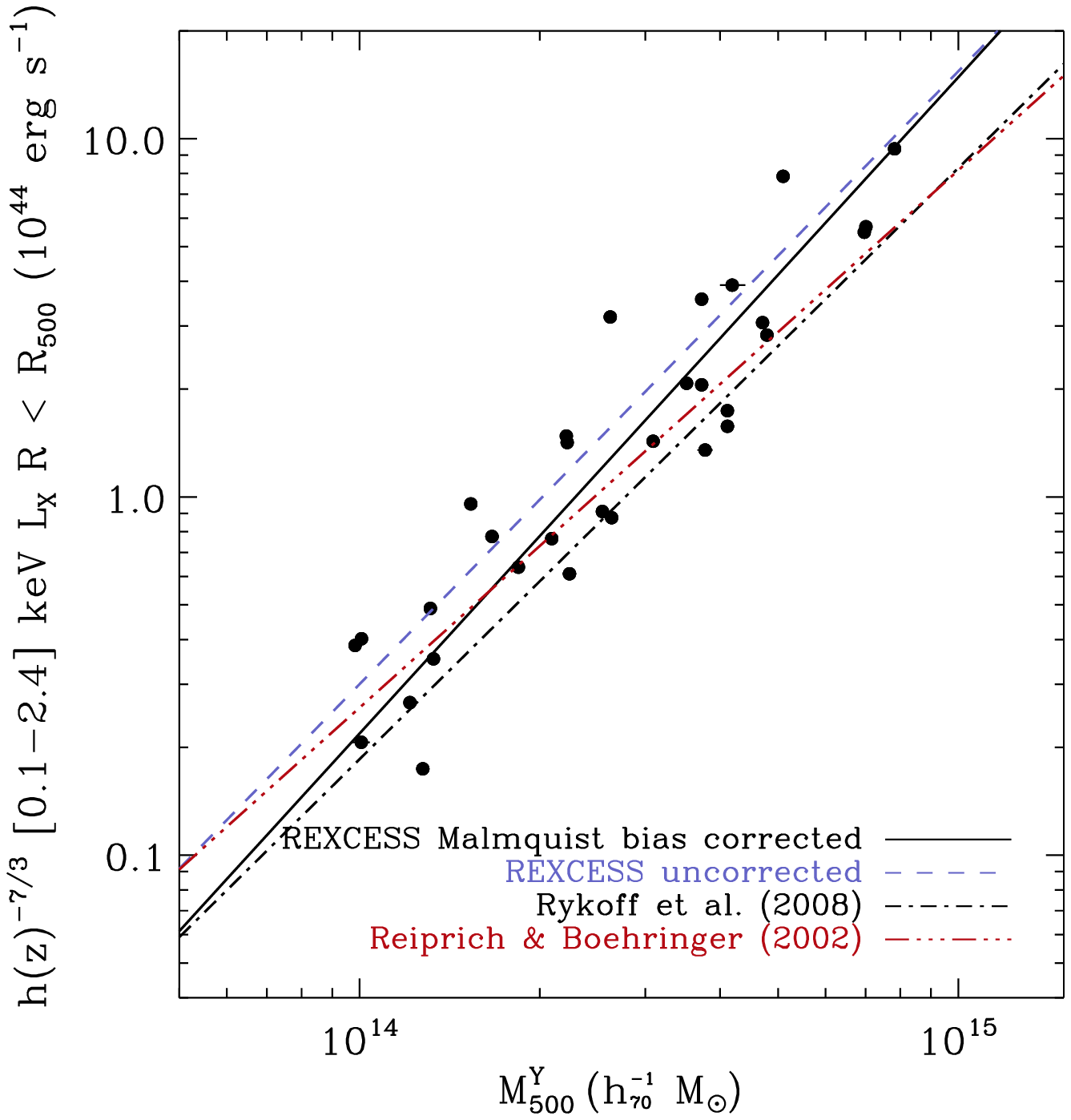}
\hfill
\includegraphics[width=0.47\textwidth]{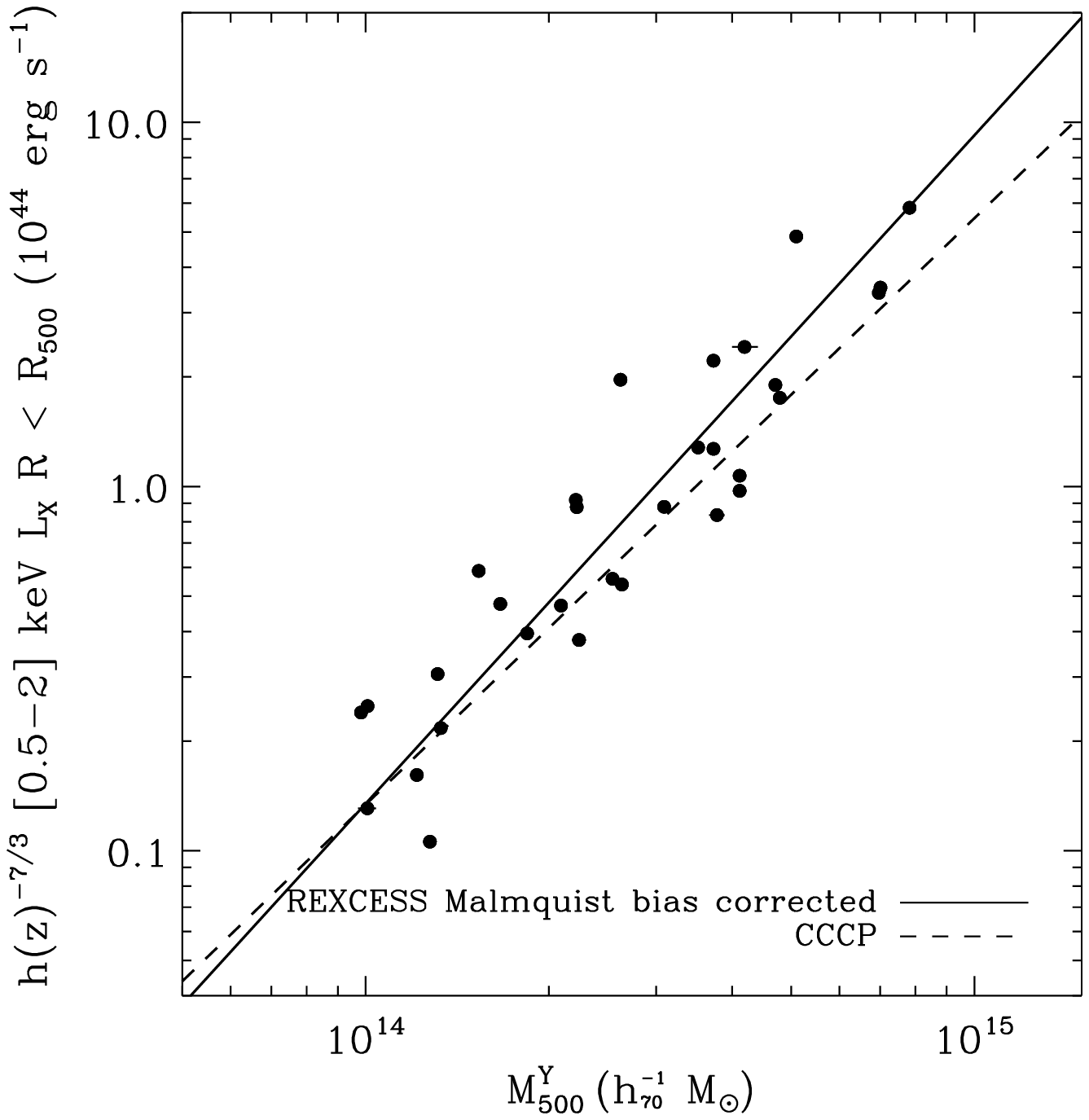}
\caption{{\footnotesize  {\it Left panel}: The $L[0.1-2.4] - M_Y$ relation compared with previous determinations from X-ray hydrostatic analysis assuming isothermality \citep{rb02} and stacked weak lensing analysis \citep{rykoff08}. The points are the bias-corrected \rexcess\ values. {\it Right panel}: Malmquist bias corrected $L[0.5-2] - M_Y$ relation compared with  the results from \citet{vikh08}. The best fitting relation is given in Table~\ref{tab:lxrelsoft}. All fits have been undertaken with the BCES orthogonal fitting method.}}\label{fig:052LxMmcorr} 
   \end{figure*}

The corrected relations for the two survey bands are also given in Table~\ref{tab:lxrelsoft}. The left hand panel of Figure~\ref{fig:052LxMmcorr} shows the raw and corrected [0.1-2.4] keV band relation compared to previous determinations from an X-ray hydrostatic analysis assuming isothermality \citep{rb02} and a stacked weak lensing analysis \citep{rykoff08}. We convert their $L-M_{200}$ relations to $M_{500}$ using a standard NFW model with a concentration parameter of $c_{500} = 3.2$, the average concentration derived from the total mass profiles of the morphologically regular cluster sample discussed in \citet{pap05}. Our corrected relation has a 25 per cent higher normalisation than that of \citeauthor{rykoff08} at our fiducial pivot point of $2 \times 10^{14}\, M_\odot$. The \citeauthor{rb02} relation has a 6 per cent lower normalisation than our corrected relation at the same mass scale.

The right hand panel of Figure~\ref{fig:052LxMmcorr} shows the corrected [0.5-2] keV band relation compared to the results derived by \citet{vikh08} using the same Malmquist bias correction procedure on a sample of clusters observed with {\it Chandra} (the Chandra Cluster Cosmology Project, CCCP). The agreement in normalisation is good at low $L/M$, but at higher $L/M$ the \citeauthor{vikh08} relation is somewhat below ours (by approximately 40 per cent at 8 keV, or $8 \times 10^{14}\, M_\odot$). 

The REXCESS and CCCP slopes are slightly different, although it is important to note that they are in agreement within their $1\sigma$ uncertainties. The bias correction itself does not play a part because there is excellent agreement in the magnitude of the scatter about the $L-M$ relation from the two samples. We  use a different $M_{500}-Y_X$ relation to estimate masses, although in practice the effect of this difference will be small since our relation is in good agreement with theirs. One partial explanation could be due to the systematic offset in measurements between {\it Chandra} and \xmm, in which, at high temperatures, {\it Chandra} overestimates the temperature\footnote{A Comparison of Cluster Temperatures Derived from Chandra and XMM-Newton \url{http://cxc.harvard.edu/cal/memos/hrma_memo.pdf}}. Since in both cases masses are derived from $Y_X = M_{\rm gas}\, T$, this will have the effect of boosting the higher mass {\it Chandra} points at a given luminosity, leading to a flatter relation than the one we find here. The effect is of order 20 per cent at 8 keV for a mass calculated from the $M-Y_X$ relation, which alleviates the difference somewhat. Finally, the samples contain different clusters. The individual samples probe slightly different mass ranges as \rexcess\ contains more lower mass systems, while the local CCCP sample contains more higher mass systems. In addition, differences in the number of cool core systems, and their distribution across the mass range, could change the slope. In particular, if \rexcess\ has more cool core systems at higher mass than CCCP, their higher luminosity would make the \rexcess\ relation slightly steeper.

\begin{table*}
\begin{center}
\caption{{\footnotesize Cluster properties in the soft X-ray band suitable for surveys. 
Luminosities are given in the $[0.1-2.4]$ keV band appropriate for {\it ROSAT\/}, and also the $[0.5-2]$ keV band. All quantities are calculated assuming $\Omega_M=0.3$, $\Omega_{\Lambda}=0.7$, and $h_0 = 0.7$.}}\label  
{tab:tabappx}
\centering
\begin{tabular}{l l l r r l r r}
\hline
\hline

\multicolumn{1}{c}{Cluster} & \multicolumn{1}{c}{$z$} &
\multicolumn{1}{c}{$T_1$} & \multicolumn{1}{c}{$L\, {[0.1-2.4]}_1$} &
\multicolumn{1}{c}{$L\, {[0.5-2]}_1$} &
\multicolumn{1}{c}{$T_2$} & \multicolumn{1}{c}{$L\, {[0.1-2.4]}_2$} &
\multicolumn{1}{c}{$L\, {[0.5-2]}_2$} \\

\multicolumn{1}{c }{(1)} & \multicolumn{1}{c}{(2)} & 
\multicolumn{1}{c}{(3)} & \multicolumn{1}{c}{(4)} & 
\multicolumn{1}{c }{(5)} & \multicolumn{1}{c}{ (6)} & 
\multicolumn{1}{c }{(7)} &
\multicolumn{1}{c}{(8)} \\
\hline
\\
RXC\,J0003+0203 & 0.0924 & $3.85_{-0.09}^{+0.09}$ & $ 1.02_{-0.01}^{+0.01}$ & $ 0.63_{-0.00}^{+0.00}$ & $3.64_{-0.09}^{+0.09}$ & $ 0.65_{-0.01}^{+0.01}$ & $ 0.40_{-0.00}^{+0.00}$ \\
RXC\,J0006-3443 & 0.1147 & $5.03_{-0.19}^{+0.19}$ & $ 1.96_{-0.02}^{+0.02}$ & $ 1.21_{-0.01}^{+0.01}$ & $4.60_{-0.16}^{+0.21}$ & $ 1.58_{-0.02}^{+0.02}$ & $ 0.97_{-0.01}^{+0.01}$ \\
RXC\,J0020-2542 & 0.1410 & $5.69_{-0.11}^{+0.11}$ & $ 2.93_{-0.02}^{+0.02}$ & $ 1.81_{-0.01}^{+0.01}$ & $5.24_{-0.15}^{+0.15}$ & $ 1.92_{-0.02}^{+0.02}$ & $ 1.18_{-0.01}^{+0.01}$ \\
RXC\,J0049-2931 & 0.1084 & $3.09_{-0.10}^{+0.10}$ & $ 1.06_{-0.01}^{+0.01}$ & $ 0.65_{-0.01}^{+0.01}$ & $2.79_{-0.11}^{+0.11}$ & $ 0.62_{-0.01}^{+0.01}$ & $ 0.38_{-0.01}^{+0.01}$ \\
RXC\,J0145-5300 & 0.1168 & $5.53_{-0.13}^{+0.13}$ & $ 2.26_{-0.02}^{+0.02}$ & $ 1.40_{-0.01}^{+0.01}$ & $5.51_{-0.16}^{+0.16}$ & $ 1.76_{-0.02}^{+0.02}$ & $ 1.09_{-0.01}^{+0.01}$ \\
RXC\,J0211-4017 & 0.1008 & $2.07_{-0.00}^{+0.07}$ & $ 0.55_{-0.00}^{+0.00}$ & $ 0.34_{-0.00}^{+0.00}$ & $2.02_{-0.06}^{+0.06}$ & $ 0.33_{-0.00}^{+0.00}$ & $ 0.20_{-0.00}^{+0.00}$ \\
RXC\,J0225-2928 & 0.0604 & $2.47_{-0.06}^{+0.15}$ & $ 0.33_{-0.01}^{+0.01}$ & $ 0.21_{-0.00}^{+0.00}$ & $2.61_{-0.16}^{+0.16}$ & $ 0.19_{-0.00}^{+0.00}$ & $ 0.12_{-0.00}^{+0.00}$ \\
RXC\,J0345-4112 & 0.0603 & $2.19_{-0.04}^{+0.04}$ & $ 0.51_{-0.01}^{+0.01}$ & $ 0.32_{-0.00}^{+0.00}$ & $2.15_{-0.08}^{+0.08}$ & $ 0.25_{-0.00}^{+0.00}$ & $ 0.15_{-0.00}^{+0.00}$ \\
RXC\,J0547-3152 & 0.1483 & $6.02_{-0.11}^{+0.11}$ & $ 3.88_{-0.02}^{+0.02}$ & $ 2.40_{-0.01}^{+0.01}$ & $5.68_{-0.11}^{+0.11}$ & $ 2.58_{-0.02}^{+0.02}$ & $ 1.59_{-0.01}^{+0.01}$ \\
RXC\,J0605-3518 & 0.1392 & $4.56_{-0.05}^{+0.05}$ & $ 4.72_{-0.02}^{+0.02}$ & $ 2.94_{-0.01}^{+0.01}$ & $4.81_{-0.12}^{+0.12}$ & $ 2.07_{-0.02}^{+0.02}$ & $ 1.28_{-0.01}^{+0.01}$ \\
RXC\,J0616-4748 & 0.1164 & $4.22_{-0.10}^{+0.10}$ & $ 1.24_{-0.01}^{+0.01}$ & $ 0.76_{-0.01}^{+0.01}$ & $4.16_{-0.12}^{+0.12}$ & $ 0.98_{-0.01}^{+0.01}$ & $ 0.60_{-0.01}^{+0.01}$ \\
RXC\,J0645-5413 & 0.1644 & $6.95_{-0.13}^{+0.13}$ & $ 7.57_{-0.04}^{+0.04}$ & $ 4.69_{-0.03}^{+0.03}$ & $6.97_{-0.19}^{+0.19}$ & $ 4.57_{-0.03}^{+0.03}$ & $ 2.83_{-0.02}^{+0.02}$ \\
RXC\,J0821+0112 & 0.0822 & $2.68_{-0.09}^{+0.09}$ & $ 0.49_{-0.01}^{+0.01}$ & $ 0.30_{-0.00}^{+0.00}$ & $2.44_{-0.12}^{+0.12}$ & $ 0.36_{-0.01}^{+0.01}$ & $ 0.22_{-0.00}^{+0.00}$ \\
RXC\,J0958-1103 & 0.1669 & $5.34_{-0.21}^{+0.21}$ & $ 5.30_{-0.07}^{+0.07}$ & $ 3.28_{-0.04}^{+0.04}$ & $5.85_{-0.40}^{+0.45}$ & $ 2.31_{-0.06}^{+0.06}$ & $ 1.43_{-0.04}^{+0.04}$ \\
RXC\,J1044-0704 & 0.1342 & $3.41_{-0.03}^{+0.03}$ & $ 4.24_{-0.01}^{+0.01}$ & $ 2.62_{-0.01}^{+0.01}$ & $3.52_{-0.05}^{+0.05}$ & $ 1.70_{-0.01}^{+0.01}$ & $ 1.04_{-0.01}^{+0.01}$ \\
RXC\,J1141-1216 & 0.1195 & $3.31_{-0.03}^{+0.03}$ & $ 2.14_{-0.01}^{+0.01}$ & $ 1.33_{-0.00}^{+0.00}$ & $3.40_{-0.06}^{+0.06}$ & $ 0.97_{-0.01}^{+0.01}$ & $ 0.60_{-0.00}^{+0.00}$ \\
RXC\,J1236-3354 & 0.0796 & $2.70_{-0.05}^{+0.05}$ & $ 0.64_{-0.01}^{+0.01}$ & $ 0.40_{-0.00}^{+0.00}$ & $2.57_{-0.03}^{+0.11}$ & $ 0.39_{-0.00}^{+0.00}$ & $ 0.24_{-0.00}^{+0.00}$ \\
RXC\,J1302-0230 & 0.0847 & $2.97_{-0.07}^{+0.06}$ & $ 0.83_{-0.01}^{+0.01}$ & $ 0.51_{-0.00}^{+0.00}$ & $2.92_{-0.07}^{+0.09}$ & $ 0.51_{-0.00}^{+0.00}$ & $ 0.31_{-0.00}^{+0.00}$ \\
RXC\,J1311-0120 & 0.1832 & $8.91_{-0.08}^{+0.08}$ & $12.48_{-0.03}^{+0.03}$ & $ 7.76_{-0.02}^{+0.02}$ & $8.24_{-0.13}^{+0.13}$ & $ 5.49_{-0.02}^{+0.02}$ & $ 3.41_{-0.01}^{+0.01}$ \\
RXC\,J1516+0005 & 0.1181 & $4.51_{-0.06}^{+0.06}$ & $ 2.08_{-0.01}^{+0.01}$ & $ 1.28_{-0.01}^{+0.01}$ & $4.18_{-0.08}^{+0.08}$ & $ 1.45_{-0.01}^{+0.01}$ & $ 0.89_{-0.01}^{+0.01}$ \\
RXC\,J1516-0056 & 0.1198 & $3.55_{-0.07}^{+0.07}$ & $ 1.30_{-0.01}^{+0.01}$ & $ 0.80_{-0.01}^{+0.01}$ & $3.40_{-0.08}^{+0.08}$ & $ 1.02_{-0.01}^{+0.01}$ & $ 0.62_{-0.01}^{+0.01}$ \\
RXC\,J2014-2430 & 0.1538 & $4.78_{-0.05}^{+0.05}$ & $10.24_{-0.03}^{+0.03}$ & $ 6.34_{-0.02}^{+0.02}$ & $5.63_{-0.11}^{+0.11}$ & $ 3.38_{-0.03}^{+0.03}$ & $ 2.09_{-0.02}^{+0.02}$ \\
RXC\,J2023-2056 & 0.0564 & $2.71_{-0.09}^{+0.09}$ & $ 0.39_{-0.01}^{+0.01}$ & $ 0.24_{-0.00}^{+0.00}$ & $2.46_{-0.12}^{+0.12}$ & $ 0.26_{-0.01}^{+0.01}$ & $ 0.16_{-0.00}^{+0.00}$ \\
RXC\,J2048-1750 & 0.1475 & $4.65_{-0.07}^{+0.13}$ & $ 2.55_{-0.01}^{+0.01}$ & $ 1.57_{-0.01}^{+0.01}$ & $4.59_{-0.08}^{+0.08}$ & $ 2.21_{-0.01}^{+0.01}$ & $ 1.36_{-0.01}^{+0.01}$ \\
RXC\,J2129-5048 & 0.0796 & $3.81_{-0.15}^{+0.15}$ & $ 0.79_{-0.01}^{+0.01}$ & $ 0.49_{-0.01}^{+0.01}$ & $3.64_{-0.12}^{+0.16}$ & $ 0.65_{-0.01}^{+0.01}$ & $ 0.41_{-0.01}^{+0.01}$ \\
RXC\,J2149-3041 & 0.1184 & $3.26_{-0.04}^{+0.04}$ & $ 2.06_{-0.01}^{+0.01}$ & $ 1.27_{-0.01}^{+0.01}$ & $3.40_{-0.08}^{+0.08}$ & $ 0.91_{-0.01}^{+0.01}$ & $ 0.56_{-0.00}^{+0.00}$ \\
RXC\,J2157-0747 & 0.0579 & $2.46_{-0.08}^{+0.08}$ & $ 0.29_{-0.00}^{+0.00}$ & $ 0.18_{-0.00}^{+0.00}$ & $2.30_{-0.06}^{+0.10}$ & $ 0.25_{-0.00}^{+0.00}$ & $ 0.15_{-0.00}^{+0.00}$ \\
RXC\,J2217-3543 & 0.1486 & $4.86_{-0.09}^{+0.09}$ & $ 2.98_{-0.01}^{+0.01}$ & $ 1.84_{-0.01}^{+0.01}$ & $4.45_{-0.09}^{+0.09}$ & $ 1.89_{-0.01}^{+0.01}$ & $ 1.16_{-0.01}^{+0.01}$ \\
RXC\,J2218-3853 & 0.1411 & $5.84_{-0.11}^{+0.11}$ & $ 4.13_{-0.03}^{+0.03}$ & $ 2.56_{-0.02}^{+0.02}$ & $5.88_{-0.15}^{+0.20}$ & $ 2.44_{-0.03}^{+0.03}$ & $ 1.51_{-0.02}^{+0.02}$ \\
RXC\,J2234-3744 & 0.1510 & $7.78_{-0.15}^{+0.15}$ & $ 7.20_{-0.04}^{+0.04}$ & $ 4.47_{-0.03}^{+0.03}$ & $6.95_{-0.14}^{+0.14}$ & $ 4.96_{-0.04}^{+0.04}$ & $ 3.07_{-0.02}^{+0.02}$ \\
RXC\,J2319-7313 & 0.0984 & $2.22_{-0.03}^{+0.03}$ & $ 1.34_{-0.01}^{+0.01}$ & $ 0.82_{-0.01}^{+0.01}$ & $2.48_{-0.08}^{+0.08}$ & $ 0.63_{-0.01}^{+0.01}$ & $ 0.38_{-0.01}^{+0.01}$ \\
\\
\hline
\end{tabular}
\end{center}

Columns: (1) Cluster name; (2) $z$: cluster redshift; (3) $T_1$: spectroscopic  
temperature of the $R < \Rv$ region in keV; (4) $L\,[0.1-2.4]_1$: $[0.1-2.4]$ keV band luminosity in the $R < \Rv$ region in units of $10^{44}$ erg s$^{-1}$; (5) $L\,[0.5-2]_1$: $[0.5-2]$ keV band luminosity in the $R < \Rv$ region in units of $10^{44}$ erg s$^{-1}$; (6) $T_2$: spectroscopic temperature in the $[0.15-1]\,\Rv$ region in keV; (7) $L\,[0.1-2.4]_2$: $[0.1-2.4]$ keV band luminosity in the $[0.15-1]\, \Rv$ region in units of $10^{44}$ erg s$^{-1}$; (8) $L\,[0.5-2]_2$: $[0.5-2]$ keV band luminosity in the $[0.15-1]\, \Rv$ region in units of $10^{44}$ erg s$^{-1}$.
    \end{table*}

\begin{table*}[]
\begin{center}
\caption{{\footnotesize Observed survey band X-ray luminosity scaling relations for the full \rexcess\ sample. For each set of observables ($L,A$), we fitted a power law relation of the form 
$h(z)^n L = C(A/A_0)^{\alpha}$, with $A_0=5$ keV and $2\times10^{14}\,
M_{\odot}$ keV, and $n= -1$, $-9/5$ and $-7/3$ for $T$, $Y_X$ and $M$, respectively. Results are given for the BCES (Y$|$X) and BCES orthogonal fitting methods (see Section~\ref{sec:fitting}). The intrinsic natural logarithmic scatter about the best fitting relation in the ln-ln plane is given in each case. $^a$ Since $M$ is derived from $Y_X$, the values of the scatter in the $L-M$ relation are identical to those for the $L-Y_X$ relation. $^b$ Relations corrected for Malmquist bias.}}\label{tab:lxrelsoft}
\begin{tabular}{r l l l  | l l l }
\hline
\hline

\multicolumn{1}{l}{Relation } &
\multicolumn{6}{c}{Fitting Method} \\

\multicolumn{1}{c}{ } &
\multicolumn{3}{c |}{BCES (Y$|$X)} & \multicolumn{3}{c}{BCES Orthogonal} \\

\cline{2-7}

\multicolumn{1}{c }{ } & \multicolumn{1}{c}{$C\,(10^{44}$ erg s$^{-1}$)} & 
\multicolumn{1}{c}{$\alpha$} & \multicolumn{1}{c |}{$\sigma_{\rm ln\,L, intrinsic}$ } & \multicolumn{1}{c}{$C\,(10^{44}$ erg s$^{-1}$)} & 
\multicolumn{1}{c}{$\alpha$} & \multicolumn{1}{c}{$\sigma_{\rm ln\,L, intrinsic}$ } \\

\hline
\\
\multicolumn{1}{c}{} & \multicolumn{6}{c}{$R < \Rv$} \\
\\
$L\,[0.1-2.4]_1$--$T_1$ & 
$2.86\pm0.27$ & $2.24\pm0.22$ & $0.665\pm0.119$ & 
$3.46\pm0.55$ & $3.00\pm0.35$ & $0.757\pm0.144$ \\ 

$L\,[0.5-2]_1$--$T_1$ & 
$1.77\pm0.17$ & $2.24\pm0.22$ & $0.666\pm0.119$ & 
$2.14\pm0.34$ & $3.01\pm0.35$ & $0.758\pm0.144$\\ 
\\
$L\,[0.1-2.4]_1$--$Y_X$ & 
$2.52\pm0.18$ & $0.84\pm0.05$ & $0.411\pm0.070$ & 
$2.60\pm0.20$ & $0.90\pm0.06$ & $0.412\pm0.071$ \\ 

$L\,[0.5-2]_1$--$Y_X$ & 
$1.56\pm0.11$ & $0.84\pm0.06$ & $0.413\pm0.07$ & 
$1.61\pm0.12$ & $0.90\pm0.06$ & $0.414\pm0.071$ \\ 
\\
$L\,[0.1-2.4]_1$--$M_Y$ & 
$1.03\pm0.08$ & $1.53\pm0.10$ & $^a$\ldots & 
$0.98\pm0.07$ & $1.71\pm0.12$ & $^a$\ldots \\ 

$L\,[0.1-2.4]_1$--$M_Y$ MB$^b$ & 
$0.83\pm0.07$ & $1.62\pm0.11$ & $^a$\ldots & 
$0.78\pm0.07$ & $1.83\pm0.14$ & $^a$\ldots \\ 

$L\,[0.5-2]_1$--$M_Y$ & 
$0.64\pm0.05$ & $1.53\pm0.10$ & $^a$\ldots & 
$0.61\pm0.05$ & $1.72\pm0.12$ & $^a$\ldots \\ 

$L\,[0.5-2]_1$--$M_Y$ MB$^b$ & 
$0.51\pm0.04$ & $1.62\pm0.12$ & $^a$\ldots & 
$0.48\pm0.04$ & $1.83\pm0.14$ & $^a$\ldots \\ 

\\
\hline
\\
\multicolumn{1}{c}{} & \multicolumn{6}{c} {$0.15 < R < \Rv$} \\
\\
$L\,[0.1-2.4]_2$--$T_2$ & 
$1.85\pm0.09$ & $2.32\pm0.13$ & $0.278\pm0.056$ & 
$1.95\pm0.12$ & $2.52\pm0.16$ & $0.293\pm0.062$ \\ 

$L\,[0.5-2]_2$--$T_2$ & 
$1.14\pm0.05$ & $2.34\pm0.13$ & $0.276\pm0.056$ & 
$1.20\pm0.07$ & $2.53\pm0.16$ & $0.291\pm0.062$\\ 
\\
$L\,[0.1-2.4]_2$--$Y_X$ & 
$1.50\pm0.04$ & $0.82\pm0.03$ & $0.174\pm0.044$ & 
$1.51\pm0.04$ & $0.83\pm0.03$ & $0.175\pm0.044$ \\ 

$L\,[0.5-2]_2$--$Y_X$ & 
$0.92\pm0.02$ & $0.82\pm0.03$ & $0.173\pm0.044$ & 
$0.93\pm0.03$ & $0.83\pm0.03$ & $0.174\pm0.044$ \\ 
\\
$L\,[0.1-2.4]_1$--$M_Y$ & 
$0.63\pm0.03$ & $1.49\pm0.05$ & $^a$\ldots & 
$0.62\pm0.03$ & $1.52\pm0.05$ & $^a$\ldots \\ 

$L\,[0.5-2]_1$--$M_Y$ & 
$0.39\pm0.02$ & $1.49\pm0.05$ & $^a$\ldots & 
$0.38\pm0.02$ & $1.53\pm0.05$ & $^a$\ldots \\ 
\\

\hline
\end{tabular}
\end{center}
$L_1/T_1$: luminosity/temperature interior to $\Rv$; \\
$L_2/T_2$: luminosity/temperature in the $[0.15-1]\,\Rv$ aperture;\\
$M_Y$: total mass estimated from the $M_{500}-Y_X$ relation of \citet{app07}.
\end{table*}

\end{document}